\documentclass[preprint, amsmath,amssymb,
 aps]{revtex4-1}

\usepackage{graphicx}
\usepackage{pslatex}
\usepackage{amsmath}
\usepackage{mathtools}
\usepackage{subfigure}
\usepackage{subfig}
\usepackage{wrapfig}
\usepackage{float}
\usepackage{tikz}
\usepackage{setspace}

\usetikzlibrary{positioning}

\ifx\pdftexversion\undefined
\DeclareGraphicsExtensions{.png,.eps,.ps,.eps.gz,.ps.gz}
\else
\usepackage{epstopdf}
\fi

\def\beq{\begin{equation}}
\def\eeq#1{\ifx\string#1\string \else\label{#1}\fi\end{equation}}

\newcommand{\nuhat}{{\hat{\nu}}}

\newcommand{\Omegabar}{\bar{\Omega}}

\usepackage{xspace}
\usepackage{bm}
\newcommand{\be}{\ensuremath{{\beta}}\xspace}

\newcommand{\mat}[1]{{\ensuremath{\bf{ #1}}}}
\definecolor{OliveGreen}{rgb}{0,0.6,0}

\begin{document}
 \title{Machine Learning-augmented Predictive Modeling of Turbulent Separated Flows over Airfoils}
 \author{Anand Pratap Singh}\affiliation{PhD Candidate, Department of Aerospace Engineering, University of Michigan, Ann Arbor, MI 48104.}
  \author{Shivaji Medida} \affiliation{Solver Development Manager, AcuSolve, Altair Engineering, Inc., Sunnyvale, CA 94086 } 
  \author{Karthik Duraisamy} \affiliation{Assistant Professor, Department of Aerospace Engineering, University of Michigan, Ann Arbor, MI 48104.}

\begin{abstract}
A modeling paradigm is developed to augment predictive models of turbulence by effectively utilizing {\em limited data} generated from {\em physical experiments}.
The key components of our approach involve {\em inverse modeling} to infer the spatial distribution of model discrepancies, 
and, {\em machine learning} to reconstruct discrepancy information from a large number of inverse problems  into corrective model forms.  We apply the methodology to 
 turbulent flows over airfoils involving flow separation. Model augmentations are developed for the Spalart Allmaras (SA) model using adjoint-based full field inference on experimentally measured lift coefficient data. When these model forms are reconstructed using neural networks (NN) and embedded within a standard solver, we show that much improved predictions in lift can be obtained for geometries and flow conditions that were not used to train the model. The NN-augmented SA model also predicts surface pressures extremely well. Portability of this approach is demonstrated by confirming that predictive improvements are preserved when the augmentation is embedded in a different commercial finite-element solver. The broader vision is that by incorporating data that can reveal the form of the innate model discrepancy, the applicability of data-driven turbulence models can be extended to more general flows.
\end{abstract}
\maketitle

\section{Introduction}
Accurate modeling and simulation of turbulent flows is critical to several applications in engineering and physics. 
From the viewpoint of affordability, turbulence closure models --
either in  Reynolds Averaged Navier--Stokes (RANS) form or
in a near-wall context in an eddy-resolving model -- will continue to be indispensable for the 
foreseeable future~\cite{Slotnick}. Existing turbulence closures have proven to be 
 quite useful in many contexts, but it is well-recognized that  complex effects such as flow separation, secondary flows, etc are poorly modeled. 

 While new and increasingly complex models are being developed~\cite{gerolymos,gatski,poroseva2014velocity} 
 and demonstrated to be accurate in some problems, 
 it can be argued that there has not been a significant improvement in predictive 
 accuracy over the past 15 years.
 As a result, the majority of the RANS models that are used in
 both industrial and academic CFD solvers were initially developed and published in the 1990's. 
A critical issue in turbulence model development is that even the most sophisticated model 
invokes radically simplifying assumptions about 
the structure of the underlying turbulence. Thus, the process of developing a practical
turbulence model combines physical intuition, empiricism and engineering judgment, 
while constrained by robustness and cost considerations. As a result, even if a model is
based on a physically and mathematically appealing idea -- for example, elliptic relaxation~\cite{ell1,ell2} 
 -- the model formulation typically devolves into the calibration of 
 a large number of free parameters or functions using a small set of canonical problems.

Against this scenario, our ability to perform detailed high-fidelity 
computations and resolved measurements has improved dramatically over the past decade. At the 
same time, data science is on the rise because of improvements in computational power and the increased availability of large data sets. 
 This has been accompanied by significant improvements in the effectiveness and scalability of data analytics and machine learning (ML) techniques.
Given these advances, we believe that data-driven modeling and machine learning will play a critical role
in improving the understanding and modeling of turbulence.

In the study of turbulent flows, machine learning techniques appear to have first been used
 to recreate the behavior of near-wall structures in a turbulent channel flow~\cite{Milano2002} 
and to extract coherent spatio-temporal structures\cite{Marusic2001}. With a view towards quantifying model errors, several researchers~\cite{Yarlanki2012, japanese, bayes1, 
arunjatesan} have used experimental data to infer model parameters. 
Cheung et al.~\cite{cheung2011,oliver_moser} employ Bayesian model averaging\cite{beck} to calibrate model coefficients. Edeling et al.~\cite{bayes1} use 
statistical inference on skin-friction and velocity data from a number of boundary layer experiments to quantify parametric model error. 
 These methods provide insight into parametric uncertainties and address some of the deficiencies of {\em a priori} 
processing of data. 

Dow and Wang~\cite{wang,wang2} made progress towards addressing non-parametric uncertainties 
by inferring the spatial structure of the discrepancy in the eddy viscosity coefficient based on 
a library of direct numerical simulation (DNS) datasets. The discrepancy  between the inferred 
and modeled eddy viscosity was represented as a Gaussian random field and propagated to obtain
uncertainty bounds on the mean flow velocities. 

The research group of Iaccarino~\cite{memory1,gorle1,memory2} introduced adhoc, but realizable perturbations to the non-dimensional Reynolds 
stress anisotropy tensor $a_{ij}$ to quantify structural errors in eddy viscosity models.
Tracey et al.~\cite{tracey1} applied 
neural networks to large eddy simulation data to learn the functional form of the discrepancy in $a_{ij}$ and injected these functional forms in a predictive simulation in an attempt to obtain improved predictions. Xiao and co-workers~\cite{xiao1}  inferred the spatial distribution of the perturbations in $a_{ij}$ and turbulent kinetic energy by assimilating DNS data. Weatheritt~\cite{weatheritt} uses evolutionary algorithms on DNS data  to  construct  non-linear  stress-strain  relationships for RANS models.

Ling and Templeton~\cite{ling} used machine learning-based classifiers to ascertain regions of the flow in which
commonly-used assumptions break down. King et al.~\cite{king} formulated a damped least squares problem at the test-filter scale to obtain coefficients
of a subgrid-scale model. In both of these works, results were demonstrated in an apriori setting.

Duraisamy and co-workers~\cite{duraisamy2015new,companion,paradigm,PoF_anand} took the first steps towards improving
  predictive model forms by defining a data-driven modeling paradigm based on field inversion and machine learning (FIML).
 The FIML approach consists of three key steps : a) Inferring the spatial (non-parametric) distribution
 of the model discrepancy in a number of problems using Bayesian inversion, b) Transforming the 
spatial distribution into a functional form (of model variables) 
using machine learning, and c) Embedding the functional form in a predictive setting. 
Predictions were demonstrated in turbulent channel flows and transitional flows with imposed
pressure gradients. Note that steps a) and b) involve off-line (training) computations, whereas step c) is on-line (prediction). Xiao and co-workers~\cite{xiao2} also use inference and machine learning on DNS data to reconstruct the functional form of the discrepancy in $a_{ij}$. This function is injected 
as a one time post-processing step to a computed baseline solution. The flow solution is then updated based on the Reynolds stress perturbation. 

In this work, we extend the paradigm of data-driven modeling to assist in the development of turbulence models and predictive simulation of turbulent flow over airfoils. In particular, we
demonstrate the ability of inverse modeling to provide quantitative modeling information based on very limited experimental data, and the use of machine learning to reconstruct this information into corrective model forms. When these model forms are embedded within a standard solver setting, it is shown that significantly improved predictions can be achieved.


 \begin{figure*}[t]
\centering
\includegraphics[width=0.4\textwidth]{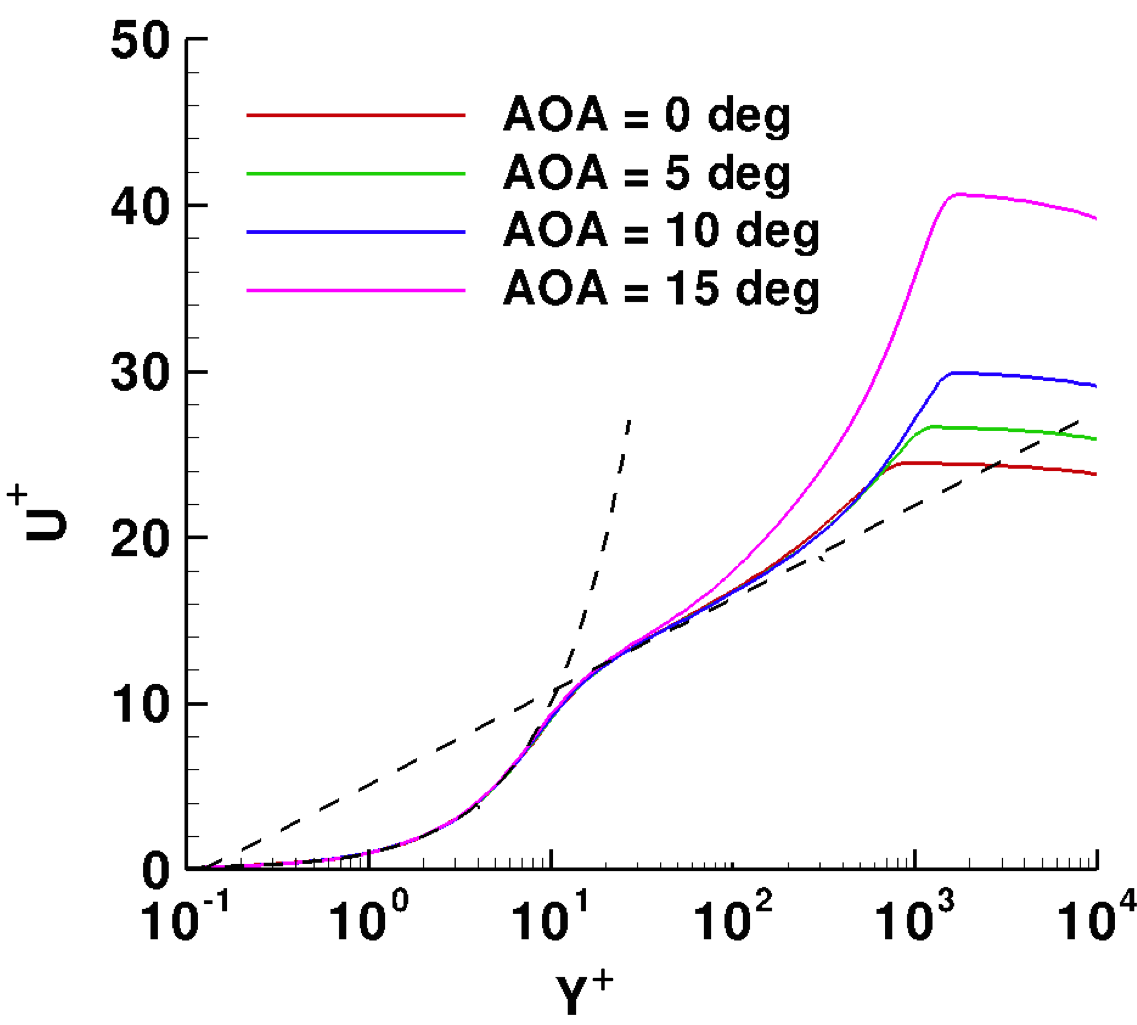}
\caption{Effect of adverse pressure gradient on the defect layer (NACA 0012 airfoil, Re = 6$\times$10$^6$). Boundary layer corresponds to the 16\% chord location on the upper surface.}
\label{uplus_yplus}
\end{figure*}

\section{Problem and Approach}

Turbulent flow separation over lifting surfaces is critical to many applications, including  high-lift systems, off-design operating envelope of new vehicles,  airframe noise, wind turbines, turbomachinery flows, and combustors. A RANS turbulence modeling capability that can confidently predict separated flows in these various contexts would be a key enabling factor in the development of aerospace and energy systems of the future. The ability  to accurately model the effects of strong adverse pressure gradients (APG) is crucial to the prediction of boundary layer separation in wall-bounded flows; however, most one- and two-equation RANS turbulence models fail to accurately predict stall onset for airfoils at high angles of attack (AoA), where strong APG is encountered. Consequently, they tend to over-predict the maximum lift and stall onset angle for a given set of flow conditions. 

Celic et al.~\cite{celic2006} compared the performance of 11 eddy-viscosity based turbulence models for aerodynamic flows with APG and concluded that none of the models perform satisfactorily for flow past airfoils near maximum lift conditions. This deficiency  can be attributed to the underlying assumptions and simplifications that are part of all eddy-viscosity based turbulence models. One such assumption implies a balance between the production and dissipation of turbulent kinetic energy. This assumption allows for the scaling of velocity profiles in the defect layer, and is instrumental in the formulation of many turbulence models. However, it is well known that practical boundary layers under strong APG are not in equilibrium.
In addition, the outer layer scaling is affected by APG, whereas the viscous sublayer and log-layer are relatively unchanged. This is illustrated in Fig. \ref{uplus_yplus} through the  velocity profiles (in wall units) on a NACA 0012 airfoil at four different angles of attack. As the  angle of attack increases, the adverse pressure gradient  on the upper surface grows steeply, and therefore, the defect layer penetrates deeper into the boundary layer. Thus, many turbulence models that assume equilibrium conditions fail to produce satisfactory behavior for strong APG flows. Certain models with stress limiters, such as the SST version of the k-$\omega$ turbulence model \cite{menter1994}, Wilcox's modified k-$\omega$~\cite{wilcox2008} model, and the strain-adaptive formulation of the Spalart-Allmaras turbulence model~\cite{rung2003, spalart}, are known to perform slightly better than the other models. Although using more sophisticated turbulence models (non-linear eddy viscosity models, second moment closure models, etc) might produce better results, poor robustness and higher computational cost associated with the usage of these methods are  major deterrents to their wider applicability for practical flows. These models are still calibrated using information from canonical configurations and applied in situations dissimilar to those in which calibrations were made. In the present work, more realistic flows are used to guide model development.
 
  \begin{figure}
\centering
\includegraphics[width=0.8\textwidth]{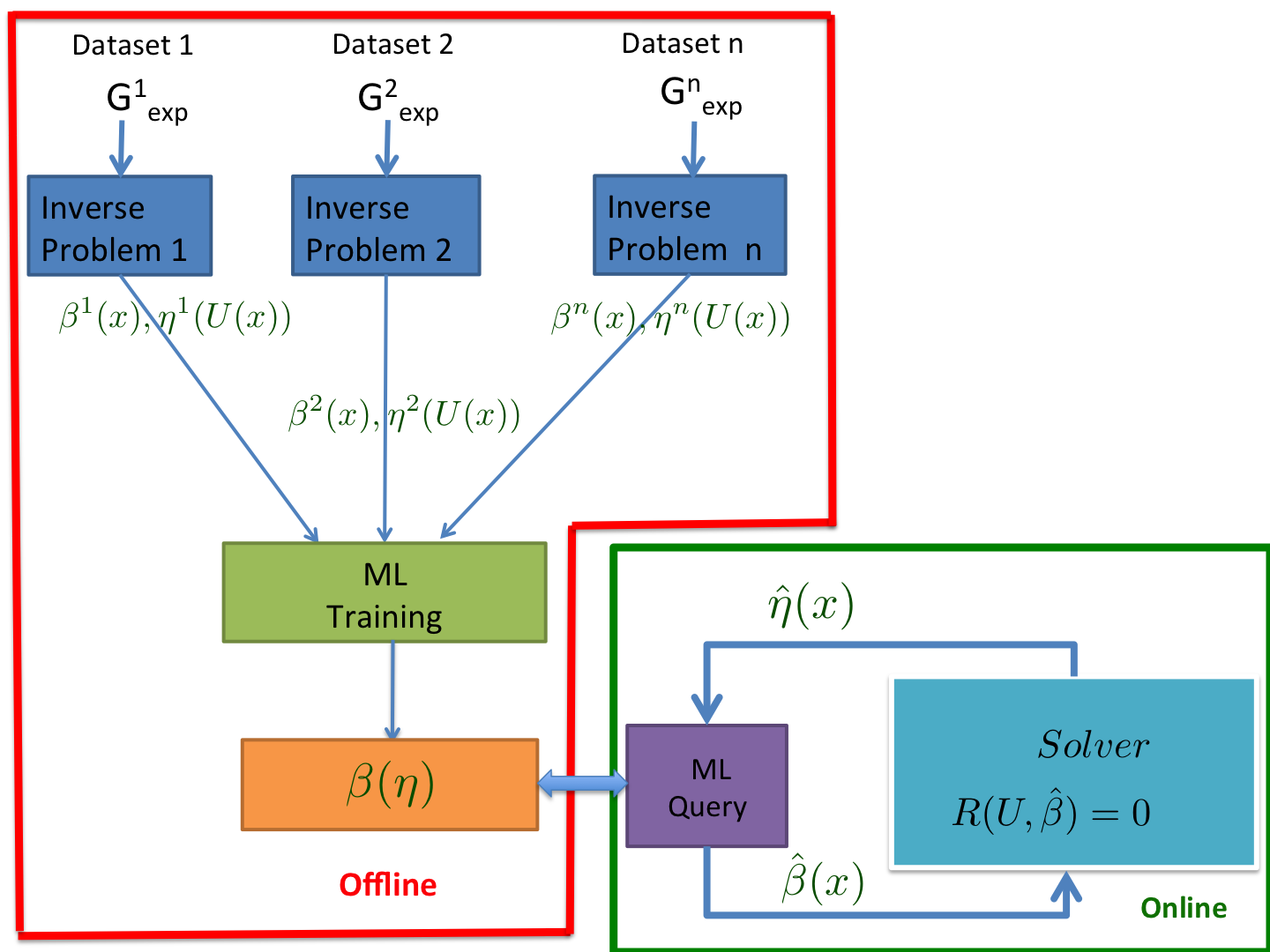}
\caption{Schematic of field inversion and machine learning framework for data-augmented turbulence modeling}
\label{fig:schematic}
\end{figure}

As proof-of-concept for the feasibility of data-assisted modeling, Tracey et al.~\cite{companion}
 applied machine learning to a database of {\em solutions} of a {\em known} turbulence model. The known turbulence model was considered to be the surrogate truth. These solutions primarily involved flat plates and airfoils. A deficient turbulent model (with deliberately 
  removed source terms) was then augmented with these machine-learned functional forms. This augmented turbulence model was able to accurately reproduce radically different flows such as transonic flow over a wing. While it was relatively easy to make  apriori (and one-time) evaluations of the trained model,
 key lessons were learnt about the formulation of the learning problem as the NN (Artificial Neural Network) had to be evaluated and injected during every iteration of a converging PDE solver.
 
 The above work demonstrated that if the underlying model form is discoverable and the data is comprehensive enough, a machine learning technique such as an artificial neural network can adequately describe it. The challenge in predictive modeling, however, is to extract an {\em optimal} model form that is sufficiently accurate. Constructing such a model and demonstrating its predictive capabilities for a class of problems is the objective of this work. 
  This data-driven framework is specifically demonstrated in predictions of turbulent, separated flows over airfoils.

A schematic of the approach is provided in Fig.~\ref{fig:schematic}. Various aspects of the schematic are organized in the paper as follows: Section III introduces the inversion framework which uses limited experimental data $G_{exp}$ to generate fields of modeling information $\be(\mat{x})$ that account for the model discrepancy. Section IV introduces the role of machine learning in transforming information from a number of inverse problems $\be^j(\mat{x})$ into model forms $\be(\boldsymbol{\eta})$, where $\boldsymbol{\eta}$ represents local field variables available in the model. Section V demonstrates that embedding model corrections $\hat{\be}$ during the simulation process can improve  predictive capabilities.  Section VI presents a summary of this work and perspectives on the extension of these techniques to general turbulence modeling.

\subsection*{Discretization}

The flow solver~\cite{duraisamy_turns1,duraisamy_turns2,vinod_turns} (ADTURNS) is based on a
cell-centered finite volume formulation of the compressible RANS equations on structured grids. The inviscid 
fluxes are discretized using the third-order MUSCL scheme~\cite{muscl} in
combination with the approximate Riemann solver of Roe~\cite{roe}. The diffusive contributions 
are evaluated using a second-order accurate central differencing scheme.
Implicit operators are
constructed using the diagonalized alternating direction implicit (D-ADI)
scheme\cite{pulliam}.

For the computations, the flow domain over airfoils is discretized using a C-grid with 291 points in the wrap–around direction and 111
points in the wall-normal direction. At this resolution, which corresponds to 200 grid points on the airfoil
surface, numerical errors are low enough  to not obscure the treatment of turbulence modeling errors. This was verified by performing a grid convergence study. The far–field boundaries are located 35 chord lengths from the airfoil surface. Characteristic freestream boundary conditions are used for the flow variables at the far–field and the eddy
viscosity is set to the fully turbulent value, $\nu_{t,\infty}/\nu_\infty = 3$.

The field inversion procedure requires gradients
with respect to every grid point. These gradients are most effectively determined using a discrete adjoint approach~\cite{giles}. The required derivatives are computed as detailed in the Appendix A.

\section{Field Inversion}
Our philosophy of inferring and reconstructing one or more corrective functional forms is general in scope with regard to data-driven modeling~\cite{PoF_anand,paradigm}. While the methodology is applicable to both eddy viscosity and Reynolds stress models, the focus of the present work is 
restricted to the Spalart-Allmaras (SA) model~\cite{spalart} (refer to Appendix C for detailed formulation). 
The baseline SA model can be written as
\begin{eqnarray}
        \frac{D\tilde{\nu}}{Dt} = P(\tilde{\nu}, \mathbf{{U}}) - D(\tilde{\nu}, \mathbf{{U}}) + T(\tilde{\nu}, \mathbf{{U}}),    \label{eq:turb_base}
\end{eqnarray}
where $\mathbf{{U}}$ represents the Reynolds averaged conserved flow variables, $P(\tilde{\nu}, \mathbf{{U}})$, $D(\tilde{\nu}, \mathbf{{U}})$, and $T(\tilde{\nu}, \mathbf{{U}})$ represent the production, destruction and transport terms respectively.
The above equation is used with a non-linear functional relationship to derive an eddy viscosity $\nu_t$ from $\tilde{\nu}$, which
is then used in a Boussinesq formulation to close the RANS equations.
The major source of modeling deficiency is the structural form of the model rather than parameters within the imposed model form. Thus, benefits  from classical parameter estimation will be limited. In other words, {\bf the functional forms of the terms in Eq.~\ref{eq:turb_base} are themselves inaccurate}, and require a reformulation.

The goal then, is to construct generalizable functional corrections to the model form in Eq.~\ref{eq:turb_base}. Accordingly, a spatially-varying term $\be(\mathbf{x})$ is introduced as a multiplier of the production term $P(\tilde{\nu}, \mathbf{{U}})$. 
\begin{eqnarray}
        \frac{D\tilde{\nu}}{Dt} = \be(\mathbf{x}) P(\tilde{\nu}, \mathbf{{U}}) - D(\tilde{\nu}, \mathbf{{U}}) + T(\tilde{\nu}, \mathbf{{U}}),    \label{eq:turb}
\end{eqnarray}

It must be recognized that the introduction of $\be(\mat{x})$ changes the entire balance of the model, (and need not  be interpreted as merely a modification of the production term). It is equivalent to adding a source term $\mat{\delta(\mat{x})} = (\be(\mat{x}) - 1)P(\mat{x})$. Inferring $\be$, however, leads to a better conditioned inverse problem, as $\be$ is non-dimensional and has a
simple initial value of unity. 

Assume a flow configuration (with a particular geometry, angle of attack, Reynolds number, etc)  consisting of $N_m$ control volumes. Given 
$N_{d}$ data points (such as wall pressure, skin-friction, etc) ${G}_{j,exp}$, we define the following inverse problem to extract the optimal field $\be \equiv \be(x_n) : 1 \le n \le N_m$:
\beq
 \min_\be \sum_{j=1}^{N_d} [{G}_{j,exp} - {G_j(\be)}]^2 + \lambda \sum_{n=1}^{N_m} [\be(x_n)-1]^2,
\eeq
where $G_j(\be)$ is the output of the RANS model.
 This inverse problem is most straightforwardly interpreted in a classical frequentist sense with Tikhonov regularization~\cite{tikhonov}, or loosely as the maximum a posteriori (MAP) estimate in a Bayesian setting assuming Gaussian distributions and a prior of unity. In the former setting, $\lambda$ is a regularization constant;  in the latter, it represents the ratio of the observational covariance to the prior covariance~\footnote{Assuming that the covariance matrices are Gaussian and diagonal}. It is to be noted that in the context of this work, the solution of a large number of inverse problems is used as a {\em means} to define corrective functions. Thus, finer-grained interpretations or formulations of the inverse problem and treatment of uncertainties - while important - are not of a primary concern in this work. A more formal treatment of observational errors and prior confidence has been pursued in previous work~\cite{paradigm}, but application was restricted to simpler problems.
 
 Nevertheless, an optimal value of $\be$ is  sought at every discrete location in the computational
domain and used in Eq.~\ref{eq:turb}, conjoined with
the conservation equations for the ensemble-averaged mass, momentum and energy.
 The resulting inverse problem is extremely high-dimensional
and  an efficient adjoint-based optimization framework is employed. For further details, please refer the Appendix A.



If experimental surface pressure coefficients $C_p$ were used as data points\cite{PoF_anand}, the  following minimization problem is formulated
\begin{equation}
 \min_\be \left[ \sum_{j=1}^{N_d} [C_{p_{j,exp}} - C_{p_j}(\be)]^2 + \lambda \sum_{n=1}^{N_m}  [\be(x_n)-1]^2 \right].
\label{eq:obj_cp}
\end{equation}

However, in the majority of experimental tests of flow over airfoils, the surface pressure is not measured. Therefore, we use the lift coefficient ($C_l$) as the observational data. Thus, the following optimization problem is formulated:

\begin{equation}
 \min_\be  \left[ \left[C_{l,exp} - C_l(\be)\right]^2 + \lambda \sum_{n=1}^{N_m} [\be(x_n)-1]^2 \right].
 \label{eq:obj_cl}
 \end{equation}

\begin{figure}[!h]
    \centering
        \subfigure[$\be(\mat{x})$ field using objective function based on $C_l$]{\includegraphics[width=0.45\textwidth]{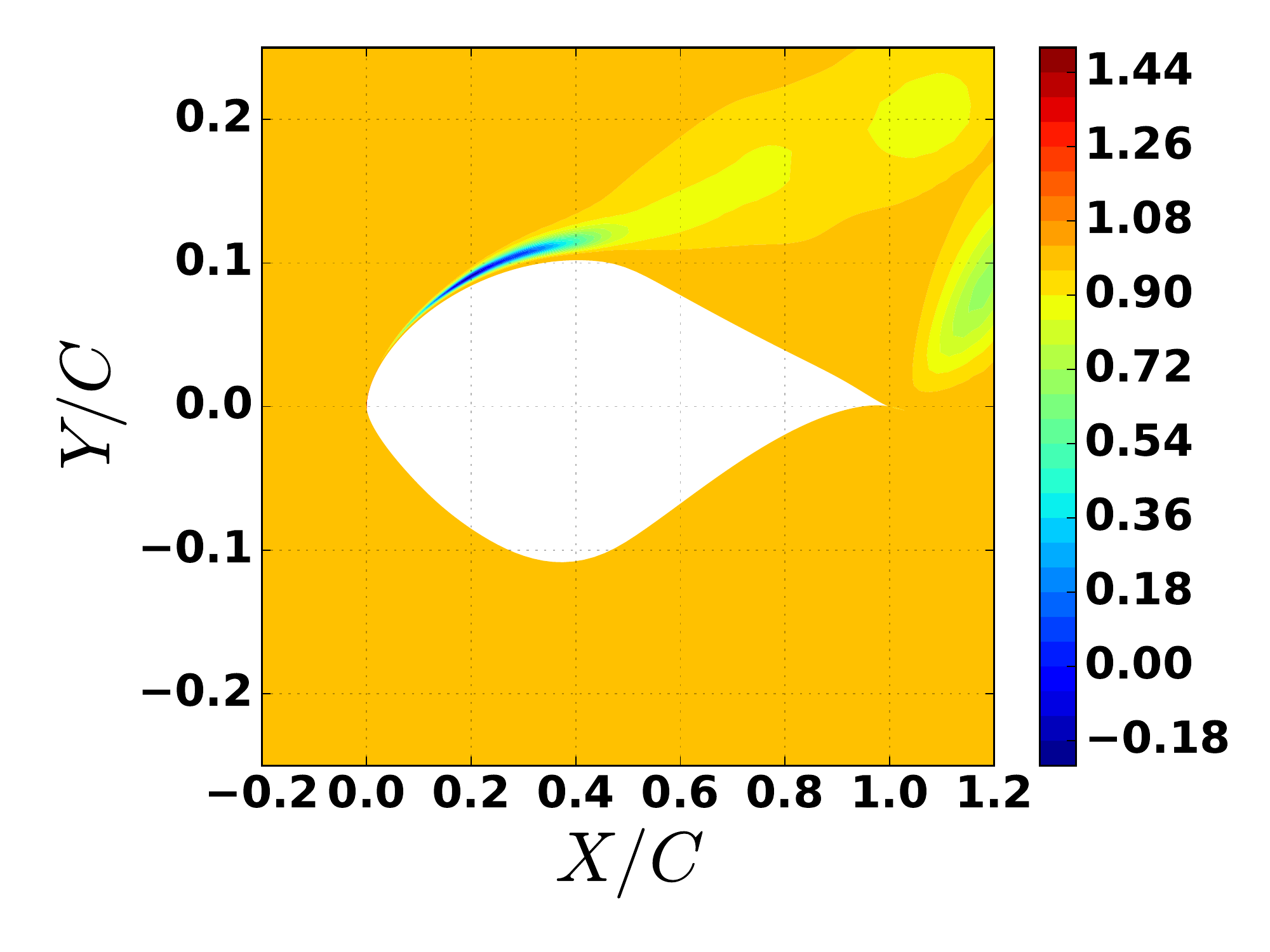}}
    \subfigure[$\be(\mat{x})$ field using objective function based on $C_p$]{\includegraphics[width=0.45\textwidth]{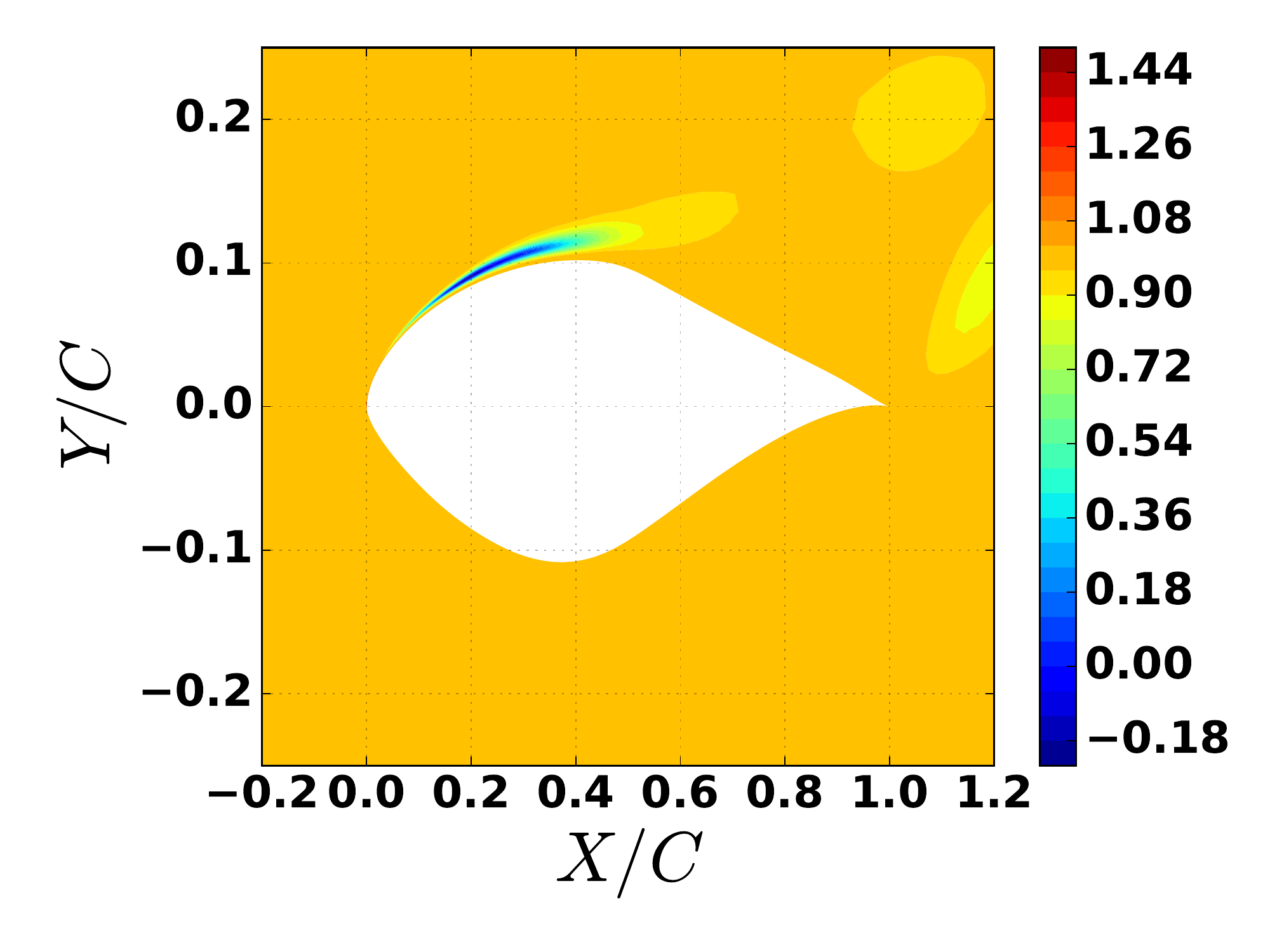}}
    \subfigure[$C_p$]{\includegraphics[width=0.45\textwidth]{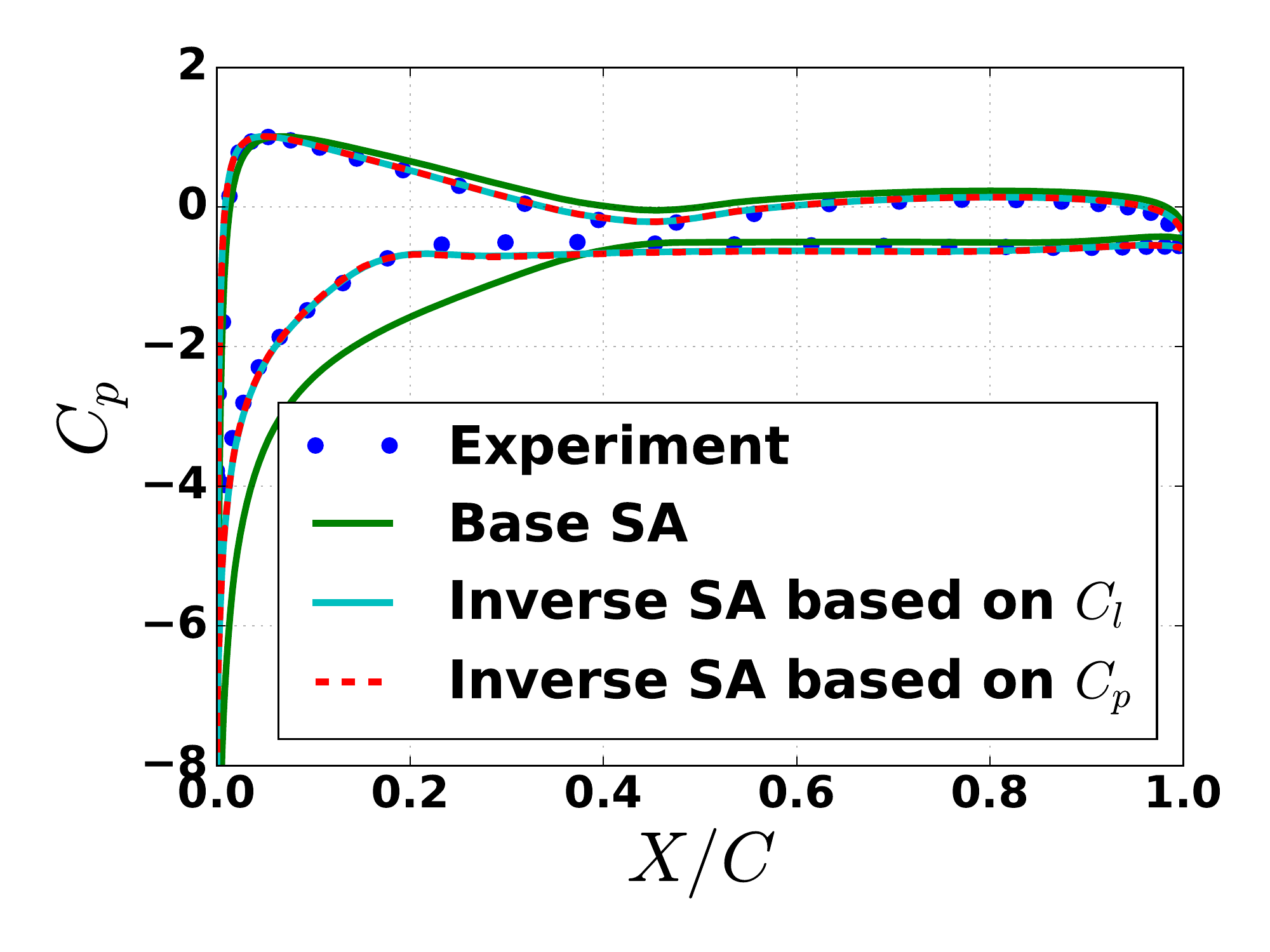}}
    \caption{Inverse solutions using objective function based on lift ($C_l$) and surface pressure ($C_p$) coefficients. $\be(\mat{x})$ in the near wall region is unaffected by the choice of objective function resulting in identical inverse $C_p$.}
    \label{ComparisonCLCP}
\end{figure}

The two objective functions (Eq. \ref{eq:obj_cp} and \ref{eq:obj_cl}) were confirmed to lead to a  similar solution to the inverse problem (Fig. \ref{ComparisonCLCP}). While there are discrepancies in the post-stall region, the near-wall features in $\be(\mat{x})$ are almost identical, resulting in indistinguishable surface pressures. The entire set of inverse problems in this work is solved for the lift-based objective function (Eq. \ref{eq:obj_cl}) with $\lambda = 4 \times 10^{-4}$. This implies a much higher level of confidence in the experimentally measured lift compared to the variability of $\be$. The optimal solution was indeed confirmed to be insensitive to order of magnitude variations in $\lambda$. 

To further probe the validity of using pressure-based information for field inversion, Appendix B presents an example in which the Reynolds stress field is available. Additional information on the characteristics of the inversion procedure can be found in Ref.~\citenum{PoF_anand}.

In section V, the ability of the lift-based model correction to  accurately predict surface pressures will be further demonstrated. The ability to utilize only the lift coefficient to generate modeling information greatly enhances the applicability of the current framework to assimilate a vast amount of available data.

\begin{figure}
\centering
\includegraphics[width=0.45\textwidth]{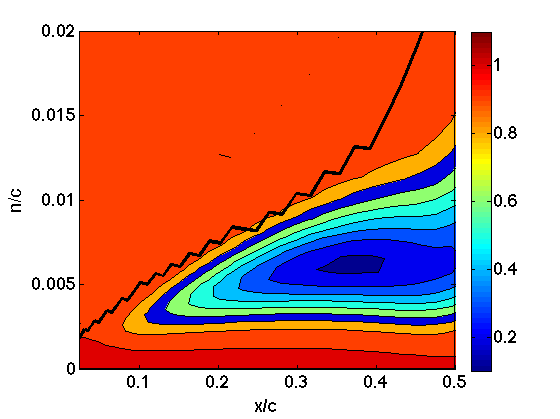}
\caption{The inferred correction function, $\be_{inverse}$, for a representative airfoil. An approximate estimate of the edge of the boundary layer is shown as black lines. $n/c$ is the normalized distance from the airfoil surface.}
\label{fig:oned:beta}
\end{figure}

\begin{figure*}
\centering
\subfigure[Skin friction coefficient]{\includegraphics[width=0.48\textwidth]{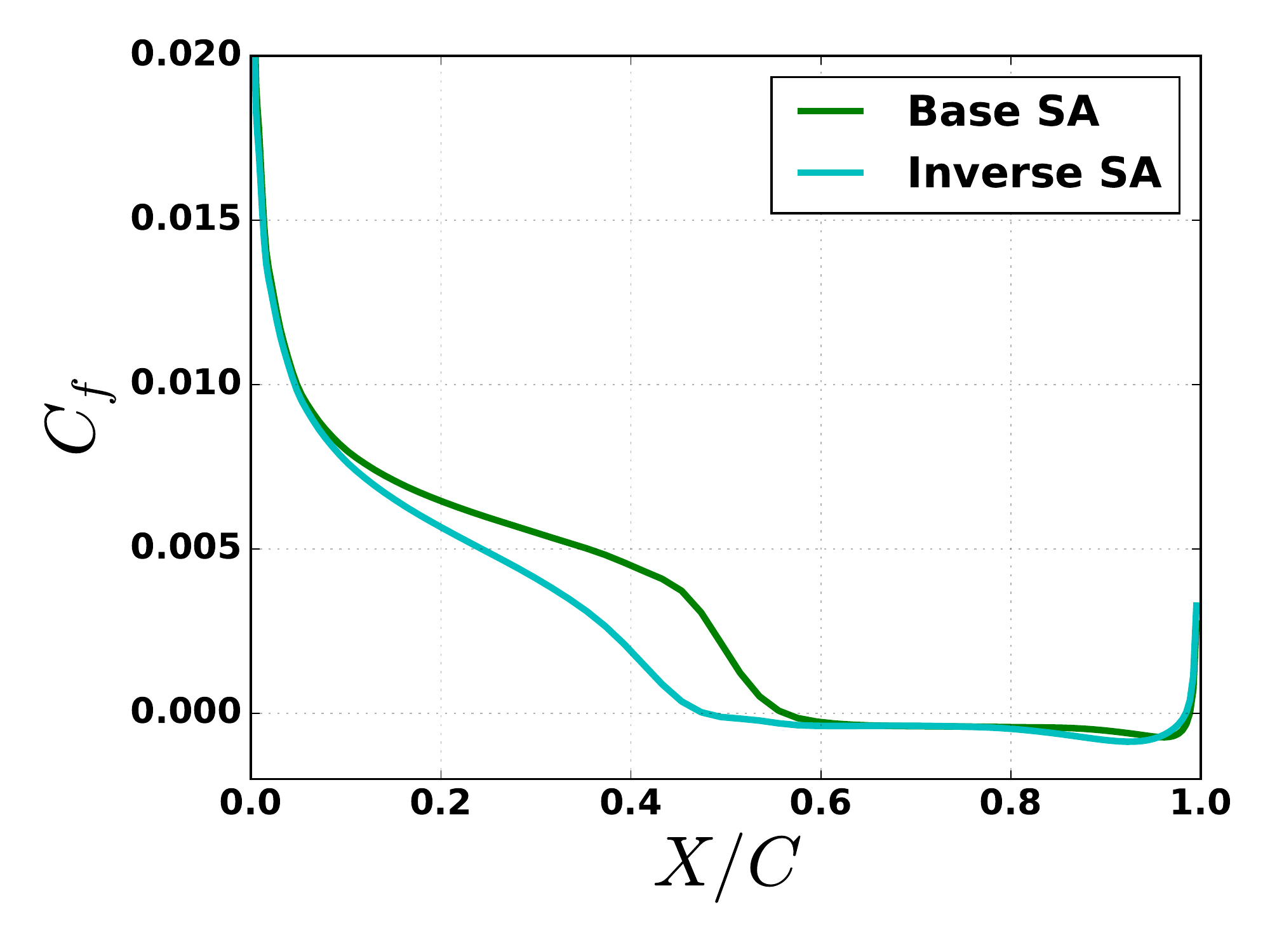}}
\subfigure[Pressure gradient parameter]{\includegraphics[width=0.48\textwidth]{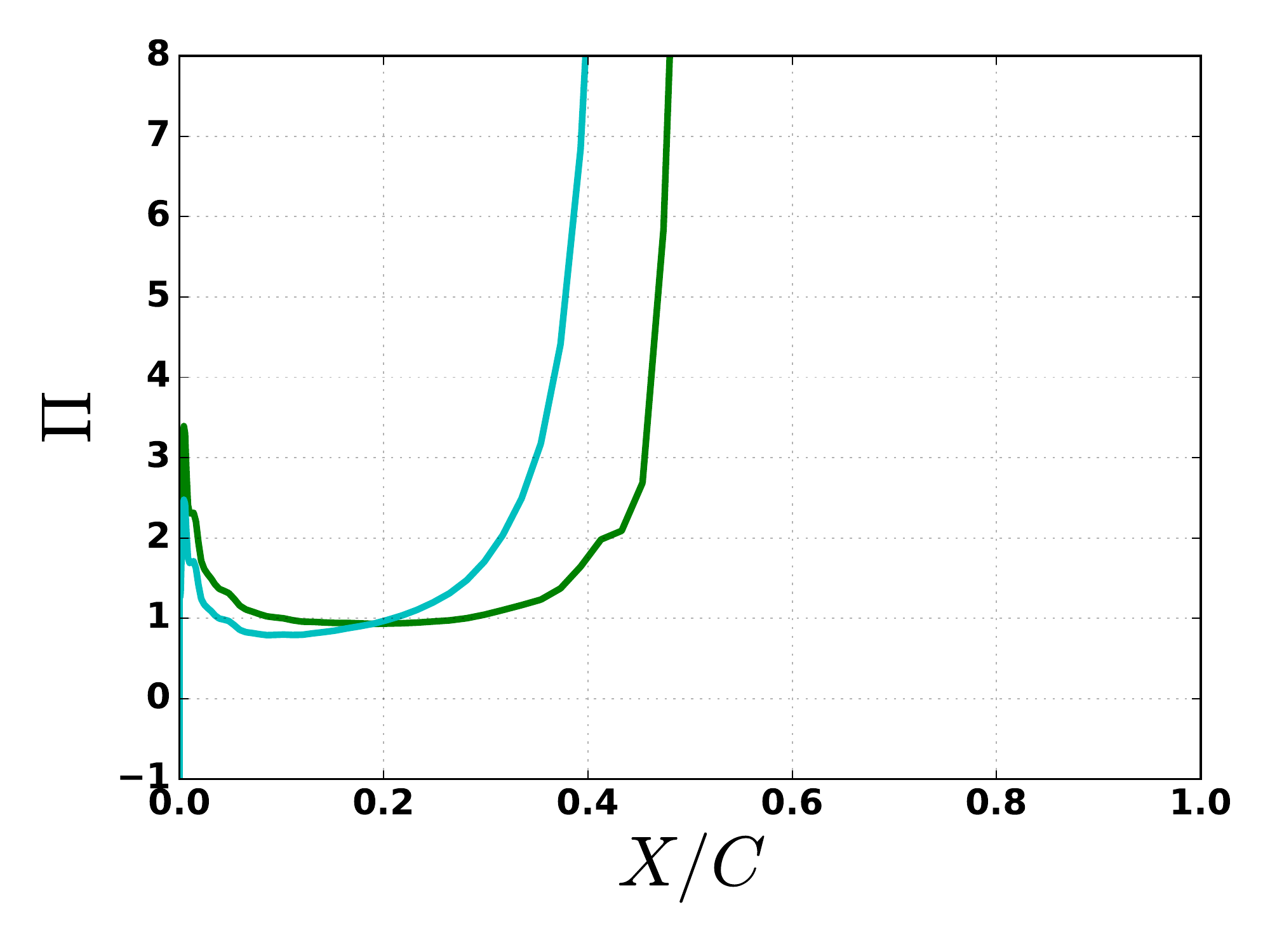}}
\caption{Prior and posterior quantities for the case in Fig \ref{fig:oned:beta}.} 
\label{fig:809CpCf}
\end{figure*}

The inverse solution serves as an input to the machine learning algorithm, while providing qualitative and quantitative insight to the modeler. It is known that eddy viscosity-based turbulence models generate very high levels of turbulence at high angles of attack resulting in delayed separation and stall~\cite{rumsey2002prediction}. The inverse solution adjusts for this deficiency by reducing the generation of turbulence in the near wall pre--separation region, {\em i.e.} the $\be(\mat{x}) < 1$ region in Fig. \ref{fig:oned:beta}. This reduced production results in early flow separation, which can be observed in the wall shear stress (Fig. \ref{fig:809CpCf}a). Furthermore, Fig.~\ref{fig:809CpCf}b  reinforces the fact that a complex relationship exists between the model corrections and the pressure gradient parameter~\footnote{$\delta^\ast$ is the displacement thickness of the boundary layer and $\frac{dP}{ds}$ is the pressure gradient.} $\Pi = \frac{\delta^*}{\tau_w}\frac{dP}{ds}$.

\section{Machine learning}
The inverse approach presented in the previous section results in an optimal correction field for a given flow condition and geometry. To be useful in {\em predictive modeling}, the problem-specific information encoded in $\beta(\mathbf{x})$ must be transformed into modeling knowledge~\cite{paradigm}. This is done by extracting the functional relationship $\beta(\mathbf{x}) \approx \beta(\boldmath{\eta})$, 
where $\boldmath{\eta} = [\eta_1,~\eta_2,~\cdots,~\eta_M]^T$ are input features derived from mean-field variables
 that will be available {\em during} the predictive solution process. 
 The functional relationship must be developed by considering the output of a number of inverse problems representative of the modeling deficiencies relevant to the predictive problem. Further, as explained below, elements of the feature vector $\boldmath{\eta}$ are chosen to be locally non-dimensional quantities such that the functional
relationship $\beta(\mathbf{\eta})$ is useful for different problems in which the $\boldmath \eta$ variables are realizable.

\subsection{Features}
To build a set of features $\boldmath{\eta}$ upon which the functional relationship $\beta(\boldmath{\eta})$ will be based, a logical place to start would be to identify the independent variables in the baseline SA model. 
The source terms in the SA model are  a function of four local flow quantities, $\nu$, $\nuhat$, $\Omega$, $d$, which represent the kinematic viscosity,
the SA working variable, the vorticity magnitude, and the distance from the wall, respectively. As 
discussed in Ref.~\citenum{companion}, these quantities do not constitute an appropriate choice for the input feature vector to the machine learning algorithm. They are dimensional quantities which may have different numeric values even when two flows are dynamically similar. Thus, the inputs are re-scaled~\cite{companion} by relevant {\em local quantities} that are representative of the state of turbulence. An obvious locally non-dimensional quantity in the baseline SA model is $\chi=\nuhat/\nu$. We define local scales, $\nu + \nuhat$ and $d$, and introduce an additional variable,
\begin{align}
\Omegabar =  \frac{d^2}{\nuhat + \nu} \Omega \enskip \text.
\end{align}

With these definitions, the non-dimensional versions ($\overline{P}, \overline{D}$) of the existing production and destruction terms ($P, D$) are given by:
\begin{align}
\nonumber \overline{P} &= \frac{d^2}{(\nuhat + \nu)^2}  s_p 
 = c_{b1} (1 - f_{t2})\bigg(\frac{\chi}{\chi + 1}\bigg) \bigg(\Omegabar + \frac{1}{\kappa^2} \frac{\chi}{\chi + 1} f_{t2}\bigg) \enskip \text{,}  \\
\nonumber \overline{D} &= \frac{d^2}{(\nuhat + \nu)^2} s_d = \bigg(\frac{\chi}{\chi + 1}\bigg)^2 c_{w1} f_w \enskip \text{,}
\end{align}
where $c_{b1}, c_{w1}$ are constants, $f_{t2}$ is a function of $\chi$ and $f_w$ 
is a function of $\Omegabar$ and $\chi$. Thus, the locally non-dimensionalized source terms in the baseline
SA model are dependent only on $\Omegabar$ and $\chi$. 


The set of features that were evaluated includes $\{\Omegabar, \chi, S/\Omega, \tau/\tau_{wall}, P/D$\}, where $S,\tau,\tau_{wall}$ represent the strain-rate magnitude, magnitude of the Reynolds stress, and the wall shear stress, respectively.


\subsection{Neural Networks}

In previous work, we have experimented with supervised learning techniques~\cite{mlpaper1} including single/multi-scale Gaussian process regression~\cite{helen_arxiv} and Artificial Neural Networks (NN)~\cite{mlbook}. In this work, we pursue NNs because of their efficiency as they can be evaluated at a computational cost that is independent of the size of the training data~\footnote{We also appreciate that other techniques such as support vector and polynomial regressors can be as scalable as NNs and thus, the choice of NNs is based on prior experience rather than on objective considerations.}
 The performance metric used in the current work for input selection is the sum squared error (SSE) 
on the validation set.

\begin{figure}[htb]
\begin{center}
\begin{tikzpicture}[x=1.5cm, y=0.8cm, >=stealth]

\tikzstyle{every pin edge}=[<-,shorten <=1pt]
\tikzstyle{neuron}=[circle,draw=black,minimum size=17pt]
\tikzstyle{input neuron}=[neuron, draw=green];
\tikzstyle{output neuron}=[neuron, draw=red];
\tikzstyle{hidden neuron}=[neuron, draw=blue];
\tikzstyle{neuron missing}=[neuron, draw=none,scale=2, text height=8pt,execute at begin node=\color{black}$\vdots$]
\tikzstyle{annot} = [text width=4em, text centered]

\foreach \m/\l [count=\y] in {1,2,3}
  \node [input neuron] (input-\m) at (0,0.5-\y) {};

\foreach \m/\l  [count=\y] in {1,2,3,missing,4}
  \node [hidden neuron, neuron \m/.try] (hidden-\m) at (2,1.5-\y) {};
  
\foreach \m [count=\y] in {1,2,3,missing,4}
  \node [hidden neuron, neuron \m/.try] (hidden2-\m) at (4,1.5-\y) {};

  \node [output neuron] (output-1) at (6,-1.5) {};

  \draw [<-] (input-1) -- ++(-1,0)
    node [above, midway] {$\eta_1$};
  \draw [<-] (input-2) -- ++(-1,0)
    node [above, midway] {$\eta_2$};
  \draw [<-] (input-3) -- ++(-1,0)
    node [above, midway] {$\eta_3$};

\foreach \l [count=\i] in {1}
  \draw [->] (output-\i) -- ++(1,0)
	node [above, midway] {$y$};
    
\foreach \i in {1,2,3}
  \foreach \j in {1,...,4}
    \draw [->] (input-\i) -- (hidden-\j);
    
\foreach \i in {1,...,4}
  \foreach \j in {1,...,4}
    \draw [->] (hidden-\i) -- (hidden2-\j);
    
\foreach \j in {1,...,4}
    \draw [->] (hidden2-\j) -- (output-1);
    
\node[annot,above of=hidden-1, node distance=0.5cm] (hl) {$z_{1,1}$};
\node[annot,below of=hidden-4, node distance=0.5cm] (hl) {$z_{1,H_1}$};
\node[annot,above of=hidden2-1, node distance=0.5cm] (hl) {$z_{2,1}$};
\node[annot,below of=hidden2-4, node distance=0.5cm] (hl) {$z_{2,H_2}$};
\end{tikzpicture}
\caption{Network diagram for a feed-forward NN with three inputs, two hidden layers, and one output.}
\label{NNdiag}
\end{center}
\end{figure}
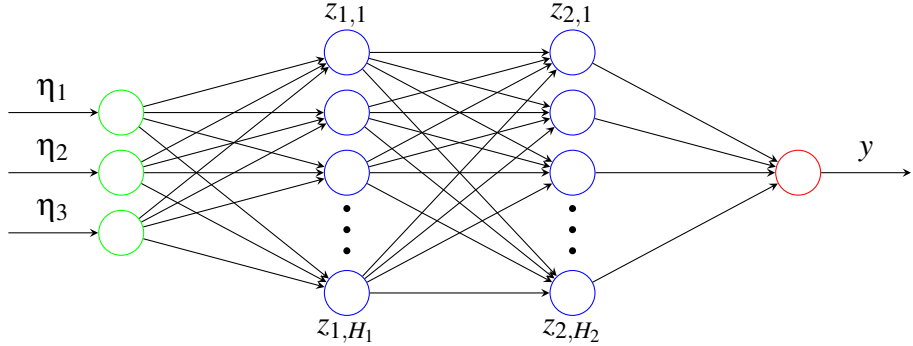

The standard NN algorithm operates by constructing linear combinations of inputs and transforming them through nonlinear activation functions. The process is repeated once for each hidden layer (marked blue in Fig. \ref{NNdiag})  in the network, until the output layer is reached. 
Fig. \ref{NNdiag} presents a sample ANN. For this sample network, the values of the hidden nodes $z_{1,1}$ through $z_{1,H_1}$ would be constructed as
\begin{equation}
z_{1,j} = a_{(1)}\left(\sum_{i=1}^3 w_{ij}^{(1)}\eta_i\right)
\end{equation}
where $a_{(1)}$ and $w_{ij}^{(1)}$ are the activation function and weights associated with the first hidden layer, respectively. Similarly, the second layer of hidden nodes is constructed as
\begin{equation}
z_{2,j} = a_{(2)}\left(\sum_{i=1}^{H_1} w_{ij}^{(2)}z_{1,i}\right)
\end{equation}
Finally, the output is
\begin{equation}
y \approx f(\boldmath{\eta}) = a_{(3)}\left(\sum_{i=1}^{H_2} w_{ij}^{(3)}z_{2,i}\right)
\end{equation}
Given training data, error back-propagation algorithms\cite{mlbook} are used to find $w_{ij}^{(n)}$.

Once the weights are found, 
computing the output depends only on the number of hidden nodes, 
and not on the volume of the training data. Hyper-parameters of the 
NN method include the number of hidden layers, the number of nodes in each hidden layer, and the forms of the activation functions. Typically, 3 layers and about 100 nodes were employed with a sigmoid activation function. The Fast Artificial Neural Network Library (FANN)\cite{fann} is used for this work.

%
%
%

\section{Results}
\begin{figure}[!h]
\centering
\includegraphics[width=0.6\textwidth]{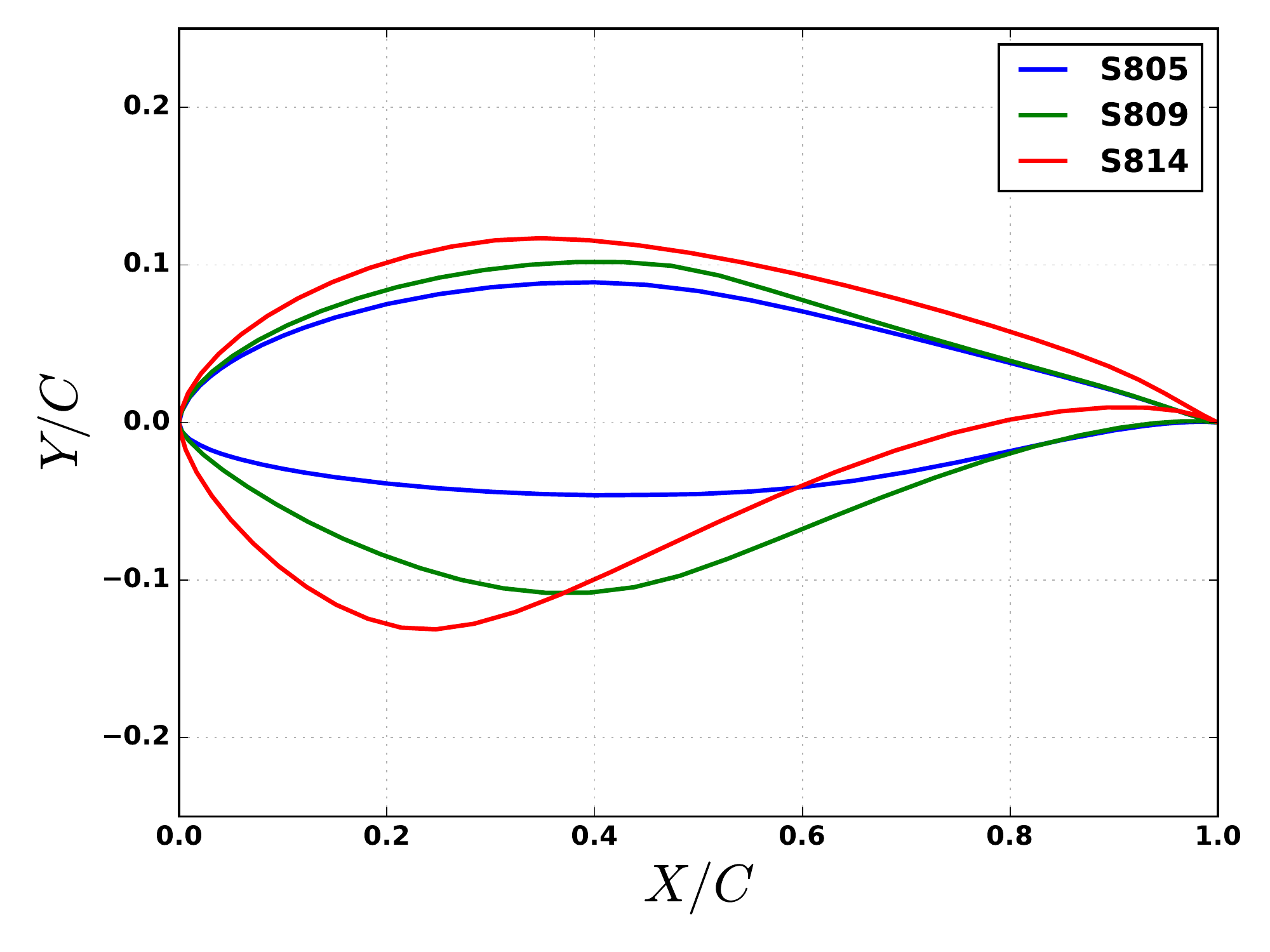}
\caption{Three different airfoils used for training and testing the neural network model. Note: axes are scaled differently.}
\label{figures:airfoils}
\end{figure}

The utility of the data-driven framework is demonstrated in three wind turbine airfoils with varying thickness: (i) S805, (ii) S809, and, (iii) S814 (Fig. \ref{figures:airfoils}). This specific set was chosen for this work because of the availability (in the open literature\cite{somers805,somers809,somers814}) of the lift and drag polar from low angles of attack through incipient and massive separation and for multiple Reynolds numbers $Re \in \{1\times10^6, 2\times 10^6, 3\times 10^6\}$. Additionally, detailed pressure measurements are available at some test points.

Full-field inversion was performed for each airfoil at different combinations of angles of attack  and Reynolds number. In all the cases, inversion was based on just the lift coefficient. For the S809 airfoil at $Re = 2 \times 10^6$, the lift-based inversion was compared to the pressure-based inversion as shown in Fig.~\ref{ComparisonCLCP}. 
Inversion is followed by employing the neural network to reconstruct model corrections. 
Neural network-augmentations are generated using the model trained on the S814 airfoil data shown in Table~\ref{table:nnet3}. This data-set was chosen because adverse pressure gradients are the largest. Later in this section, ensemble comparisons based on different training data-sets will also be shown. 

\begin{figure*}
\centering
\includegraphics[width=0.5\textwidth]{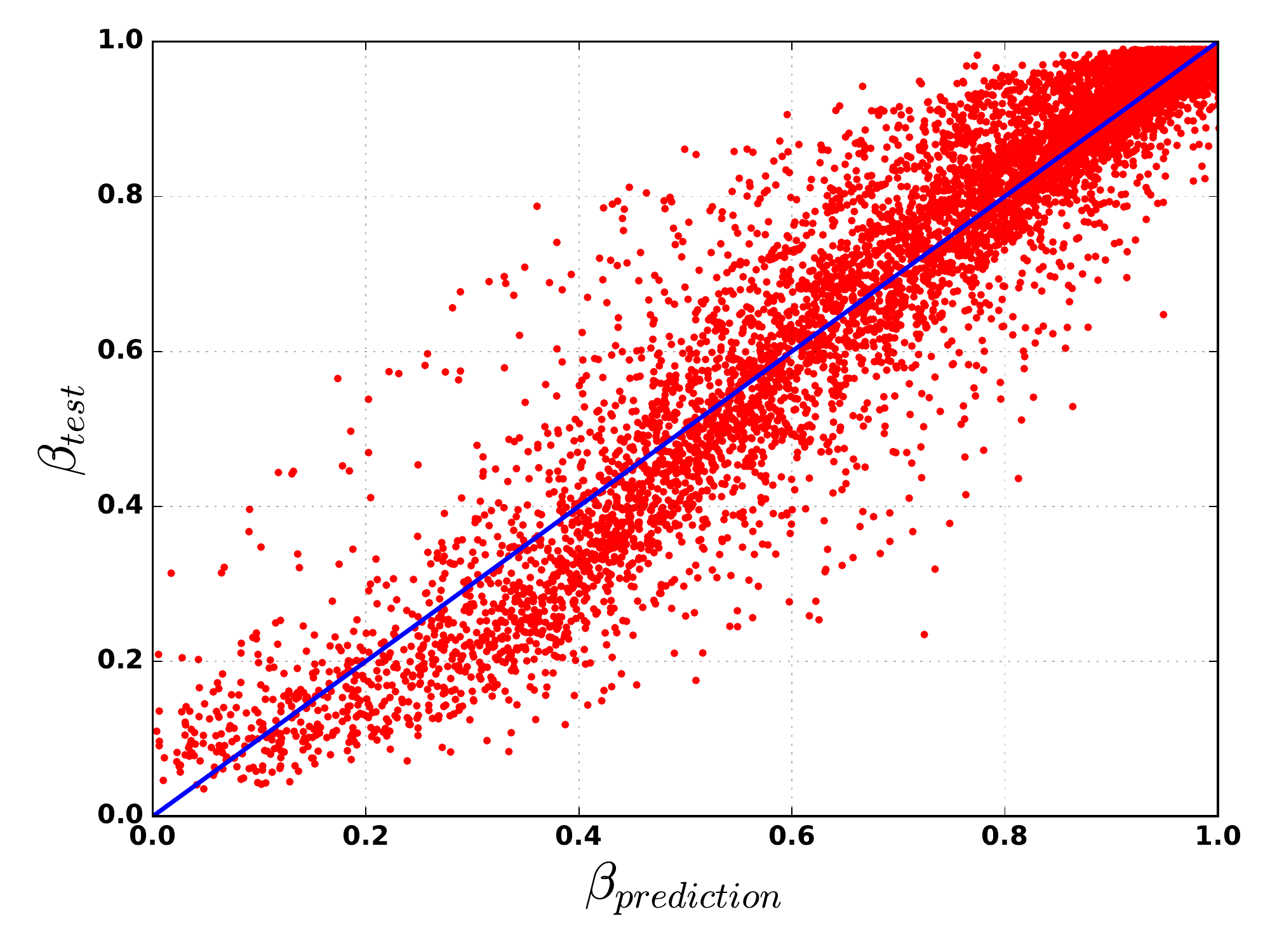}
\caption{Neural network testing and training on  data--set P. Solid blue line represents the perfect predictions and red dots represents the NN predictions.}
\label{figures:nnet:training}
\end{figure*}

As schematized in Fig.~\ref{fig:schematic}, the mapping $\beta(\eta)$ built during the training process is queried for input features $\hat{\eta}$ at every iteration of the flow solver to obtain outputs $\hat{\beta}$ which are embedded into the predictive model. This process is repeated until convergence. Thus, consistency is enforced between the underlying flowfield and the model augmentations.
Fig. \ref{figures:nnet:training} shows the testing and training on the data-set P.

\begin{table}
\centering
\begin{tabular}{|l|l|}\hline
Model label & Training data\\\hline
{\bf P } & {\bf S814} at ${\mathbf Re = 1 \times 10^6, 2 \times 10^6}$\\\hline
\end{tabular}
\caption{Training set to generate predictive model.}
\label{table:nnet3}
\end{table}

\subsection{Predictions}
\begin{figure*}[!h]
\centering
\subfigure[\be(\mat{x}) from inverse SA]{\includegraphics[width=0.32\textwidth]{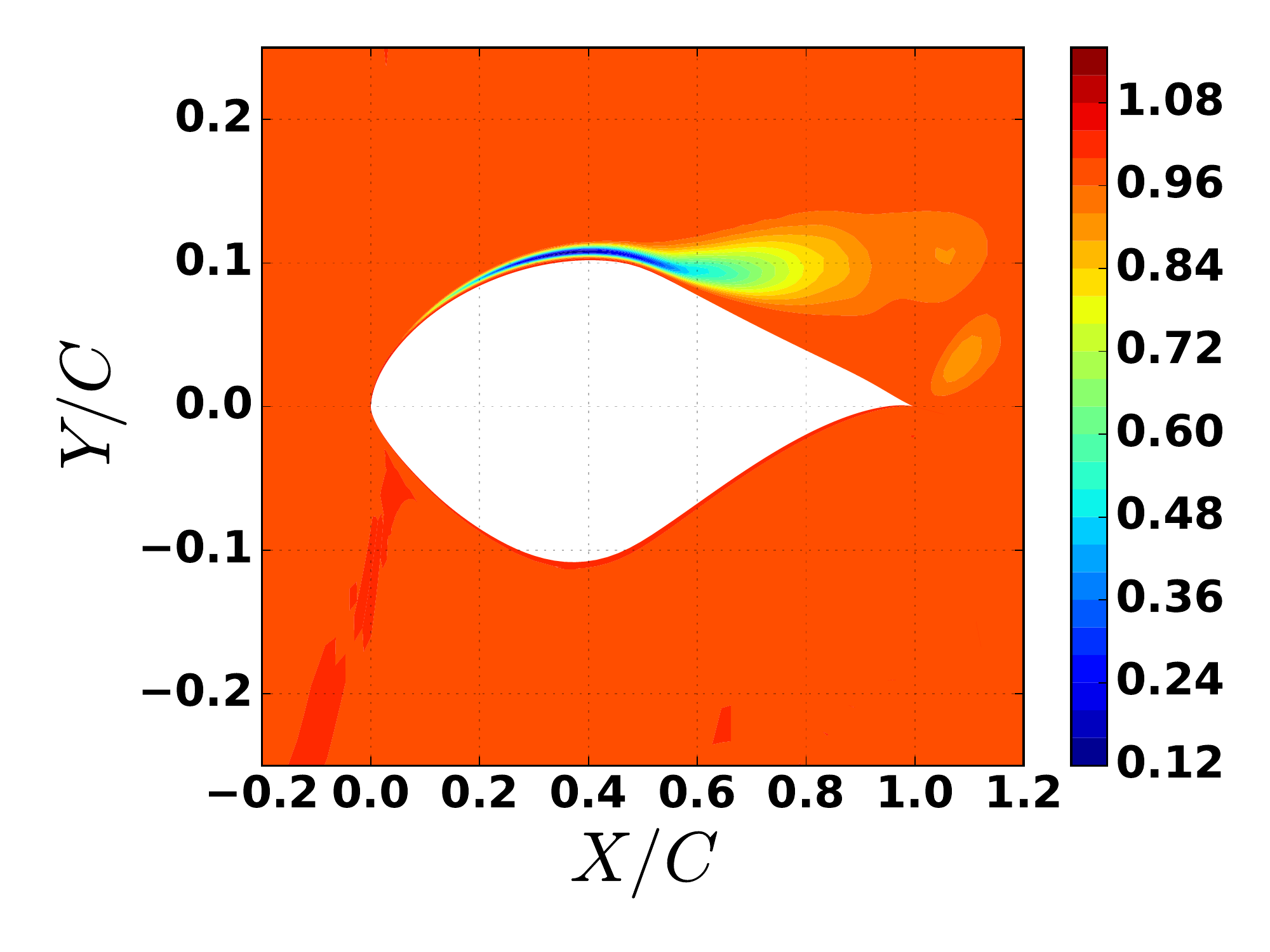}}
\subfigure[\be(\mat{U}) from NN-augmented SA (prediction) ]{\includegraphics[width=0.32\textwidth]{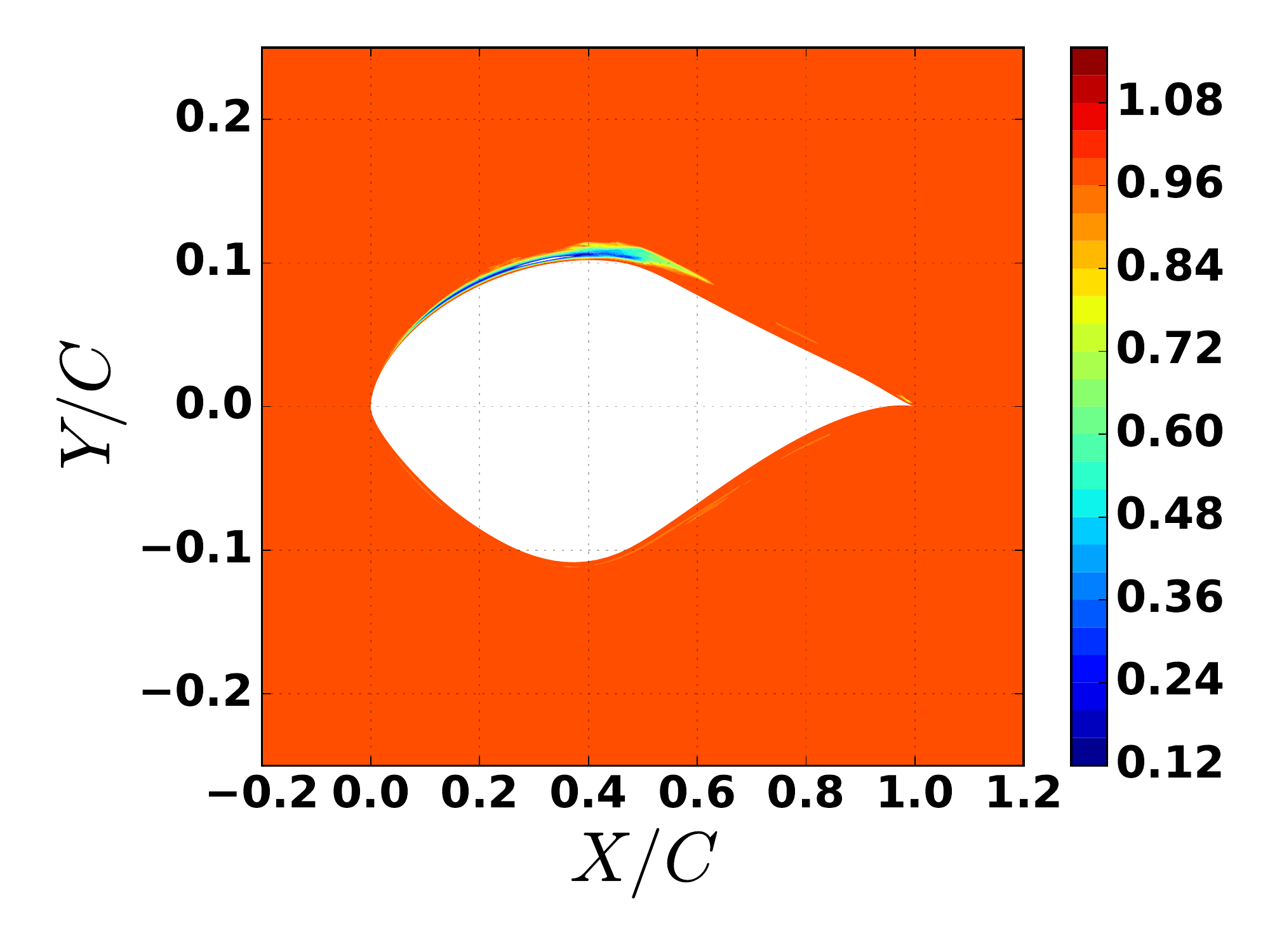}}
\subfigure[Pressure coefficient]{\includegraphics[width=0.32\textwidth]{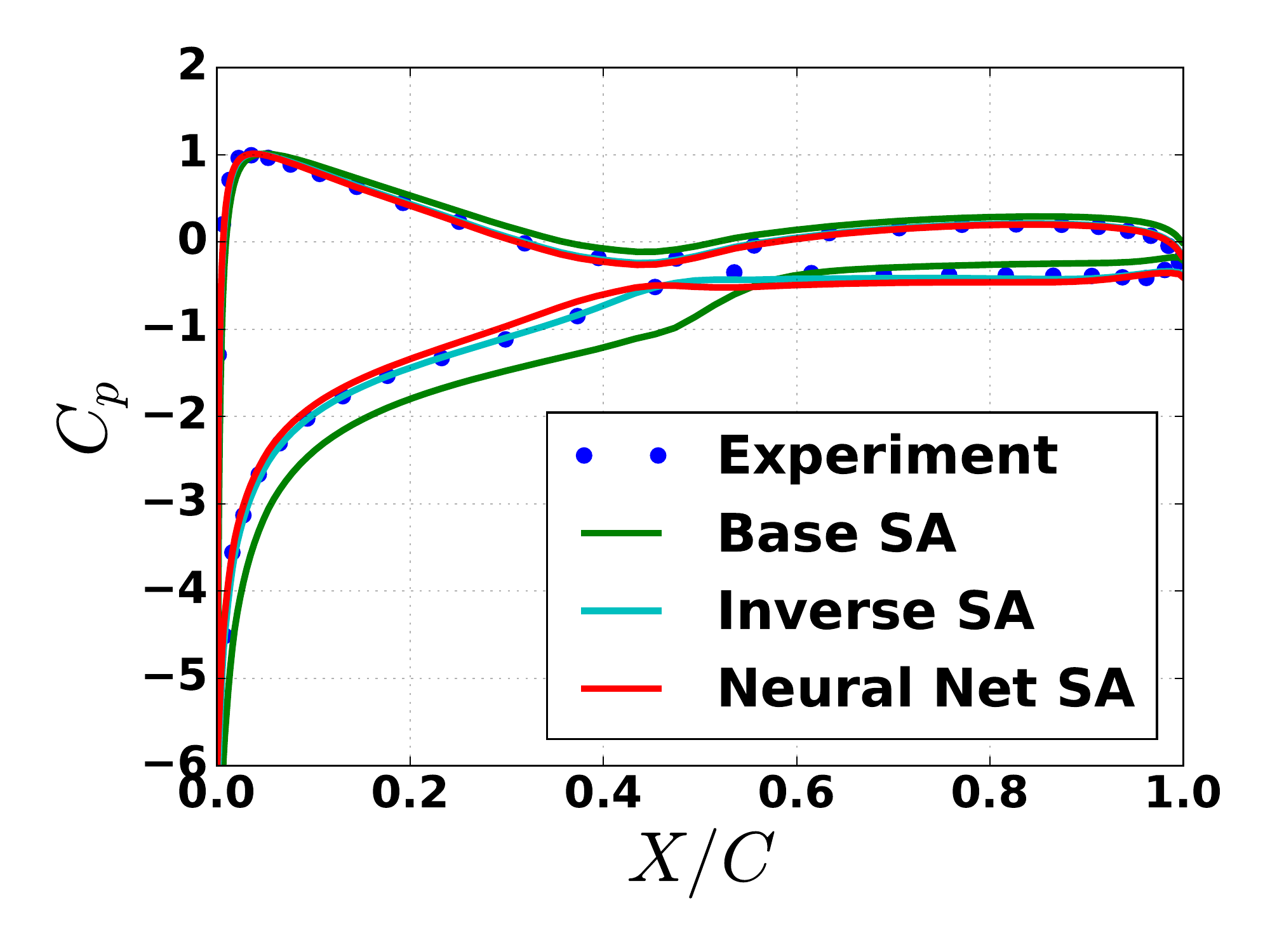}}
\caption{Comparison of inverse and NN-augmented predictions (using data-set P) for S809 airfoil at $\alpha = 14^{\circ}$ and $Re = 2\times 10^6$. }
\label{figures:s809:comparison}
\end{figure*}

\begin{figure*}[!h]
\centering
\subfigure[Base SA]{\includegraphics[width=0.32\textwidth,trim={.5cm .5cm .5cm 4.5cm},clip]{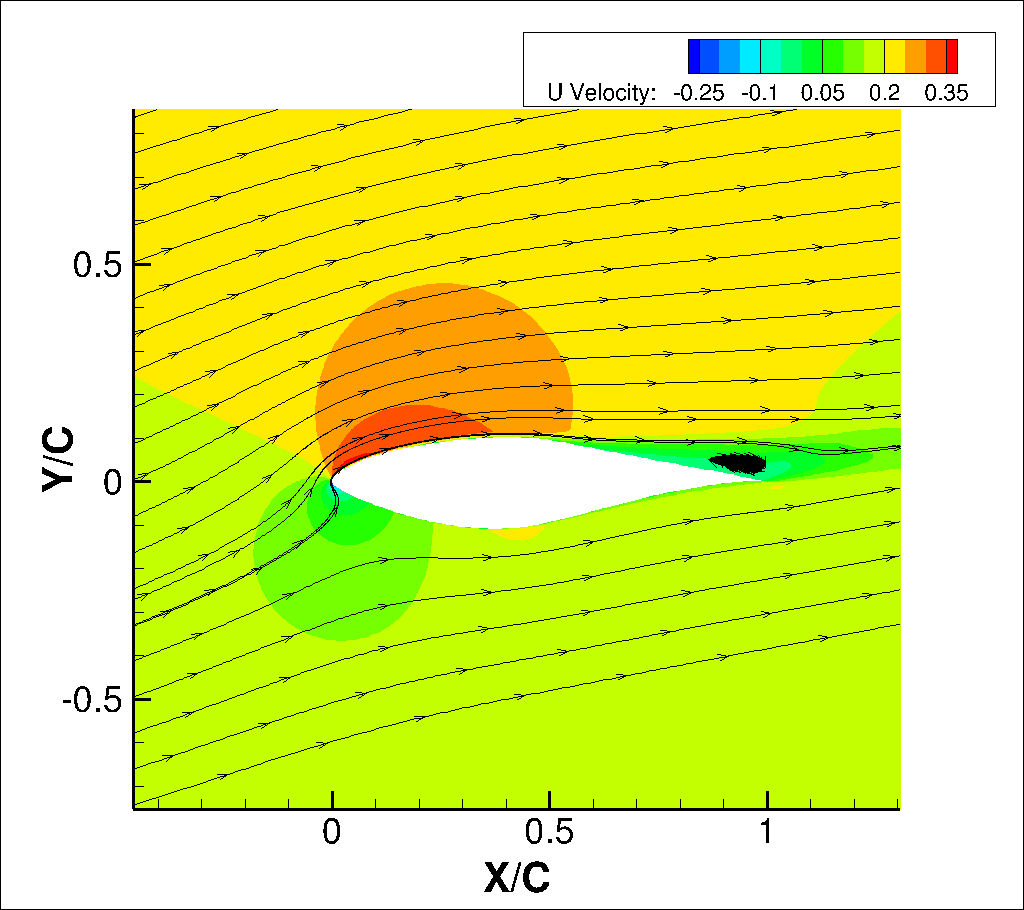}}
\subfigure[Inverse SA]{\includegraphics[width=0.32\textwidth,trim={.5cm .5cm .5cm 4.5cm},clip]{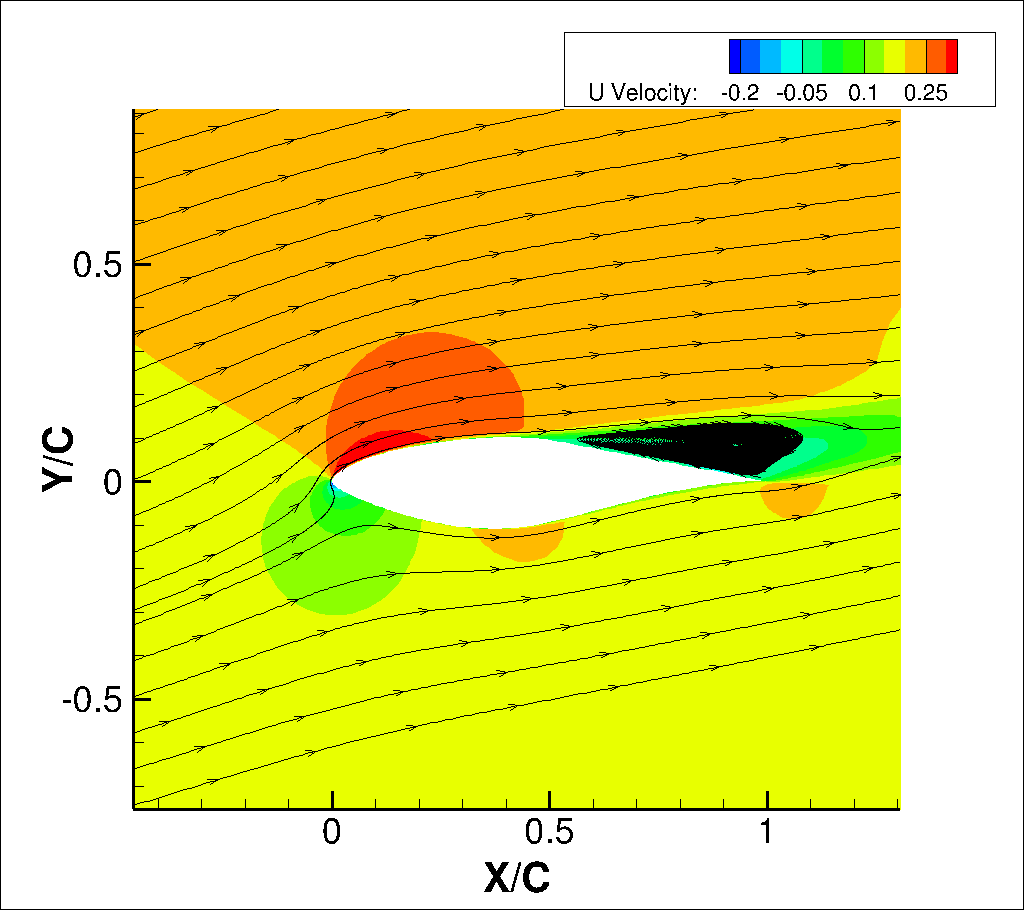}}
\subfigure[NN-augmented SA (prediction)]{\includegraphics[width=0.32\textwidth,trim={.5cm .5cm .5cm 4.5cm},clip]{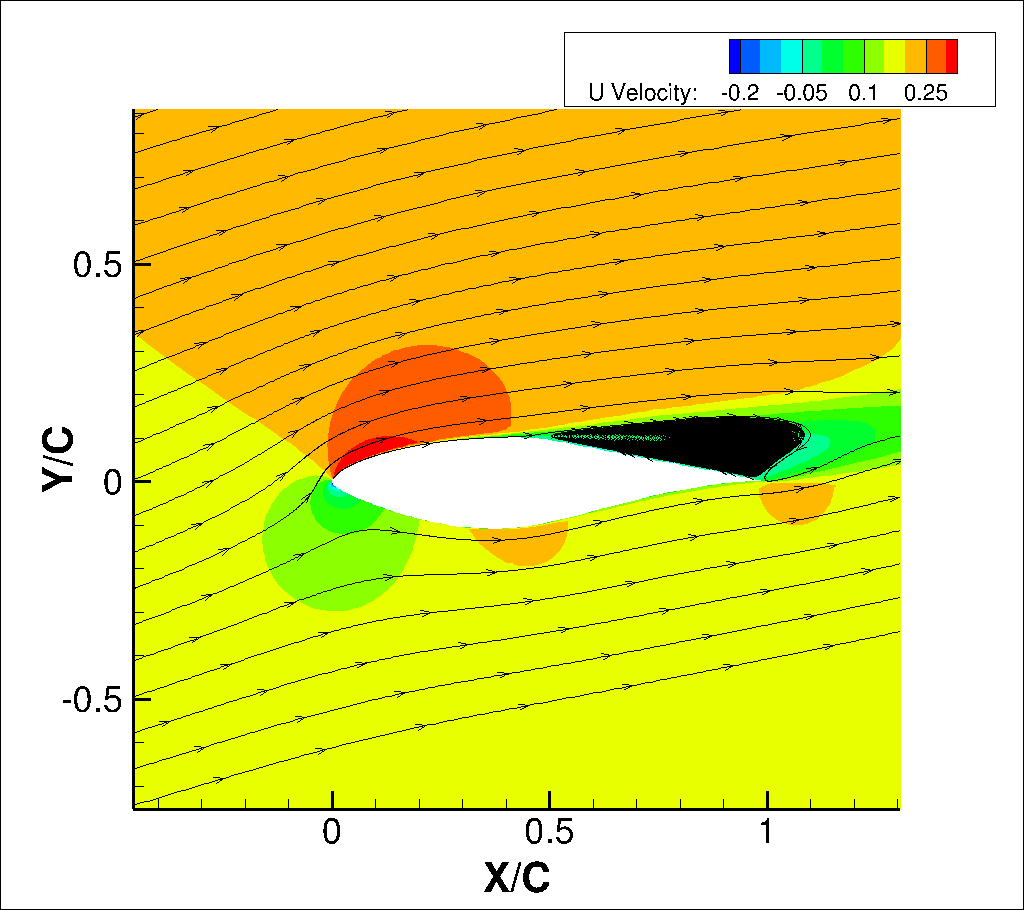}}
\caption{Streamlines and X-velocity contour for S809 airfoil at $Re = 2\times 10^6$ and $\alpha = 14^{\circ}$.}
\label{figures:s809:streamlines}
\end{figure*}

 The effectiveness of the inversion and learning is apparent in Figs.~\ref{figures:s809:comparison} and ~\ref{figures:s809:streamlines}, where the predictions based on model P are compared to the ideal scenario of direct inference on the S809 airfoil based on experimental data. It has to be mentioned that the training data-set was based on assimilating lift information only.

\begin{figure*}[!h]
\centering
\subfigure[$C_p$]{\includegraphics[width=0.42\textwidth]{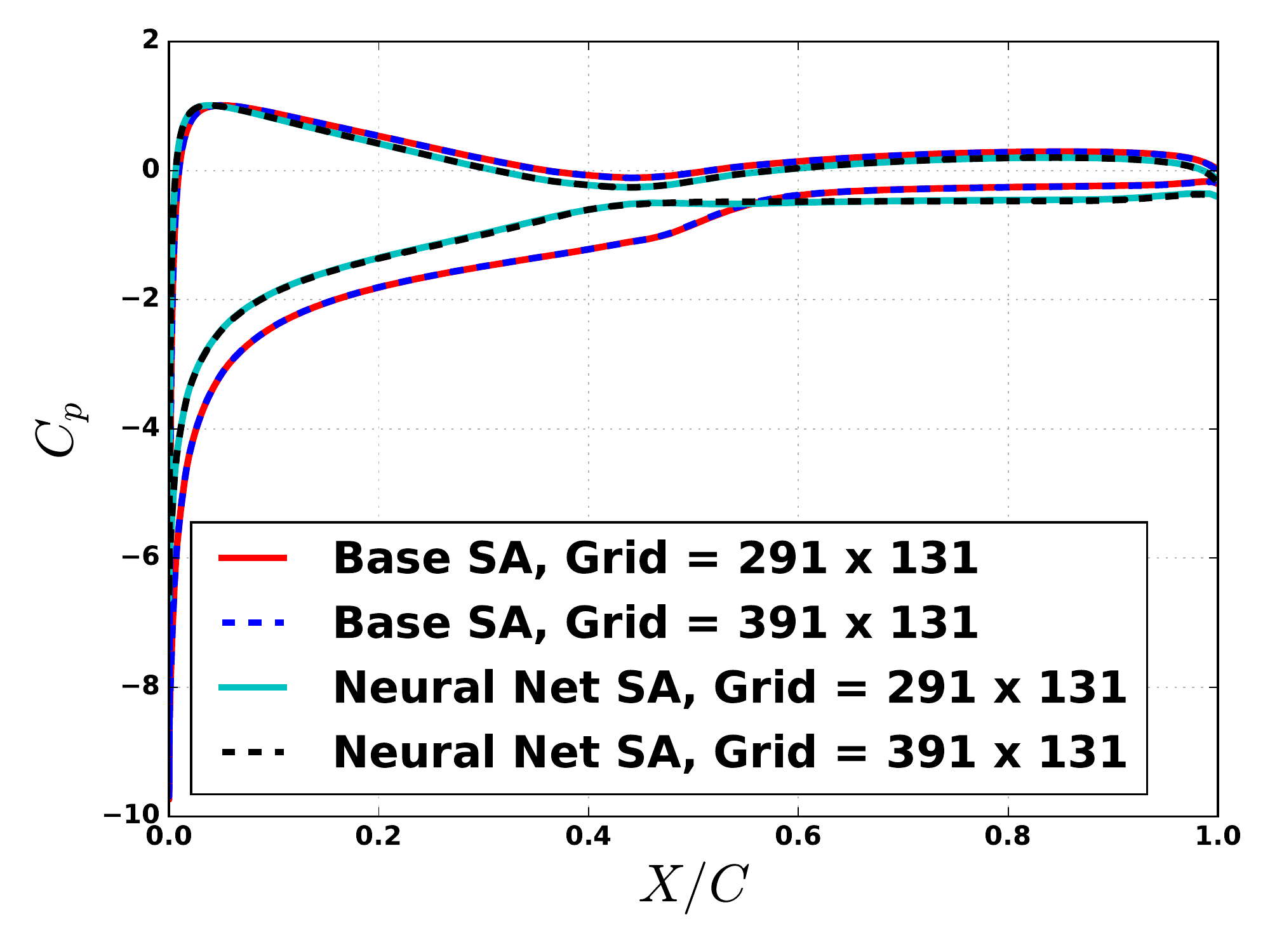}}
\subfigure[$C_f$]{\includegraphics[width=0.42\textwidth]{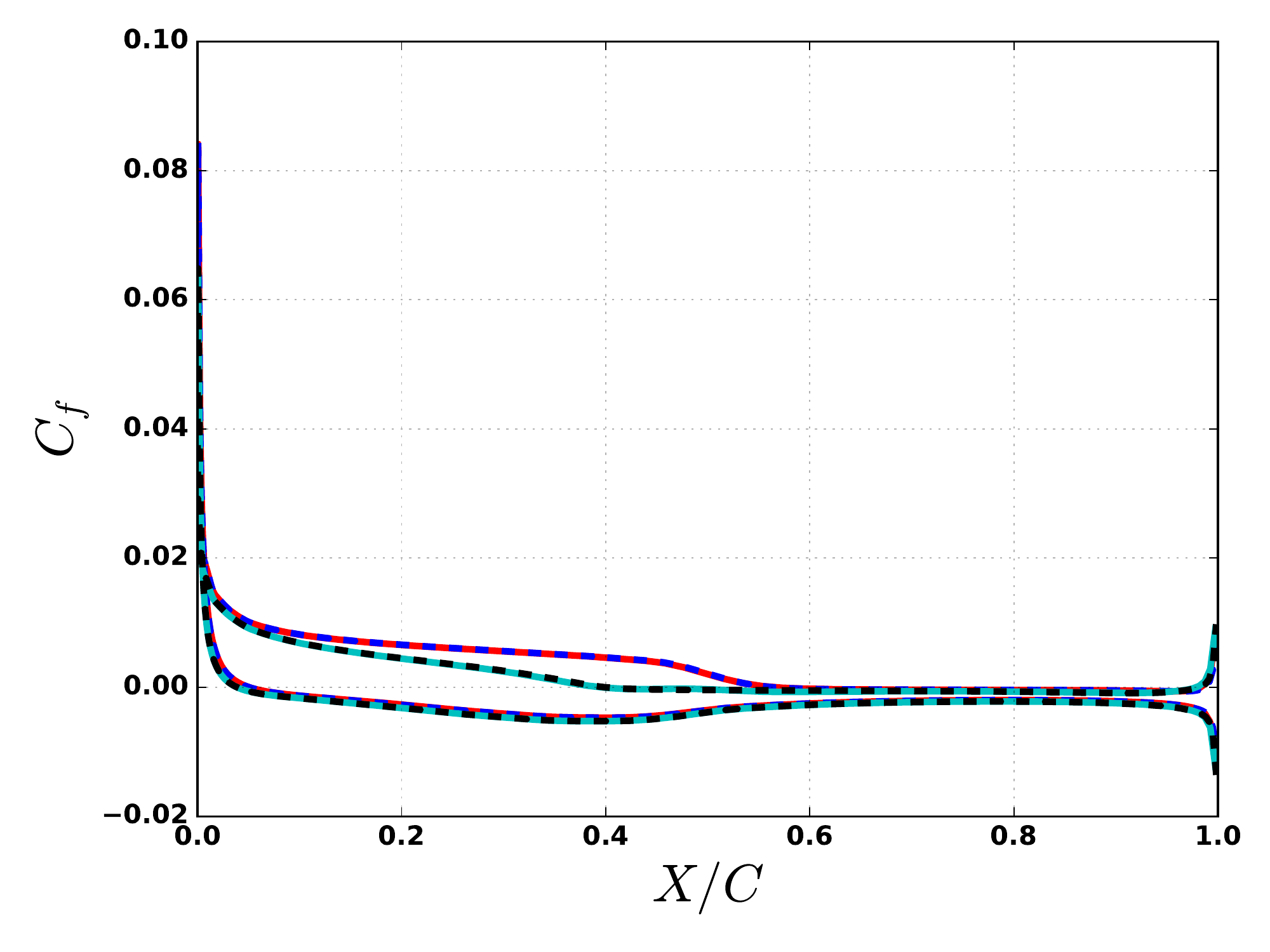}}
\caption{Pressure and skin friction (using data-set P) for S809 airfoil at $Re = 2\times 10^6$ and $\alpha = 14^{\circ}$ using grids of different spatial resolutions. Solutions of both the base SA model and the neural network augmented SA are grid converged.}
\label{figures:s809:grid_convergence}
\end{figure*}

\begin{figure*}[!h]
\centering
\subfigure[$Re=1 \times 10^6$]{\includegraphics[width=0.32\textwidth,trim={0 0 0 1.40cm},clip]{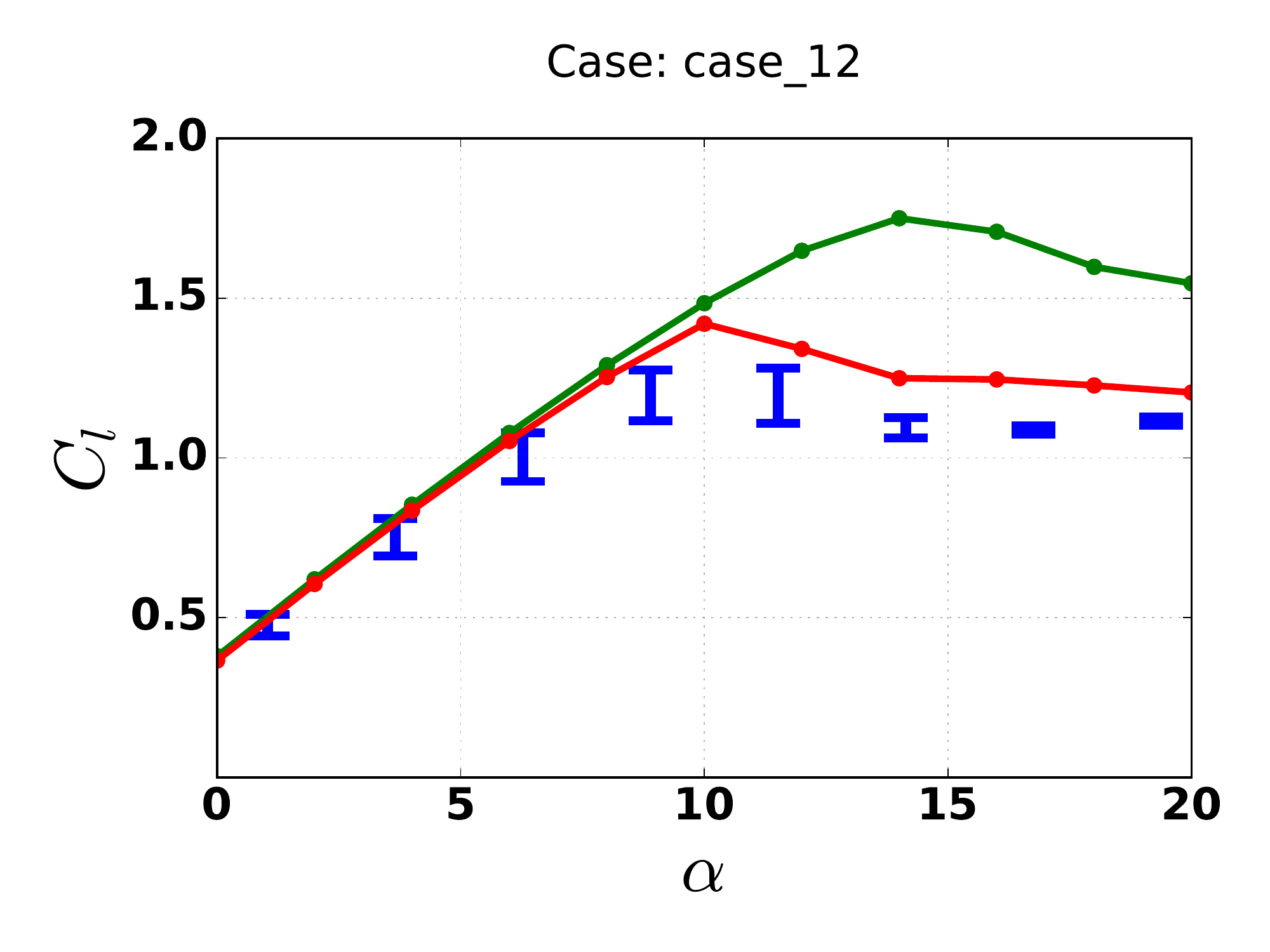}}
\subfigure[$Re=2 \times 10^6$]{\includegraphics[width=0.32\textwidth,trim={0 0 0 1.40cm},clip]{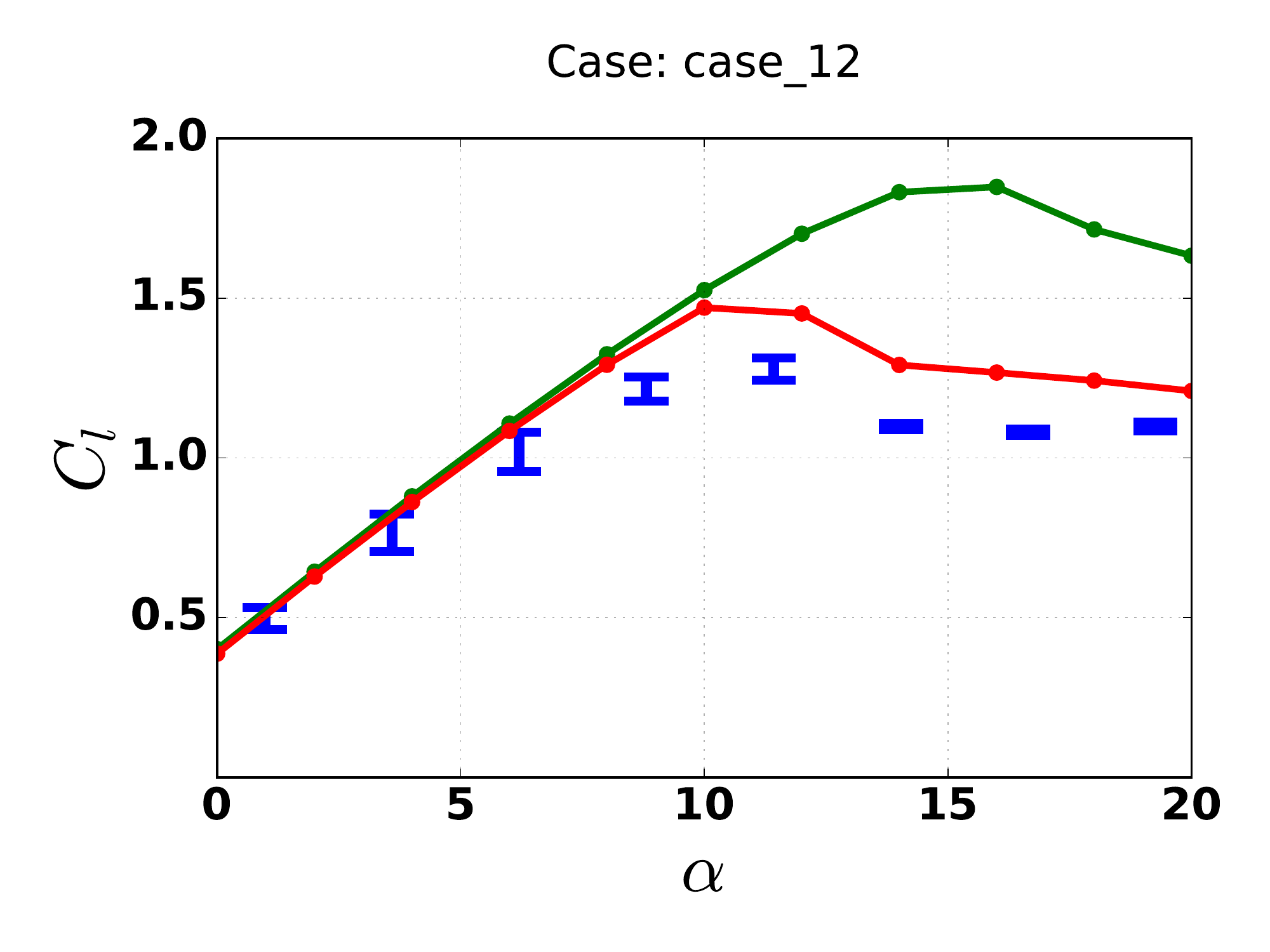}}
\subfigure[$Re=3 \times 10^6$]{\includegraphics[width=0.32\textwidth,trim={0 0 0 1.40cm},clip]{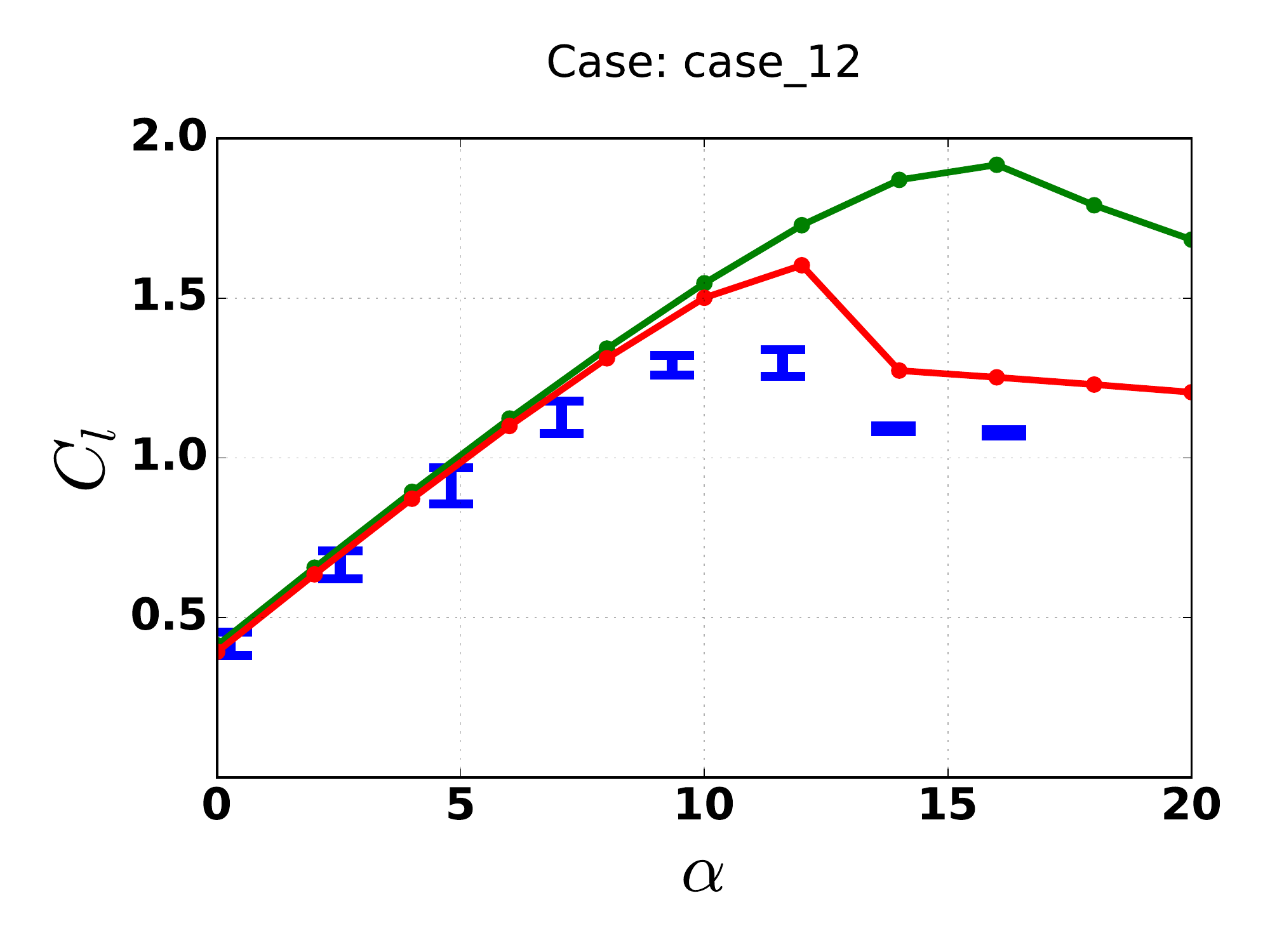}}
\subfigure[$Re=1 \times 10^6$]{\includegraphics[width=0.32\textwidth,trim={0 0 0 1.40cm},clip]{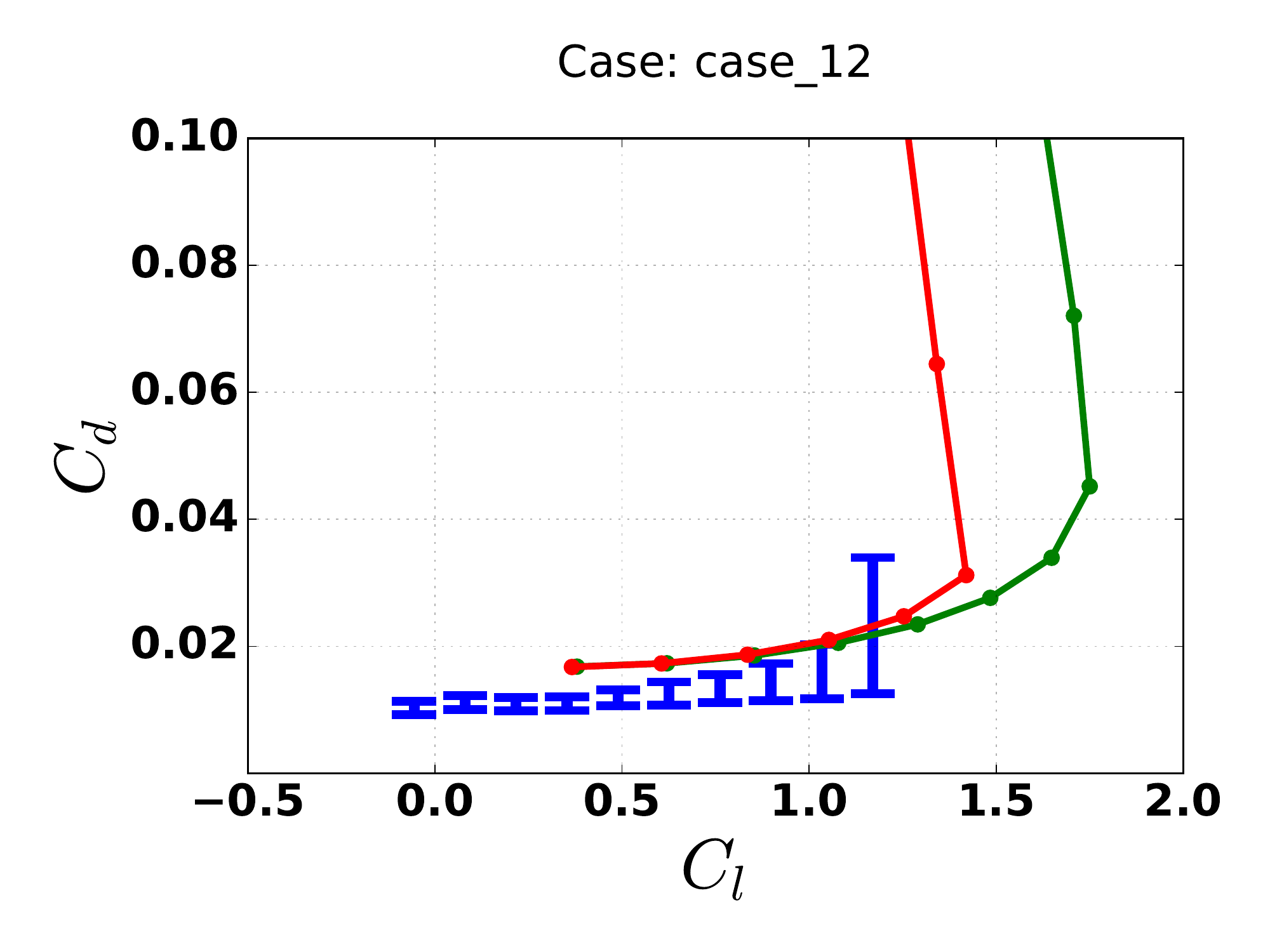}}
\subfigure[$Re=2 \times 10^6$]{\includegraphics[width=0.32\textwidth,trim={0 0 0 1.40cm},clip]{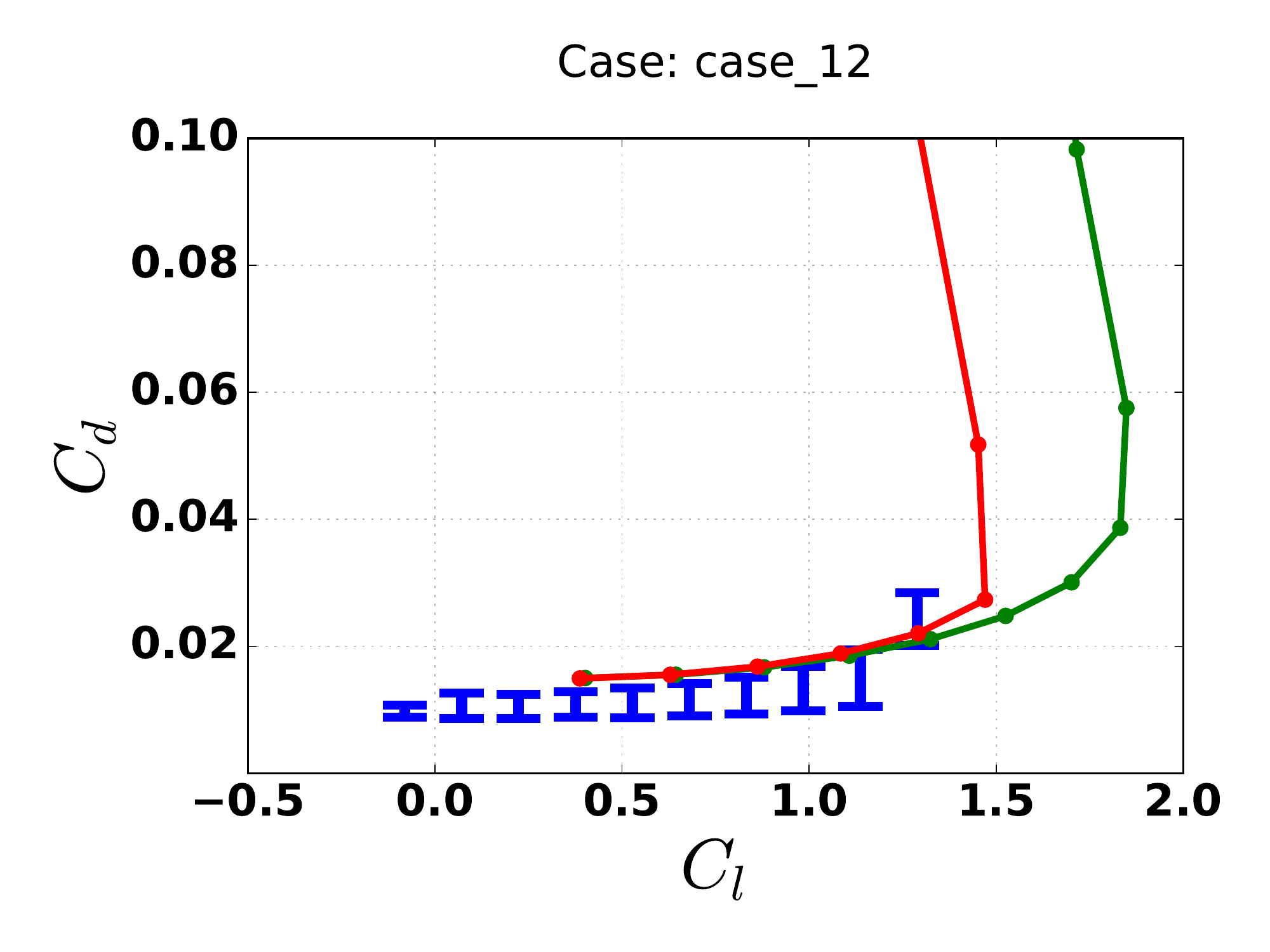}}
\subfigure[$Re=3 \times 10^6$]{\includegraphics[width=0.32\textwidth,trim={0 0 0 1.40cm},clip]{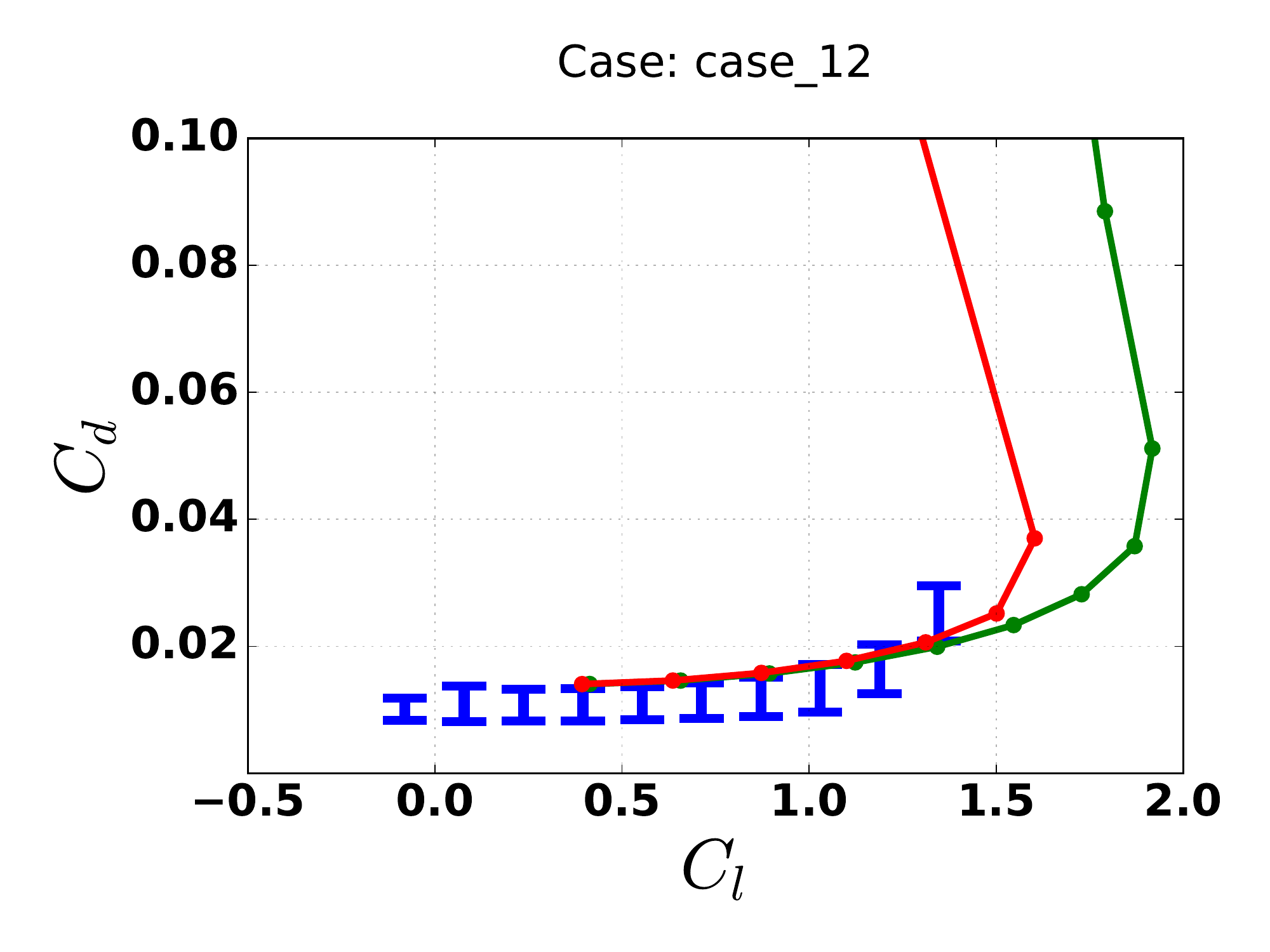}}
\caption{NN-augmented SA prediction for S814 airfoil using data-set P. {{\color{blue}\textbf{---}} Experiment}, {{\color{OliveGreen}\textbf{---}} base SA} and {{\color{red}\textbf{---}} neural network}.}
\label{figures:nnet:s814} 
\end{figure*}

\begin{figure*}[!h]
\centering
\subfigure[$Re=1 \times 10^6$]{\includegraphics[width=0.32\textwidth,trim={0 0 0 1.40cm},clip]{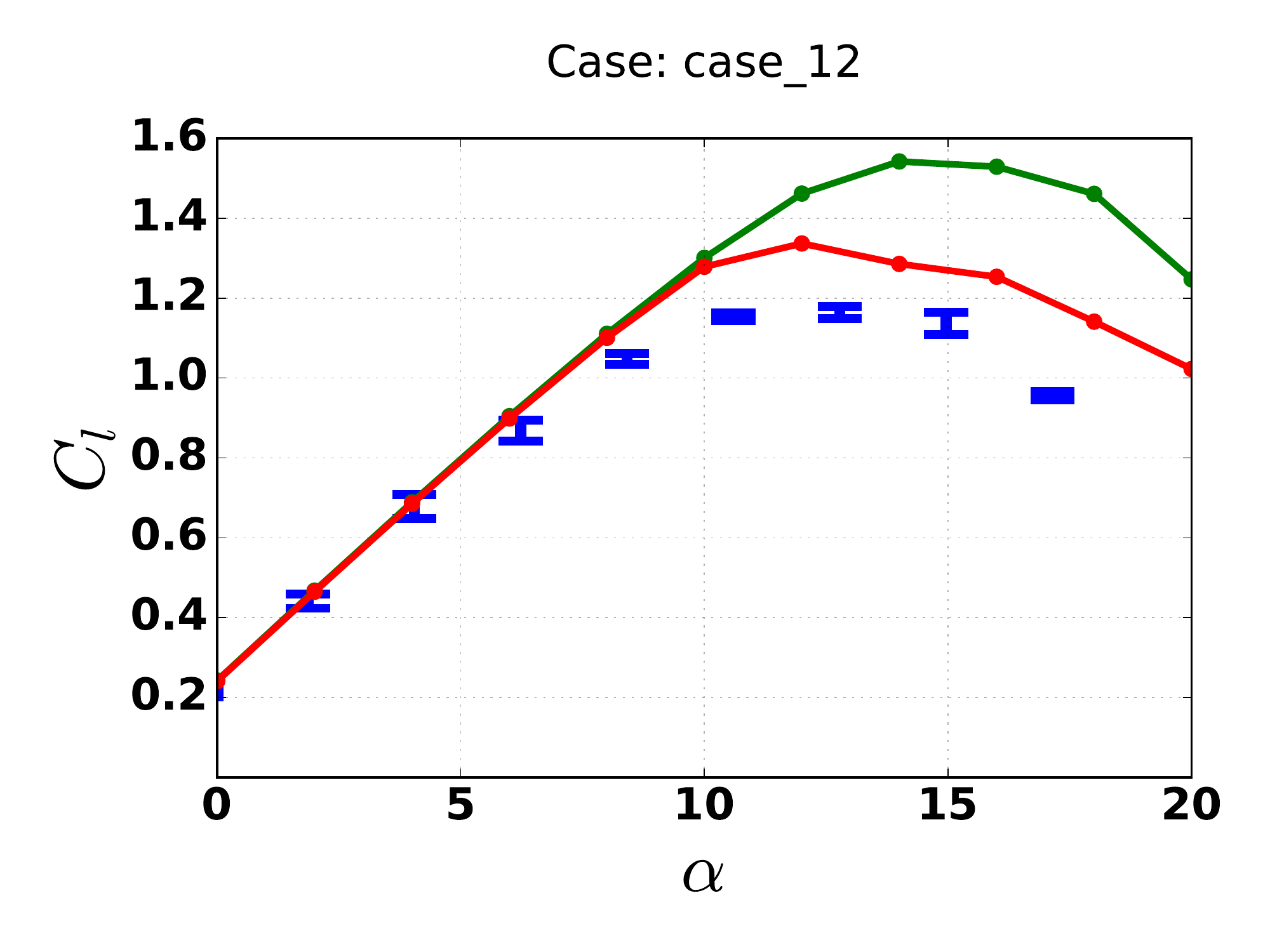}}
\subfigure[$Re=2 \times 10^6$]{\includegraphics[width=0.32\textwidth,trim={0 0 0 1.40cm},clip]{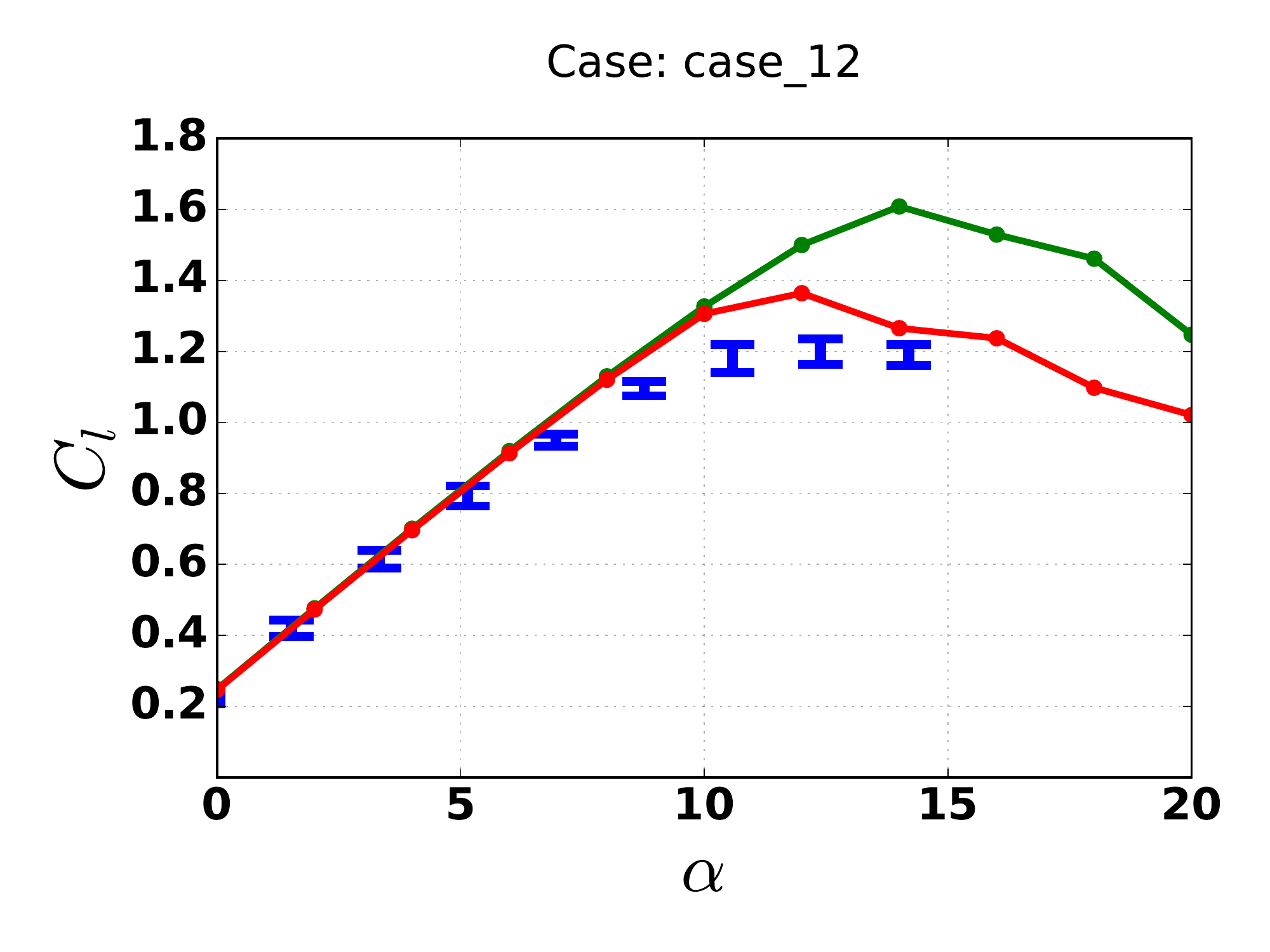}}
\subfigure[$Re=3 \times 10^6$]{\includegraphics[width=0.32\textwidth,trim={0 0 0 1.40cm},clip]{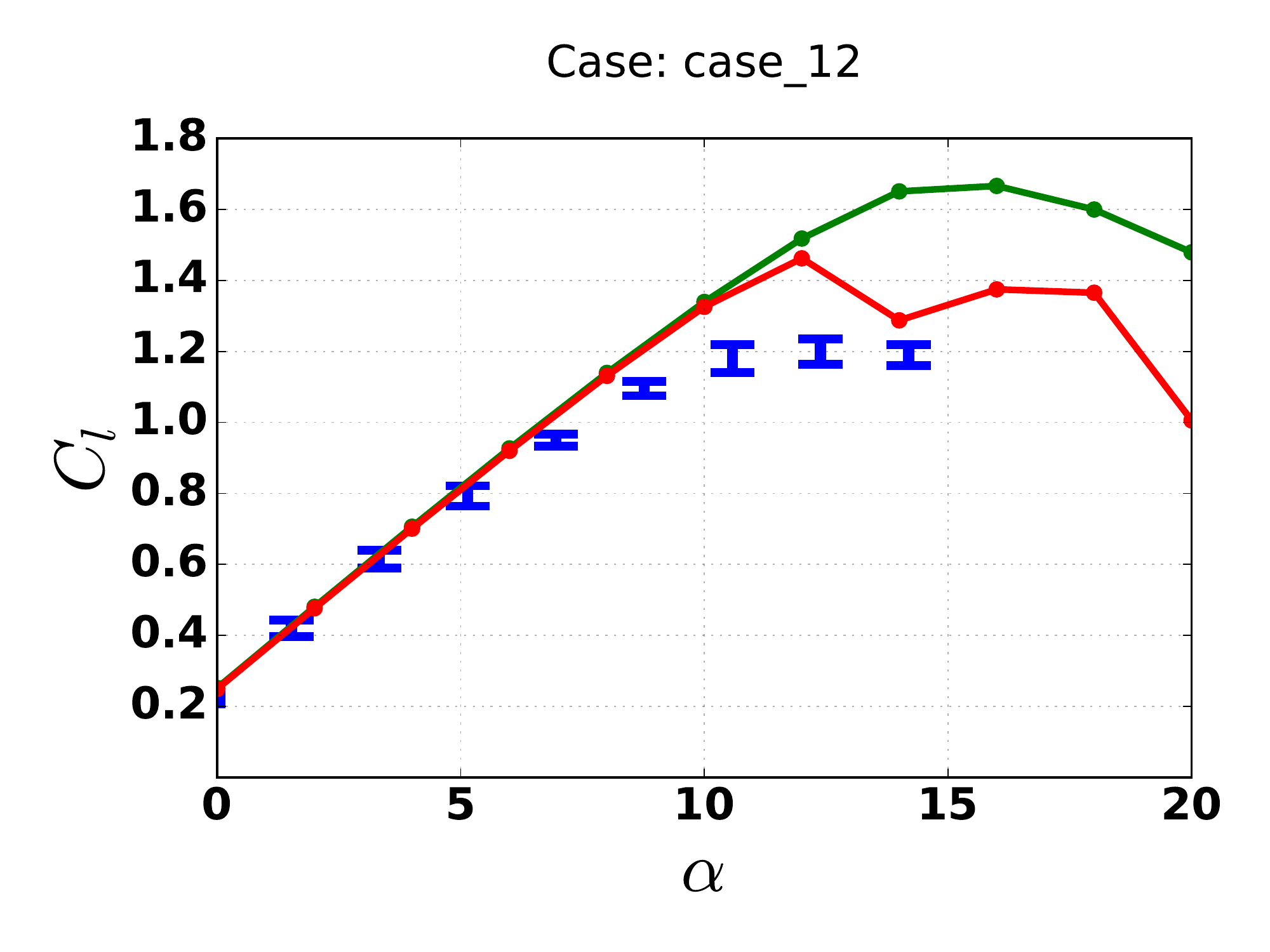}}
\subfigure[$Re=1 \times 10^6$]{\includegraphics[width=0.32\textwidth,trim={0 0 0 1.40cm},clip]{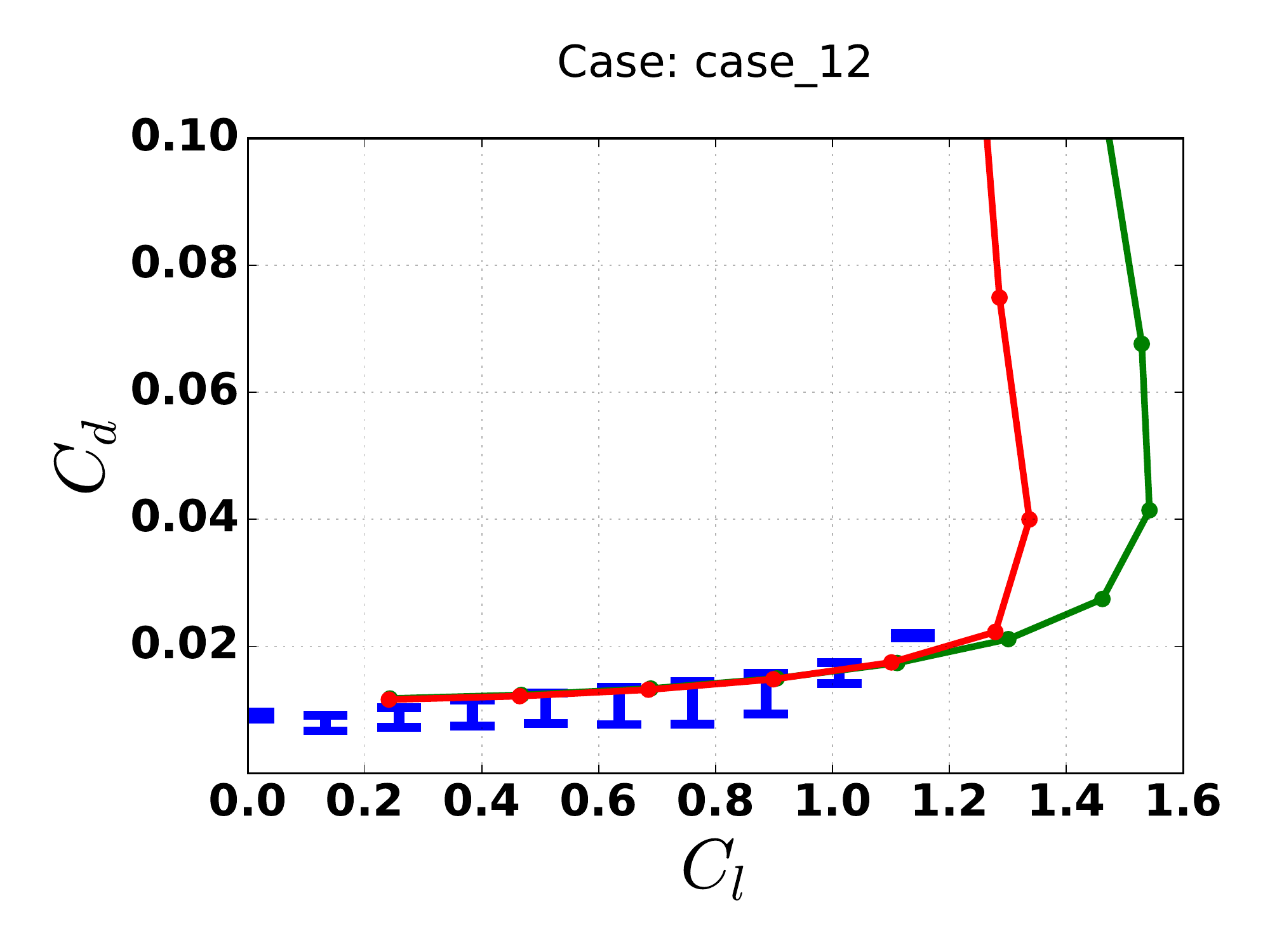}}
\subfigure[$Re=2 \times 10^6$]{\includegraphics[width=0.32\textwidth,trim={0 0 0 1.40cm},clip]{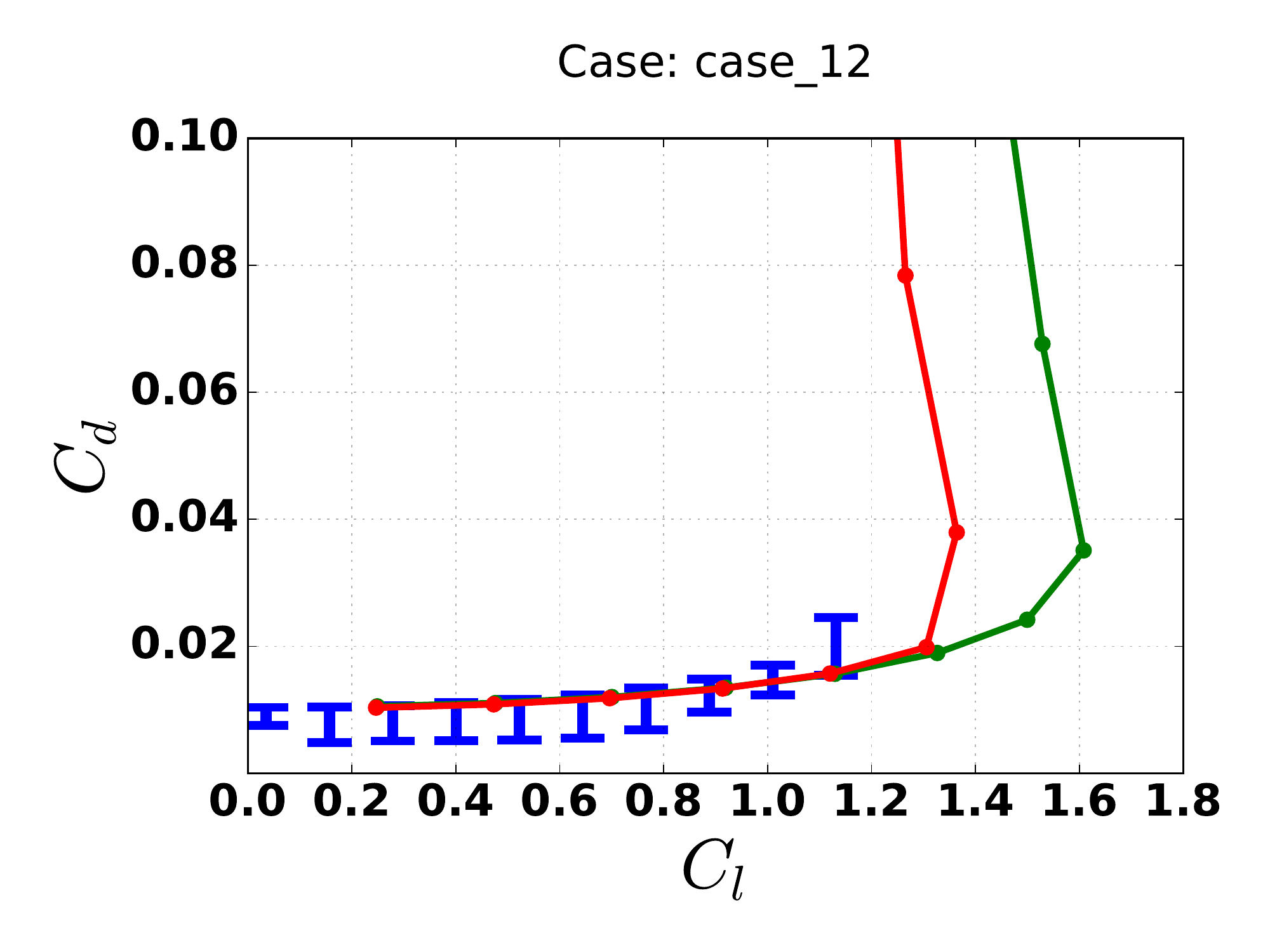}}
\subfigure[$Re=3 \times 10^6$]{\includegraphics[width=0.32\textwidth,trim={0 0 0 1.40cm},clip]{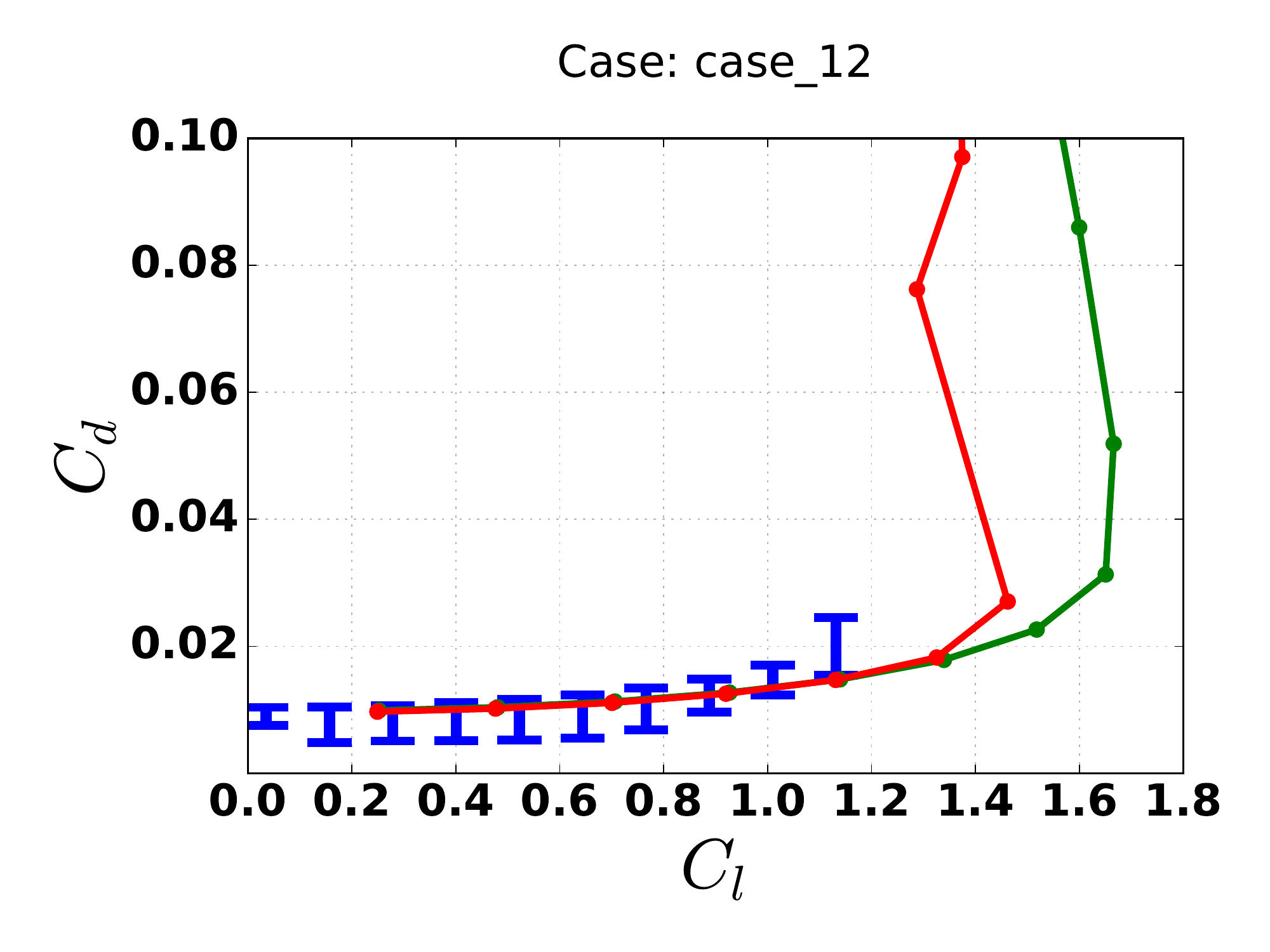}}
\caption{NN-augmented SA prediction for S805 airfoil using data-set P. {{\color{blue}\textbf{---}} Experiment}, {{\color{OliveGreen}\textbf{---}} base SA} and {{\color{red}\textbf{---}} neural network}.}
\label{figures:nnet:s805}
\end{figure*}

\begin{figure*}[!h]
\centering
\subfigure[$Re=1 \times 10^6$]{\includegraphics[width=0.32\textwidth,trim={0 0 0 1.40cm},clip]{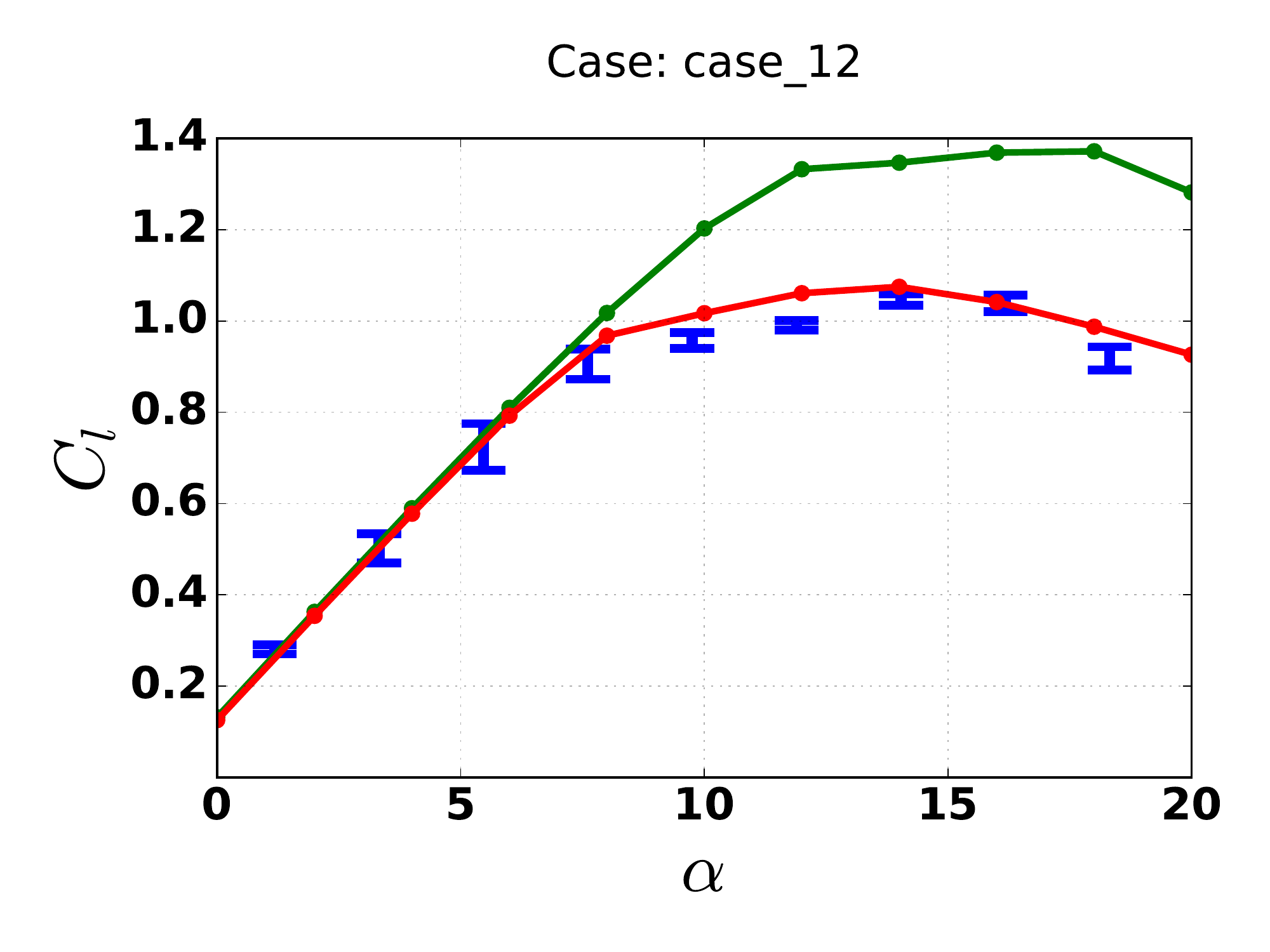}}
\subfigure[$Re=2 \times 10^6$]{\includegraphics[width=0.32\textwidth,trim={0 0 0 1.40cm},clip]{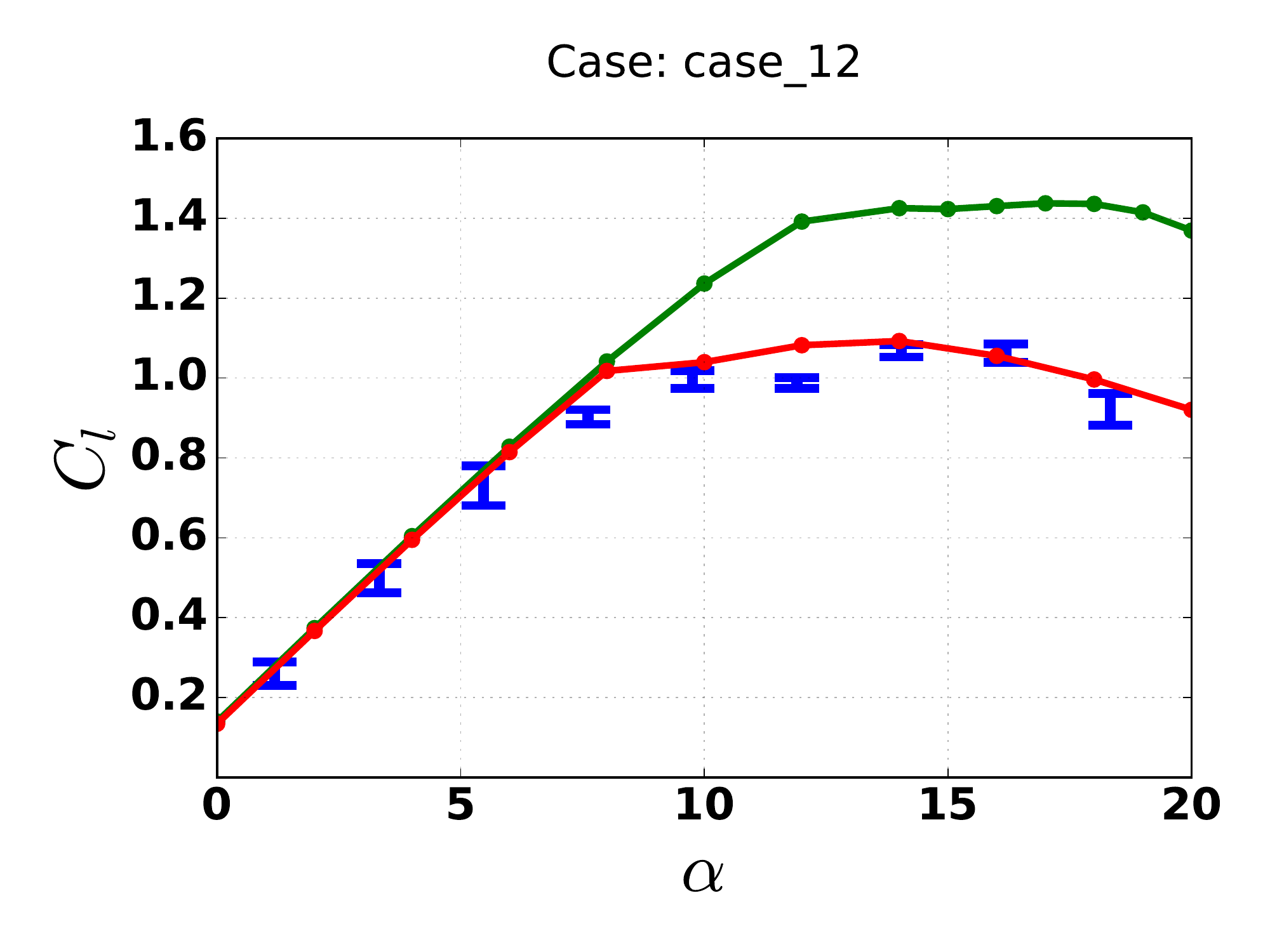}}
\subfigure[$Re=3 \times 10^6$]{\includegraphics[width=0.32\textwidth,trim={0 0 0 1.40cm},clip]{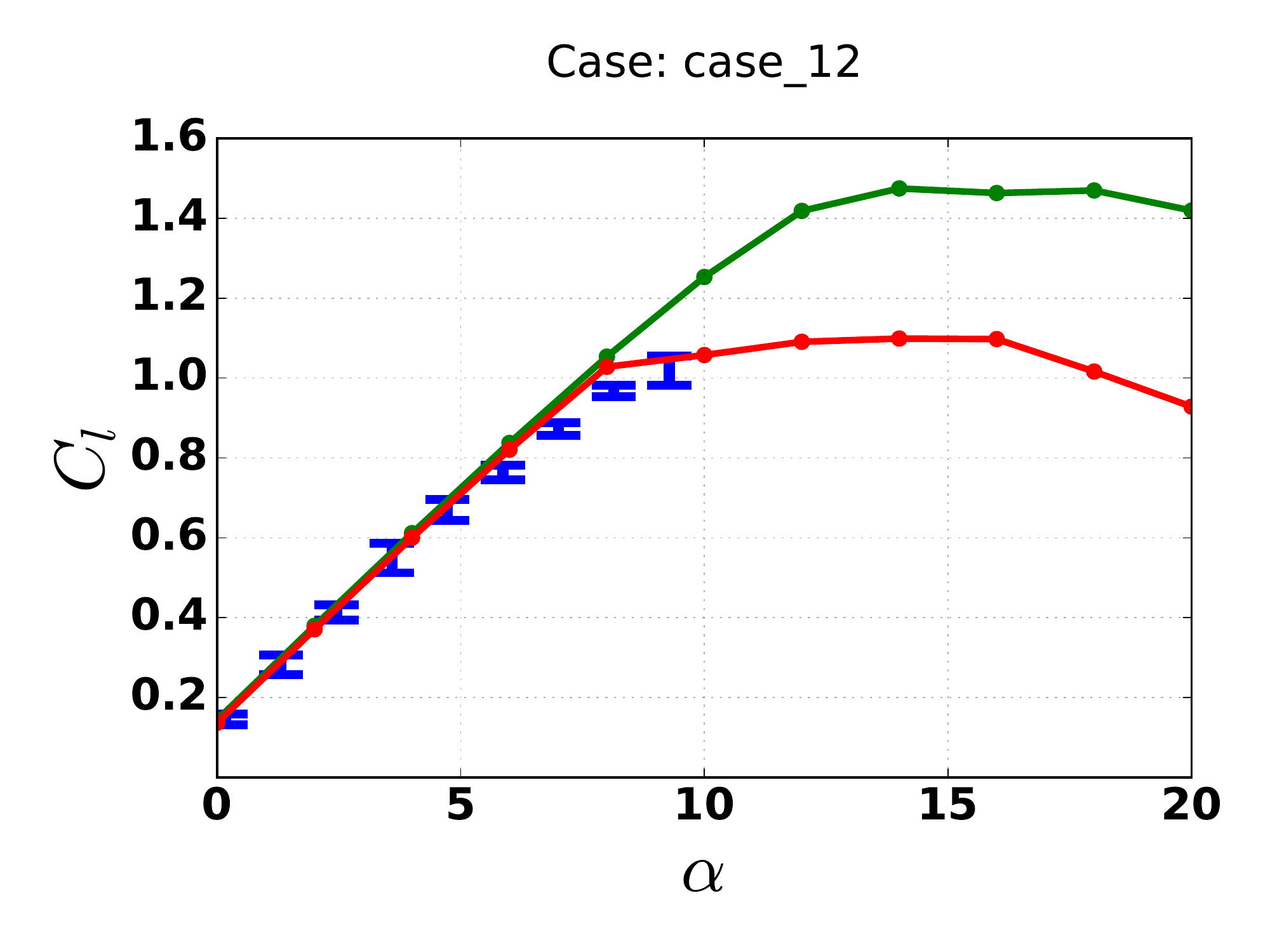}}
\subfigure[$Re=1 \times 10^6$]{\includegraphics[width=0.32\textwidth,trim={0 0 0 1.40cm},clip]{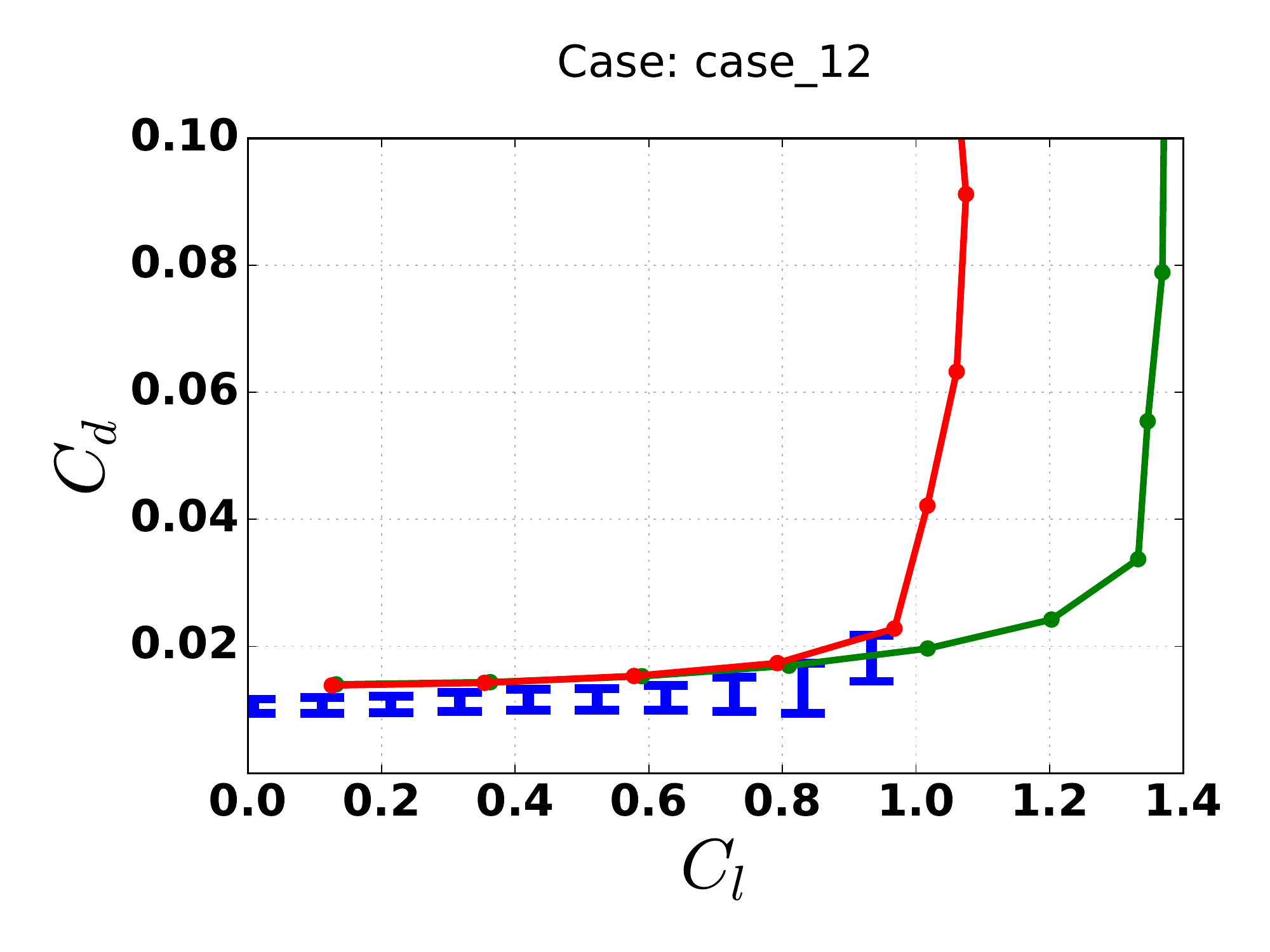}}
\subfigure[$Re=2 \times 10^6$]{\includegraphics[width=0.32\textwidth,trim={0 0 0 1.40cm},clip]{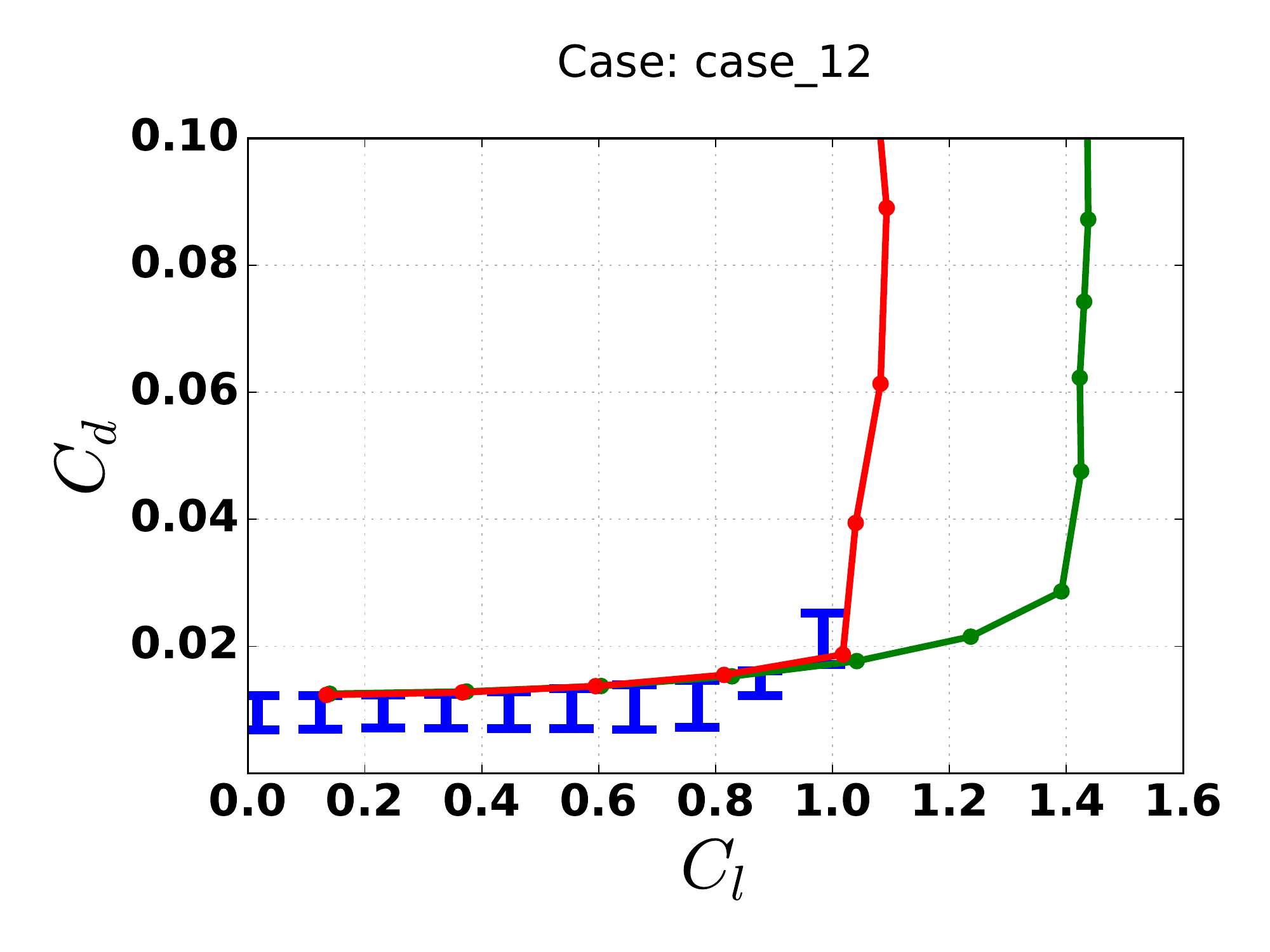}}
\subfigure[$Re=3 \times 10^6$]{\includegraphics[width=0.32\textwidth,trim={0 0 0 1.40cm},clip]{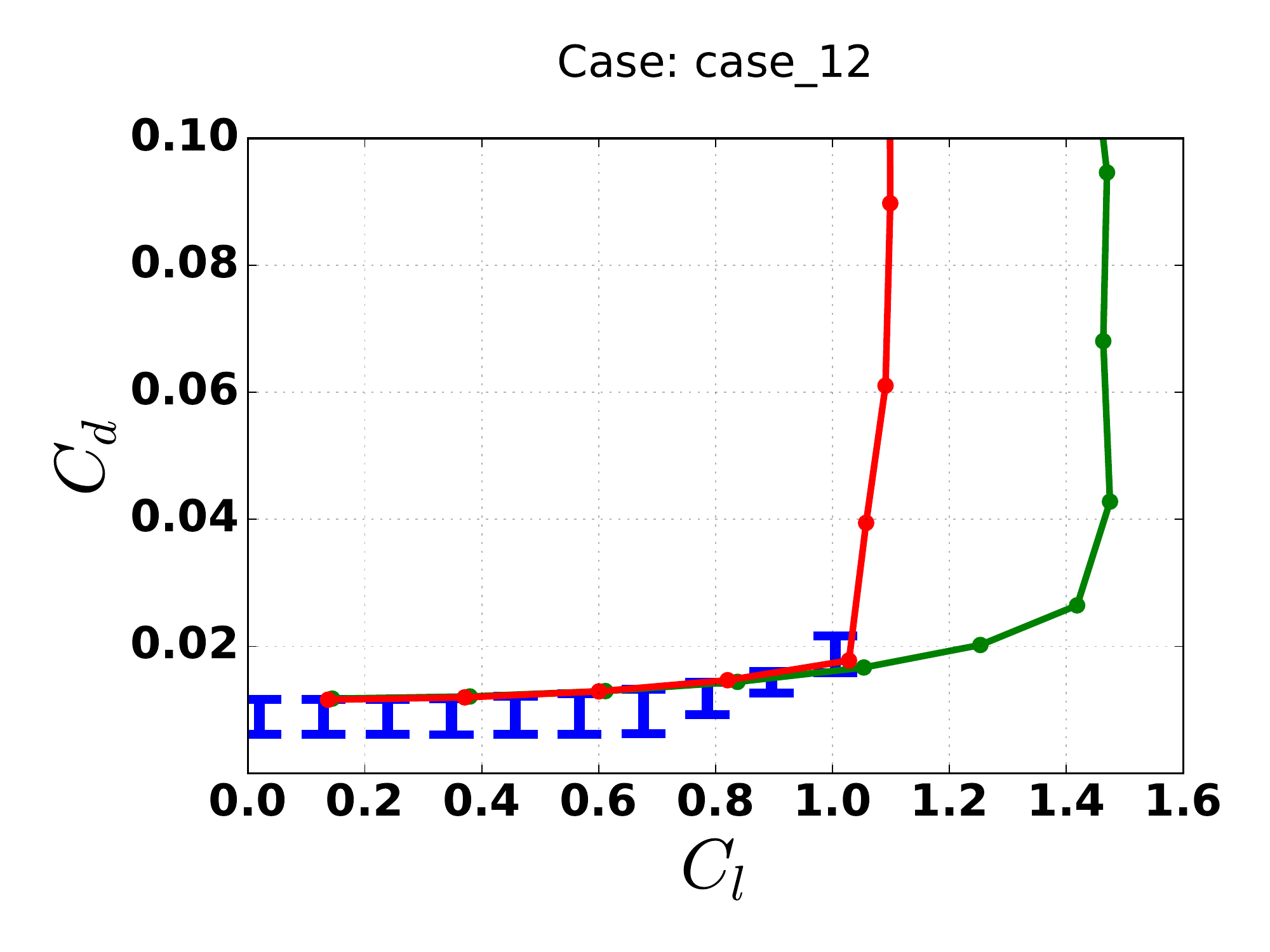}}
\caption{NN-augmented SA prediction for S809 airfoil using data-set P. {{\color{blue}\textbf{---}} Experiment}, {{\color{OliveGreen}\textbf{---}} base SA} and {{\color{red}\textbf{---}} neural network}.}
\label{figures:nnet:s809}
\end{figure*}

\begin{figure*}[!h]
\centering
\includegraphics[width=0.8\textwidth]{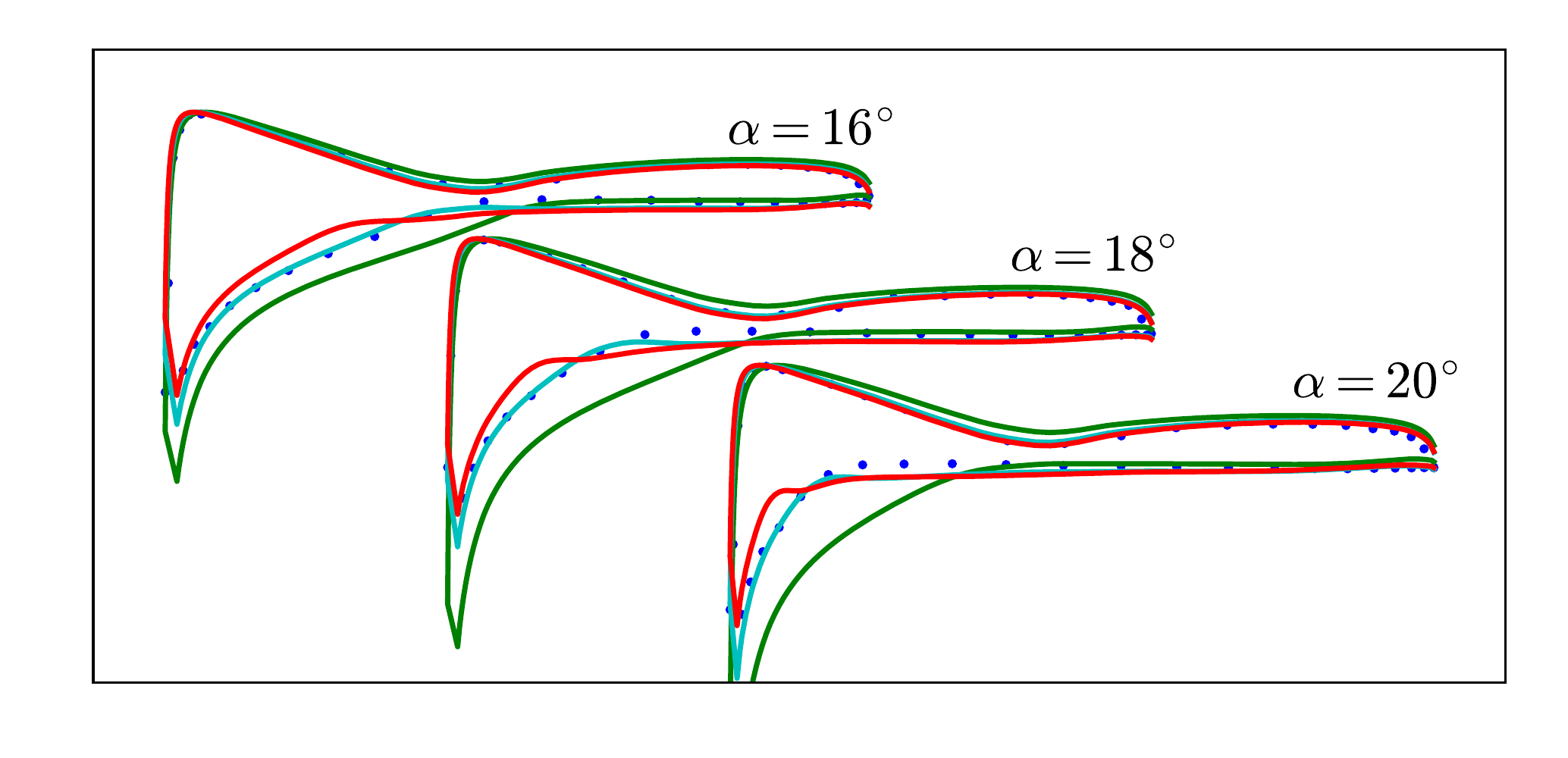}
\caption{Surface pressure coefficient for S809 airfoil at $Re = 2 \times 10^6$ and $\alpha = \{16^{\circ}, 18^{\circ}, 20^{\circ}\}$. Refer Fig. \ref{figures:s809:comparison}(c) for legend. Not to scale.}
\label{fig:nnet:s809_cp}
\end{figure*}

\begin{figure*}[!h]
\centering
\includegraphics[width=0.8\textwidth]{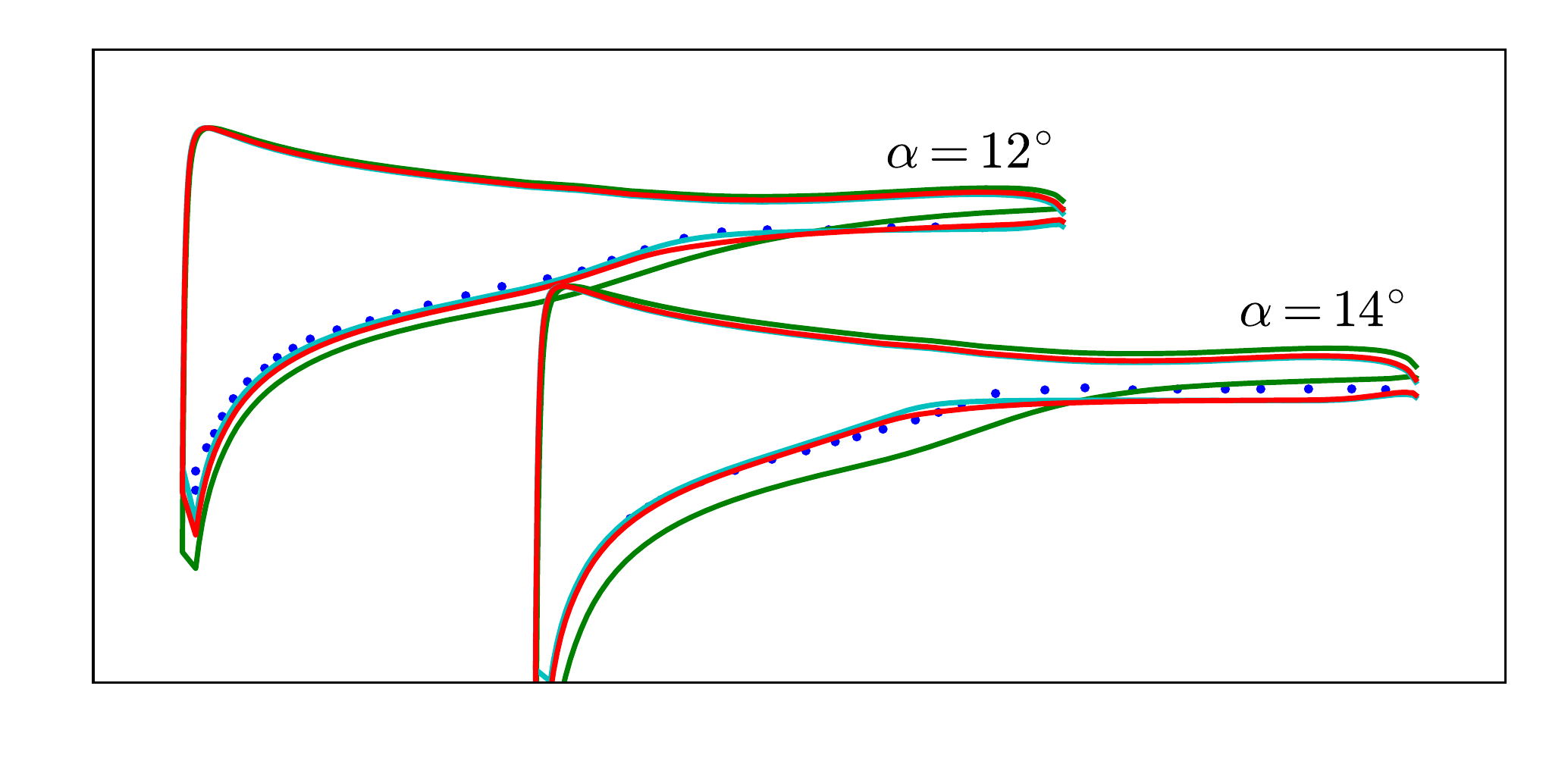}
\caption{Surface pressure coefficient for S805 airfoil at $Re = 1 \times 10^6$ and $\alpha = \{12^{\circ}, 14^{\circ}\}$. Refer Fig. \ref{figures:s809:comparison}(c) for legend. Experimental pressure is shown only for the upper surface. Not to scale.}
\label{fig:nnet:s805_cp}
\end{figure*}

\begin{figure*}[!h]
\centering
\includegraphics[width=0.8\textwidth]{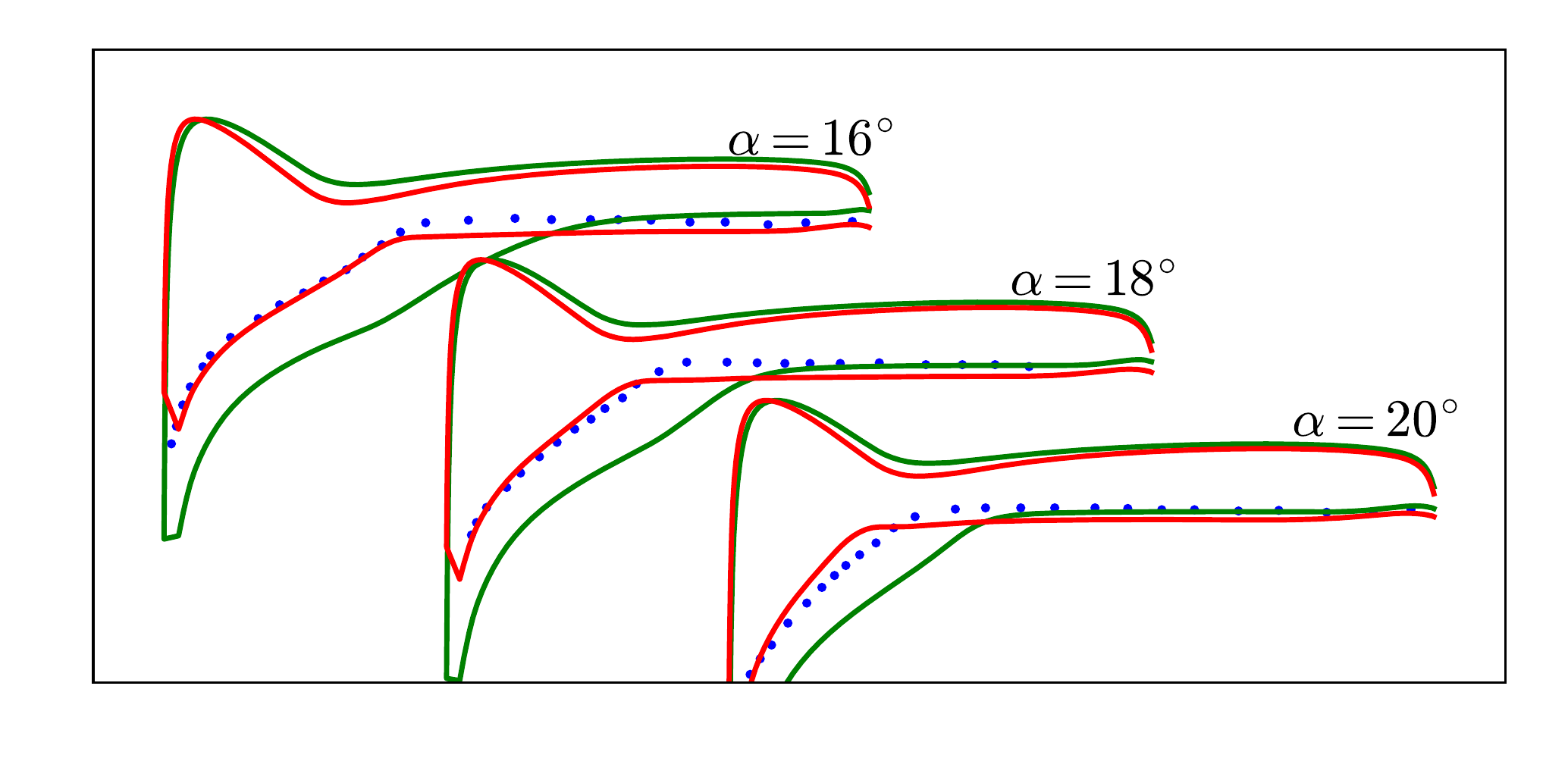}
\caption{Surface pressure coefficient for S814 airfoil at $Re = 1.5 \times 10^6$ and $\alpha = \{16^{\circ}, 18^{\circ}, 20^{\circ}\}$. Refer Fig. \ref{figures:s809:comparison}(c) for legend. Experimental pressure is shown only for the upper surface. Inversion is not performed for this case. Not to scale. }
\label{fig:nnet:s814_cp}
\end{figure*}

Fig. ~\ref{figures:nnet:s814} shows the lift and drag coefficients for all Reynolds numbers, including $Re=3 \times 10^6$, which was not used in the training set. Clearly, significant improvement in stall prediction is evident in the lift prediction. As a consequence, the drag rise is predicted  to occur at lower angles of attack than in the baseline model, a trend that is qualitatively correct. Further, there is no evidence of  deterioration of accuracy in the low angle of attack regions, where the original model is already accurate. The model performs equally well for airfoil shapes not used in the training set, i.e. S805 and S809 (Figs. \ref{figures:nnet:s805}, \ref{figures:nnet:s809}). The improvement in the quality of the predictions is further emphasized in Figs.~\ref{fig:nnet:s809_cp},~\ref{fig:nnet:s805_cp},~\ref{fig:nnet:s814_cp}. These results confirm that the NN-augmented model offers considerable predictive improvements in surface pressure distributions. Fig. ~\ref{figures:s809:grid_convergence} shows the base SA and the NN augmented SA solutions for two different grid sizes. The solutions,  using both the models, are sufficiently grid converged for the grid resolution used in this work.


\begin{table}
\centering
\begin{tabular}{|l|l|}\hline
Model label & Training data\\\hline
{\bf P } & {\bf S814} at ${\mathbf Re = 1 \times 10^6, 2 \times 10^6}$\\\hline
1 & S805 at $Re = 1 \times 10^6$\\\hline
2 & S805 at $Re = 2 \times 10^6$\\\hline
3 & S809 at $Re = 1 \times 10^6$\\\hline
4 & S809 at $Re = 2 \times 10^6$\\\hline
5 & S805 at $Re = 1 \times 10^6, 2 \times 10^6$\\\hline
6 & S809 at $Re = 1 \times 10^6, 2 \times 10^6$\\\hline
7 & S805, S809, S814 at $Re = 1 \times 10^6, 2 \times 10^6$\\\hline
\end{tabular}
\caption{List of data-sets used to study the impact of variability of the training. The main predictive model
is constructed based on data-set P. Note that $Re=3 \times 10^6$ is not included in any of the data-sets.}
\label{table:nnet}
\end{table}

\subsection{Predictive variability}

\begin{figure*}[!h]
\centering
\subfigure[$\alpha=0^\circ$]{\includegraphics[width=0.32\textwidth]{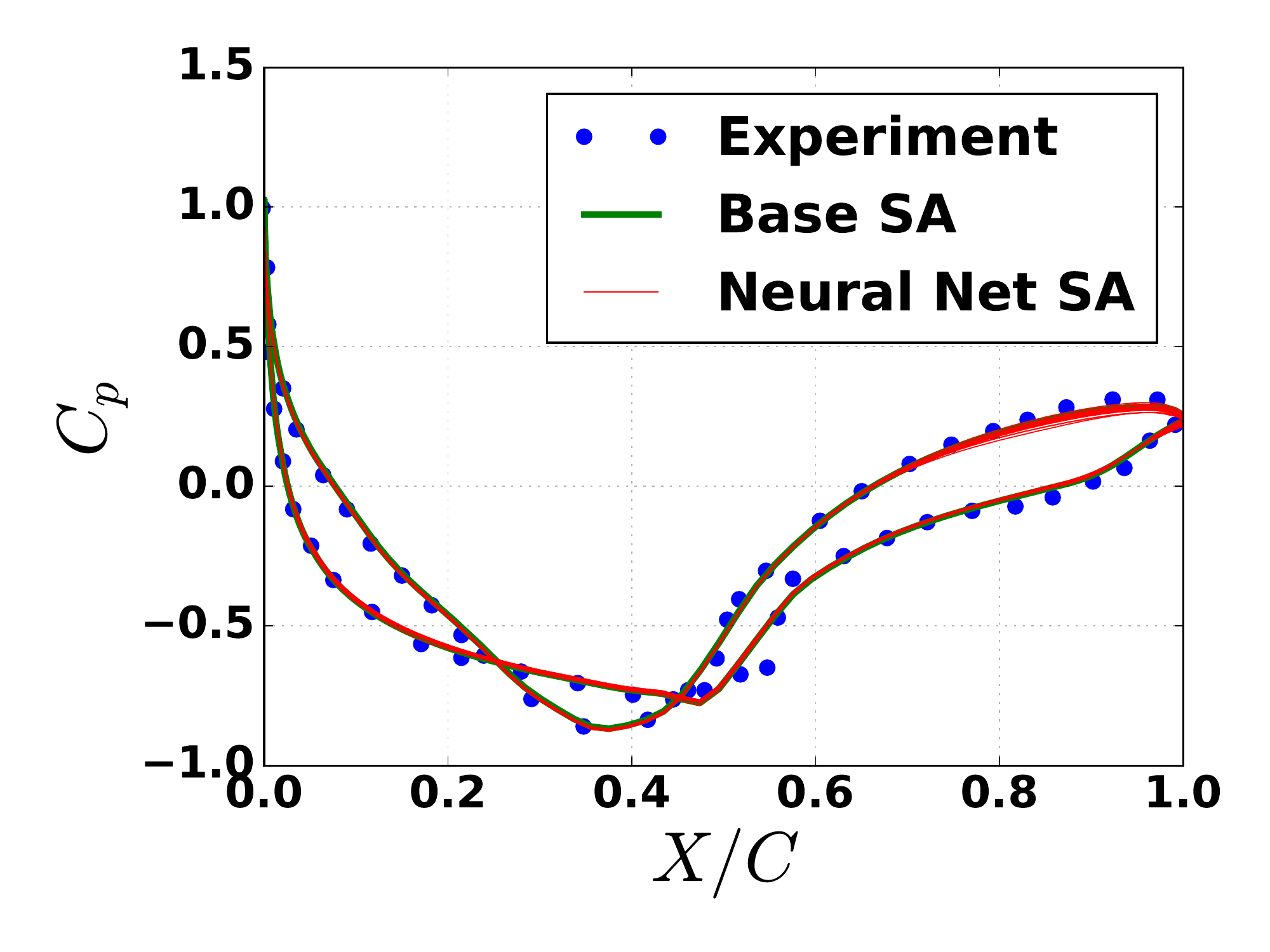}}
\subfigure[$\alpha=14^\circ$]{\includegraphics[width=0.32\textwidth]{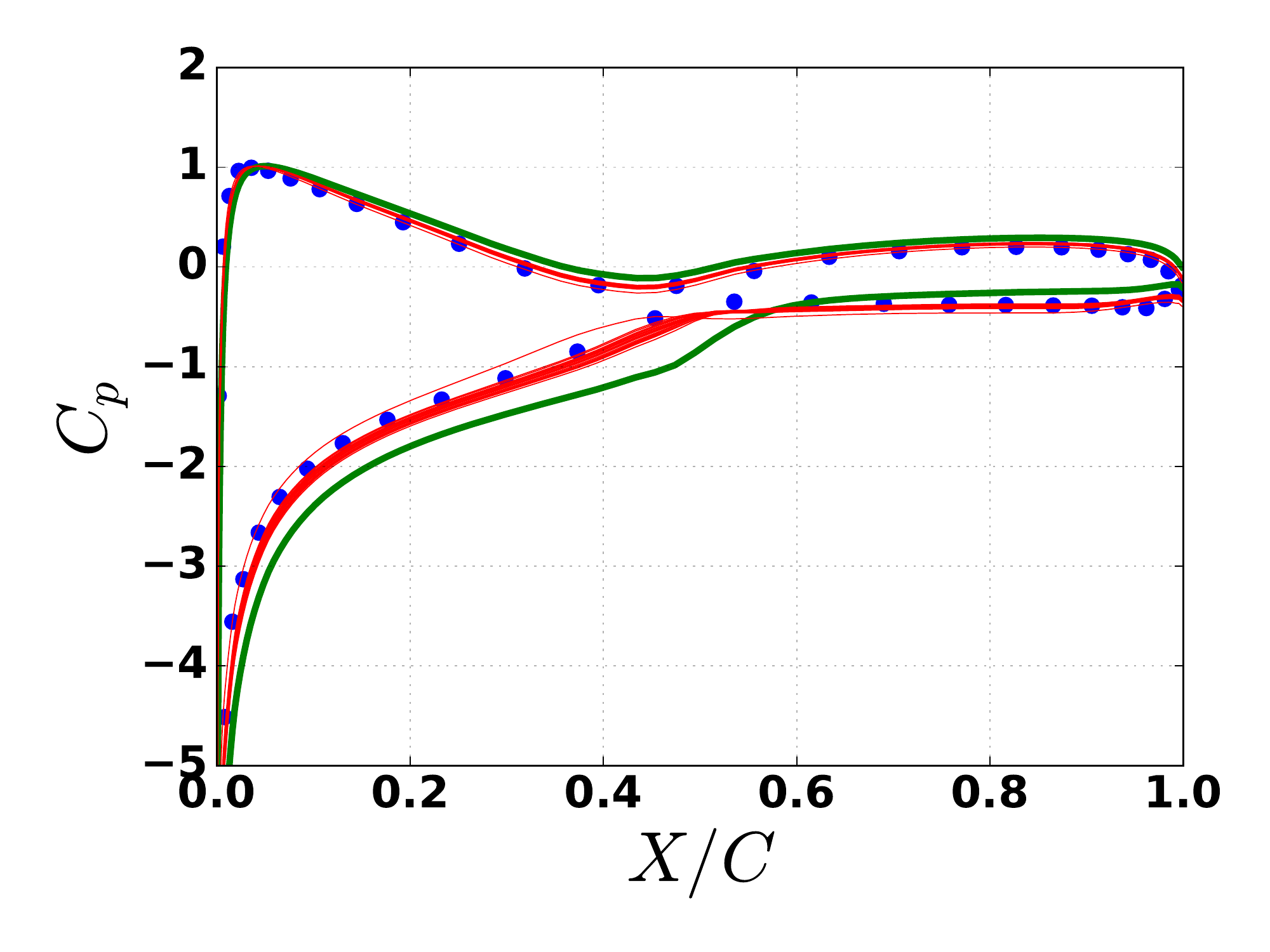}}
\subfigure[$\alpha=20^\circ$]{\includegraphics[width=0.32\textwidth]{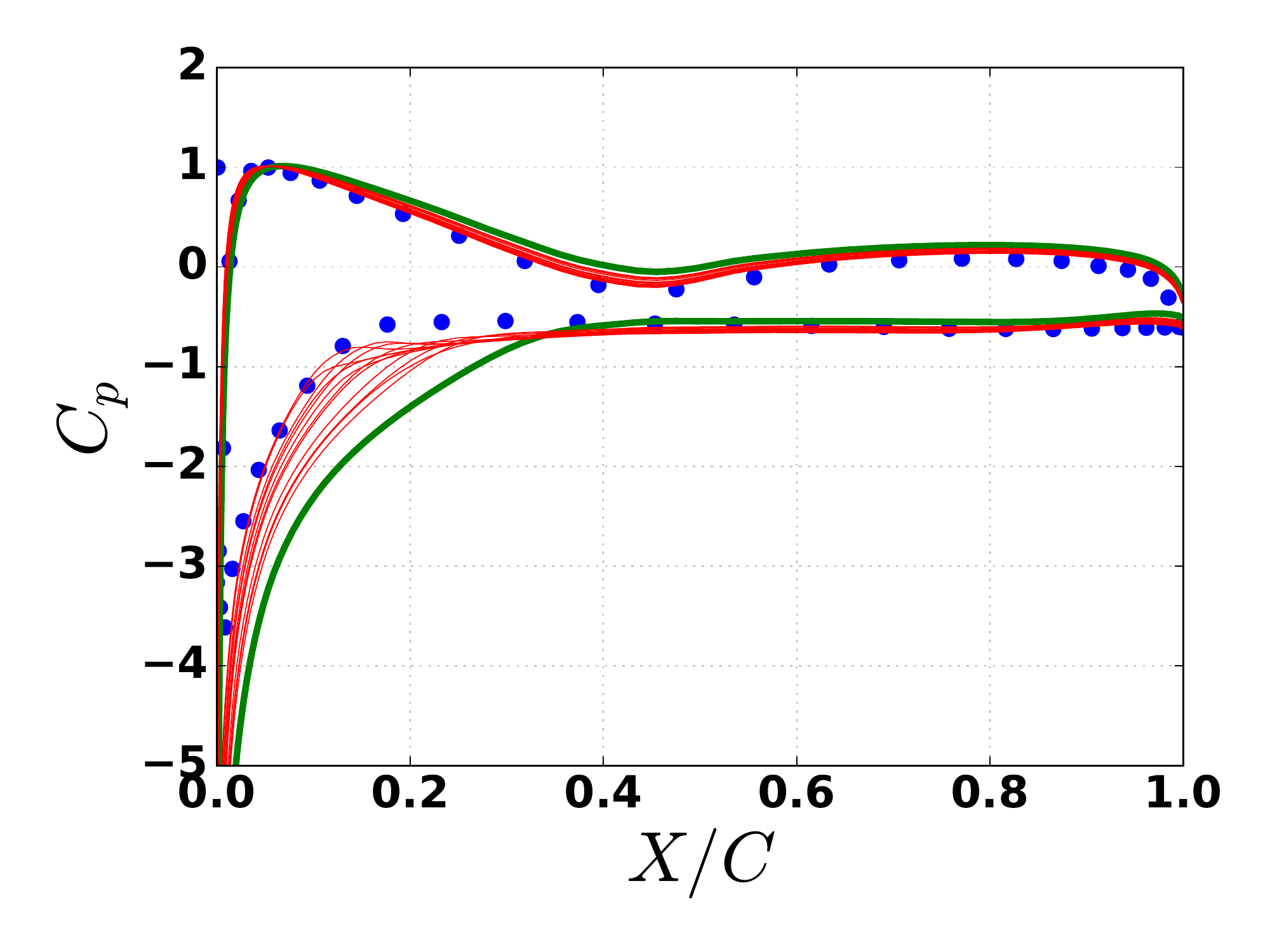}}
\caption{Predicted pressure coefficients for S809 at $Re = 2 \times 10^6$ using 8 different NN-based  models listed in table \ref{table:nnet}.}
\label{fig:nnet:samples}
\end{figure*}

It is desirable that any new modifications introduced into a turbulence model do not affect the solution to problems for which the base model is accurate. The results suggest that the NN-augmented SA model satisfies this requirement. Fig. \ref{fig:nnet:samples} showcases this feature for the S809 airfoil at a Reynolds number of $2\times 10^6$. The predicted surface pressure using neural networks trained on different data sets listed in Table \ref{table:nnet} is shown in red lines. Clearly model augmentations show variability as is apparent in Figs.~\ref{fig:nnet:samples} (b) and (c). Overall, the neural network-augmented models are more accurate than the base SA model for all the cases, and more importantly, none of the NN-augmented predictions  diverge from the base SA model at $\alpha = 0^{\circ}$. While this ensemble approach does not qualify as a formal uncertainty quantification technique, it is nevertheless a useful test to ascertain the sensitivity of the model output to the training set. If significant variabilities are revealed in the model predictions, it serves as a warning to the user that models may be operating far from conditions in which they were trained. 

Further, Fig. ~\ref{fig:nnet:samples} shows that the quality of the NN-augmented model is sensitive to the selection of the training-data. In this work, the best model ``P'' is selected by exploring several combinations of the data--sets. This observation is subjected to the uncertainty involved with the intermediate steps (feature selection, machine learning algorithm, etc.)

\subsection{Portability and Convergence}



The entire modeling framework was developed and tested on ADTURNS, a structured finite-volume flow solver augmented with adjoint optimization and neural networks. To demonstrate the portability of this approach, the NN-augmented SA model based on the data-set P is implemented into {\em AcuSolve} , a commercially available unstructured flow solver based on the Galerkin/Least-Squares (GLS) stabilized finite-element method~\cite{Hughes1989,Shakib1991}. While ADTURNS implements non-dimensionalized RANS equations, AcuSolve implements the dimensional form of RANS equations. Therefore, developing the neural network model based on a feature set consisting of locally non-dimensional flow variables, as presented in this work, is essential for portability across flow solvers.

AcuSolve is a general purpose solver that is used in a wide variety of applications such as wind power, automotive, off-shore engineering, electronics cooling, chemical mixing, bio-medical, consumer products, national laboratories, and academic research~\cite{Corson2012, AcuSolve2016,Godo2010,Lyons2009,Bagwell2006,Johnson2002}. The GLS formulation with linear shape functions provides second order accuracy for spatial discretization of all variables and utilizes tightly controlled numerical diffusion operators to obtain stability and maintain accuracy. The semi-discrete generalized-alpha method is used to integrate the equations implicitly in time for steady-state and transient simulations~\cite{Jansen2000}. The resulting system of equations is solved as a fully coupled pressure/velocity matrix system using a preconditioned iterative linear solver.

\begin{figure*}[!h]
\centering
\subfigure[$Re=1 \times 10^6$]{\includegraphics[width=0.32\textwidth,trim={0 0 0 1.40cm},clip]{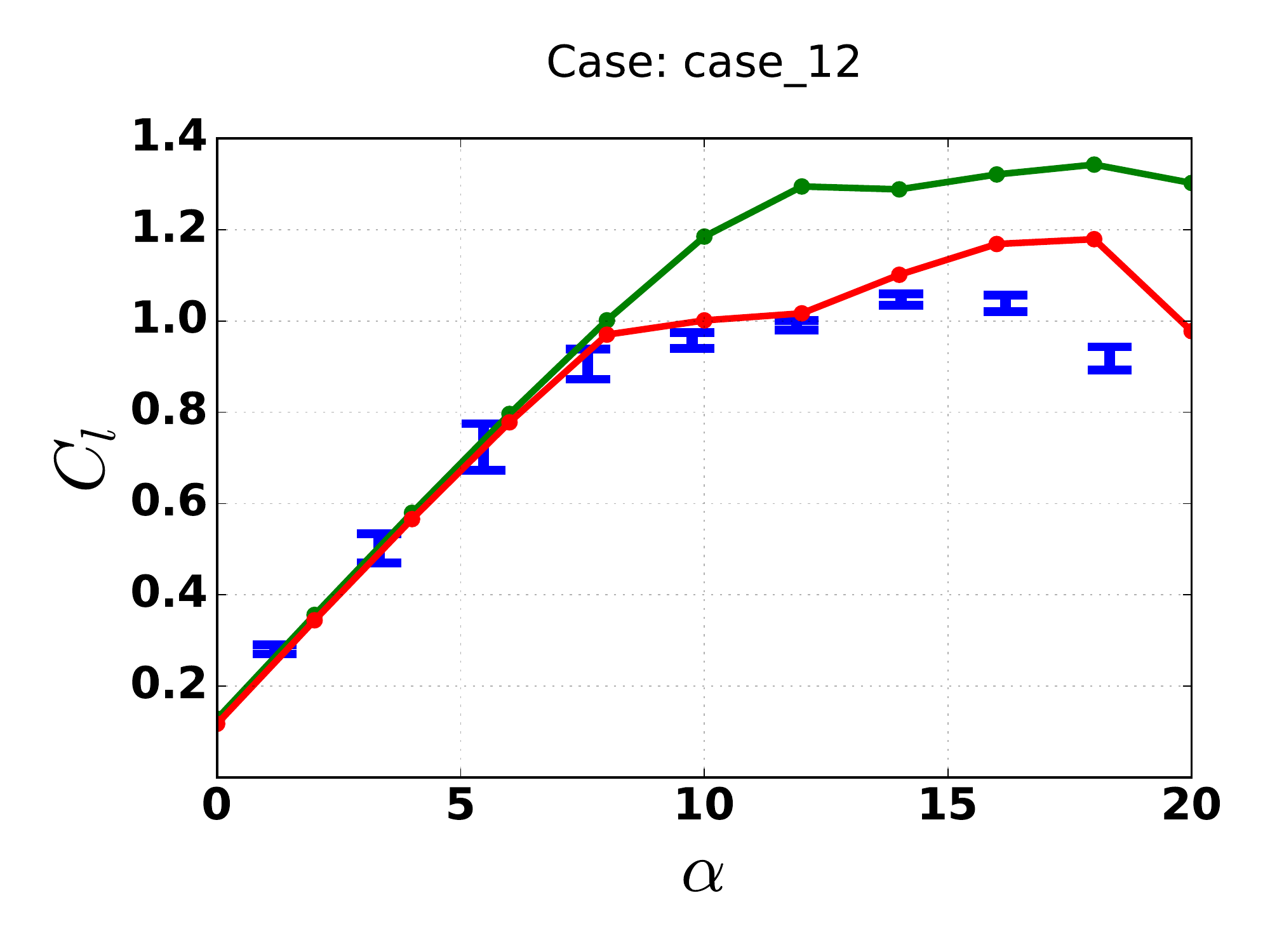}}
\subfigure[$Re=2 \times 10^6$]{\includegraphics[width=0.32\textwidth,trim={0 0 0 1.40cm},clip]{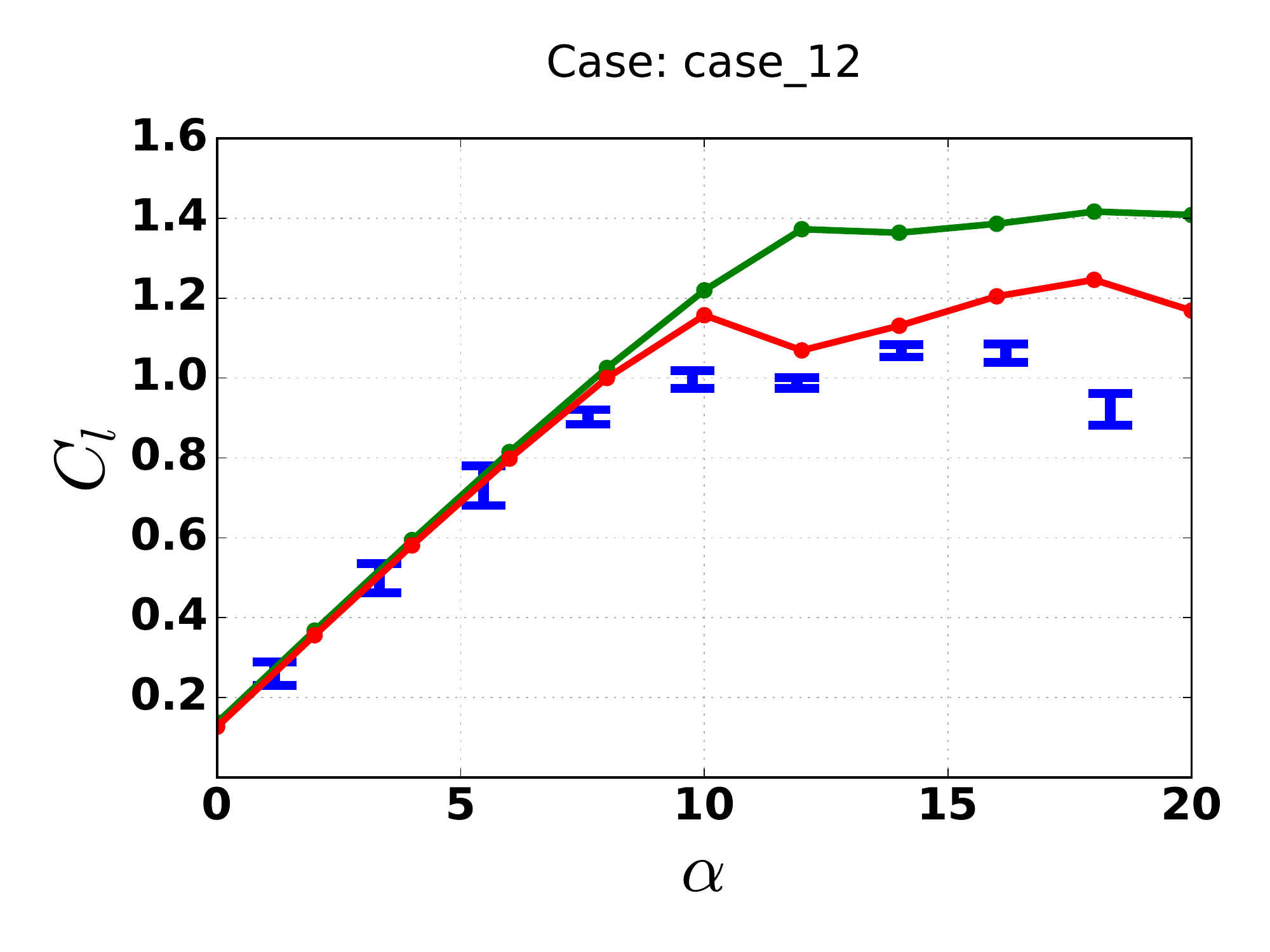}}
\subfigure[$Re=3 \times 10^6$]{\includegraphics[width=0.32\textwidth,trim={0 0 0 1.40cm},clip]{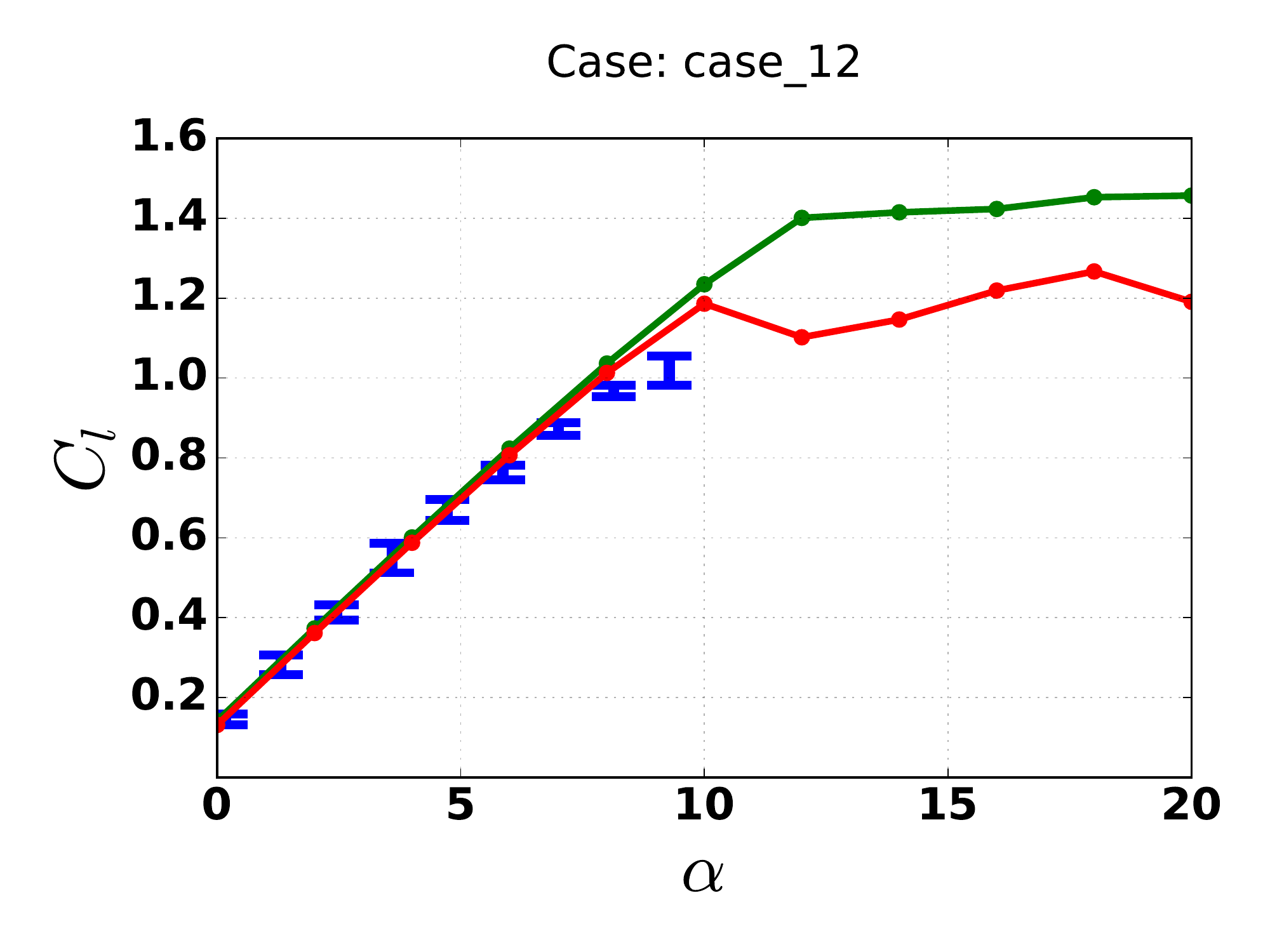}}
\subfigure[$Re=1 \times 10^6$]{\includegraphics[width=0.32\textwidth,trim={0 0 0 1.40cm},clip]{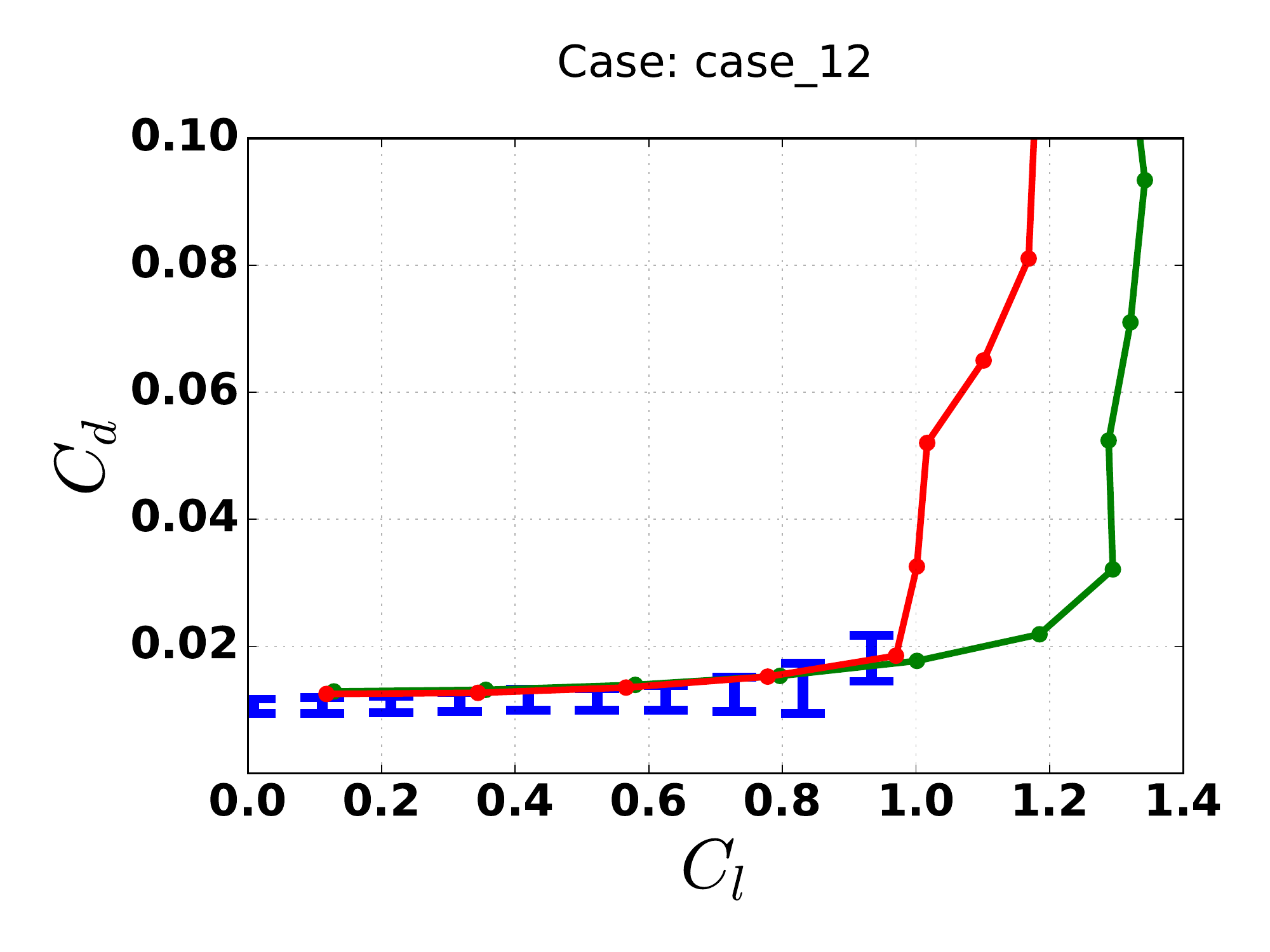}}
\subfigure[$Re=2 \times 10^6$]{\includegraphics[width=0.32\textwidth,trim={0 0 0 1.40cm},clip]{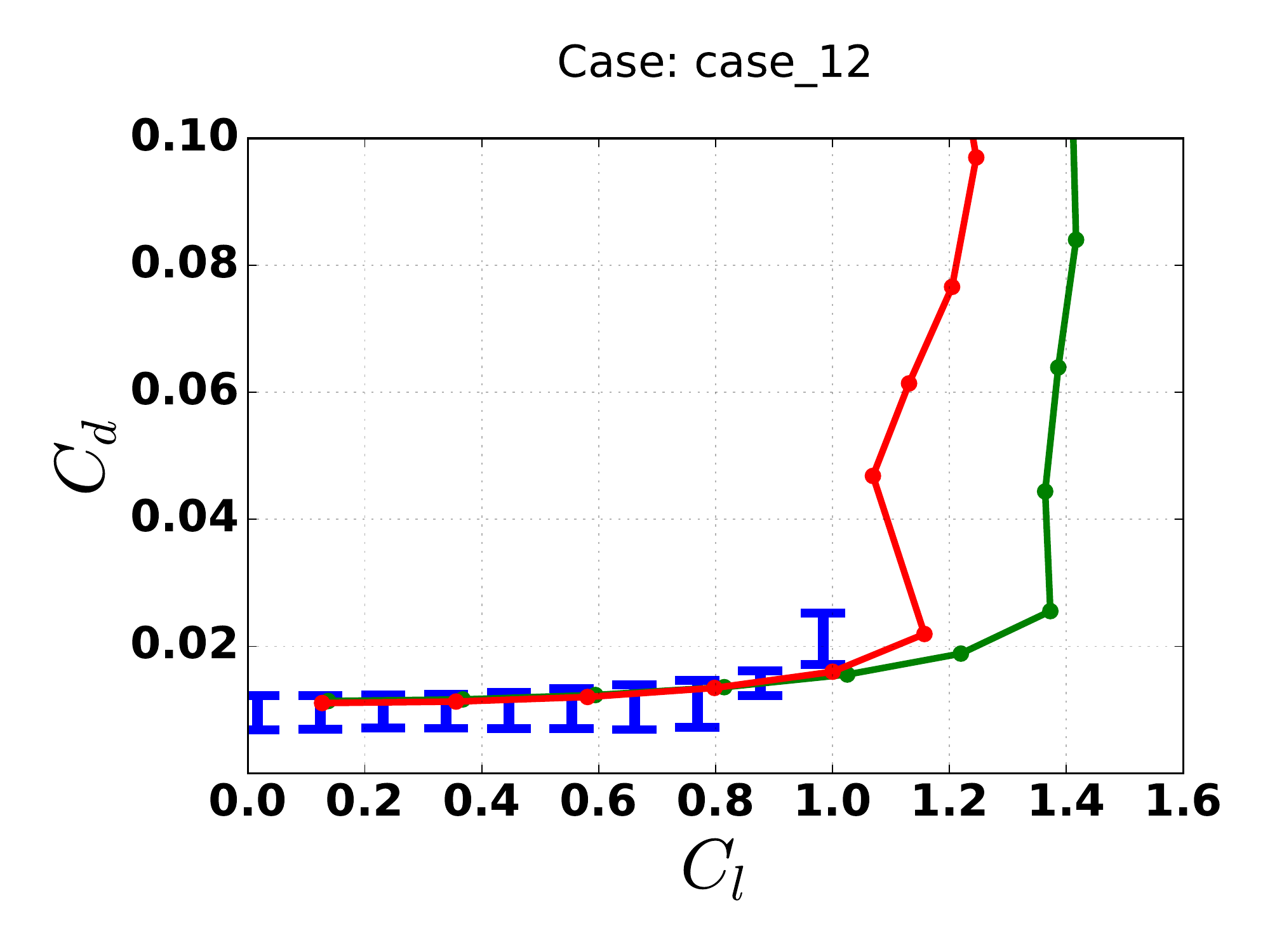}}
\subfigure[$Re=3 \times 10^6$]{\includegraphics[width=0.32\textwidth,trim={0 0 0 1.40cm},clip]{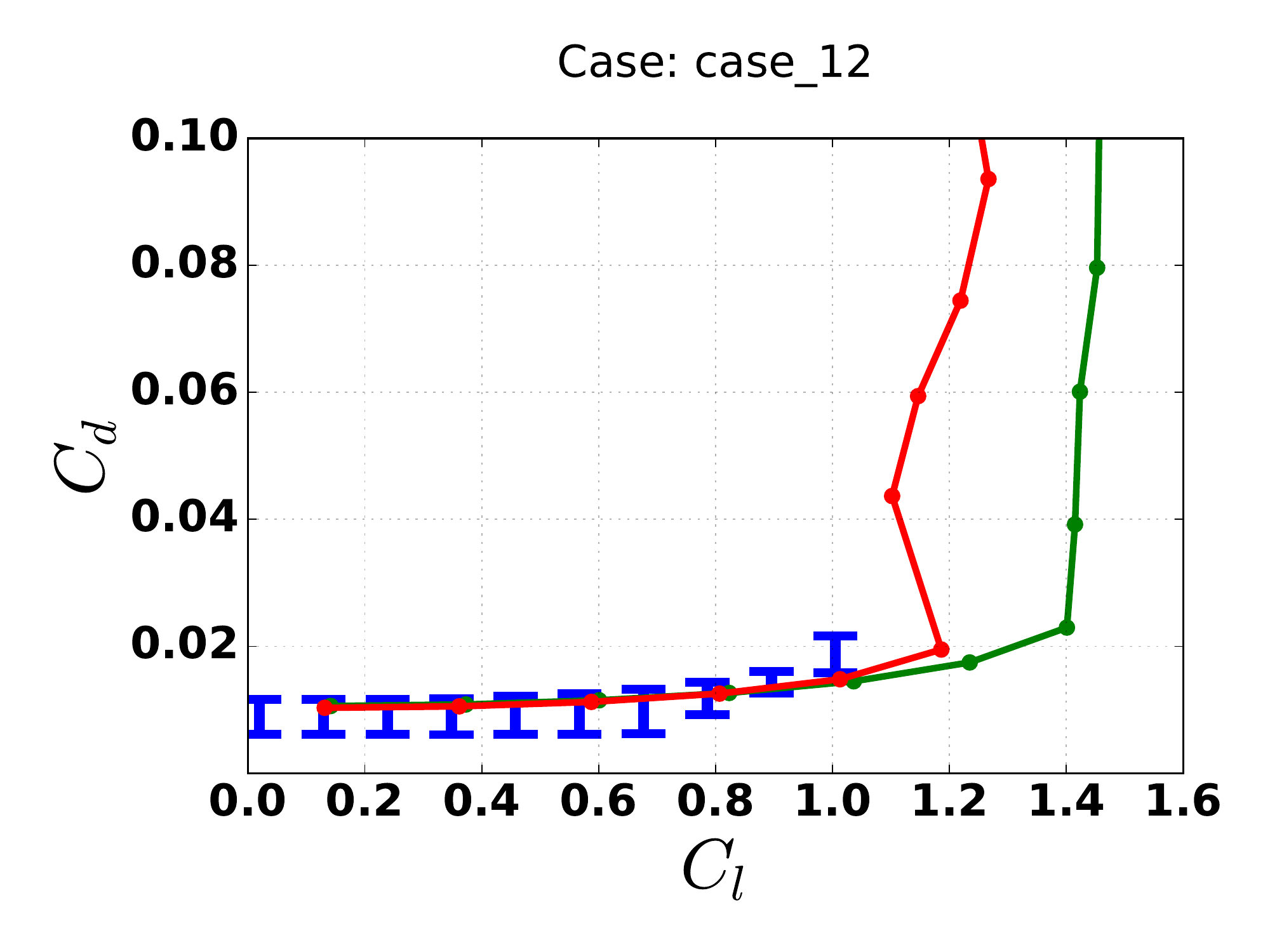}}
\caption{NN-augmented SA prediction using AcuSolve for S809 airfoil using data-set P. {{\color{blue}\textbf{---}} Experiment}, {{\color{OliveGreen}\textbf{---}} base SA} and {{\color{red}\textbf{---}} neural network}.}
\label{figures:acusolve:s809_forces}
\end{figure*}


Fig. \ref{figures:acusolve:s809_forces} shows lift and drag coefficient predictions from AcuSolve for the S-809 airfoil at three Reynolds numbers. The NN-augmentation shows significant improvement in predictions and its effectiveness is comparable to that observed in the ADTURNS solver framework. It should be noted that the AcuSolve uses a variation of the SA model which corrects for the rotation and the curvature effects. These corrections are not used in the ADTURNS code and therefore the solutions from these two codes are not expected to be identical, even for the baseline model.

\begin{figure*}[!h]
\centering
\subfigure[Lift coefficient]{\includegraphics[width=0.4\textwidth]{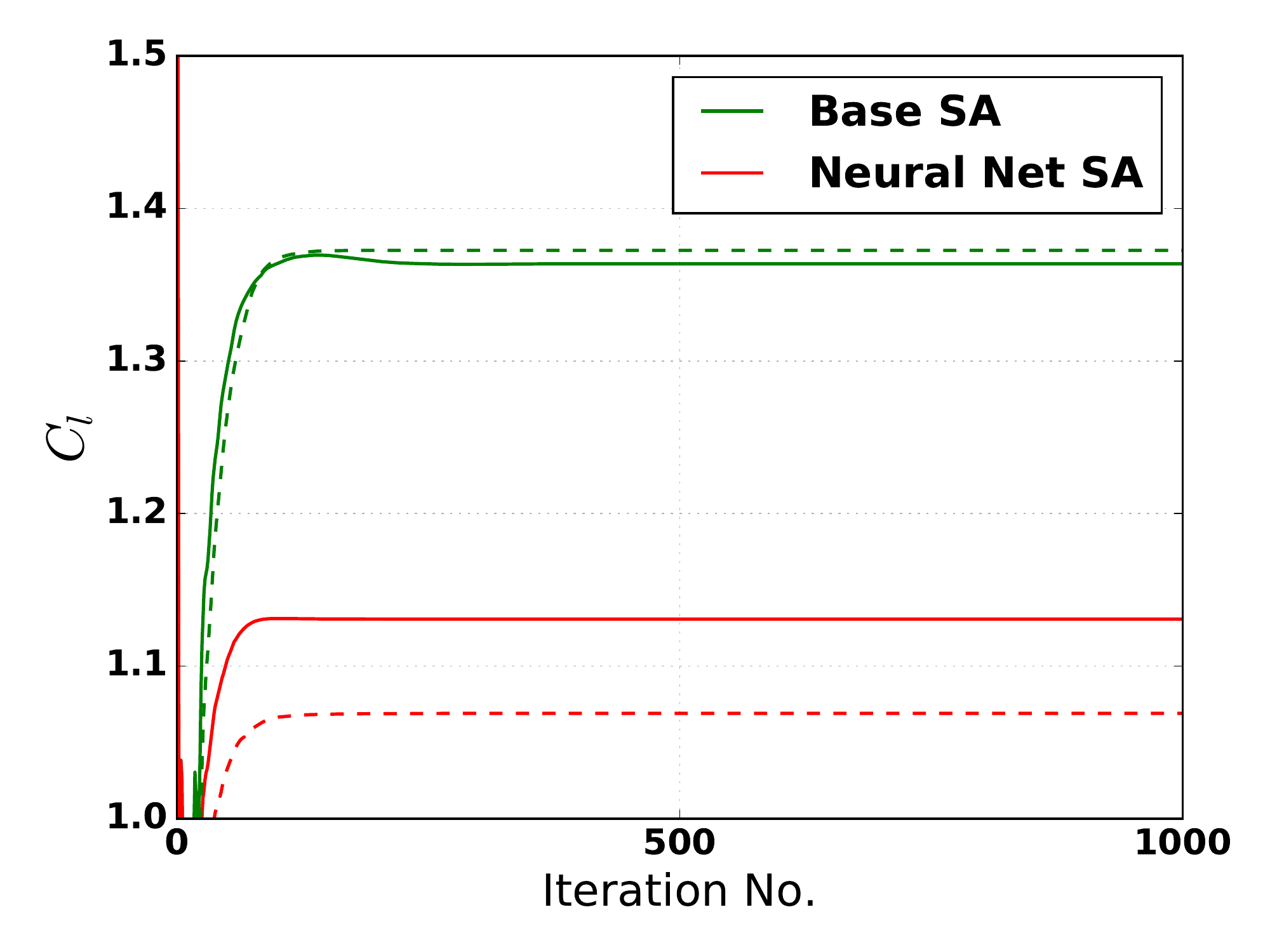}}
\subfigure[L2 norm of solver residual]{\includegraphics[width=0.4\textwidth]{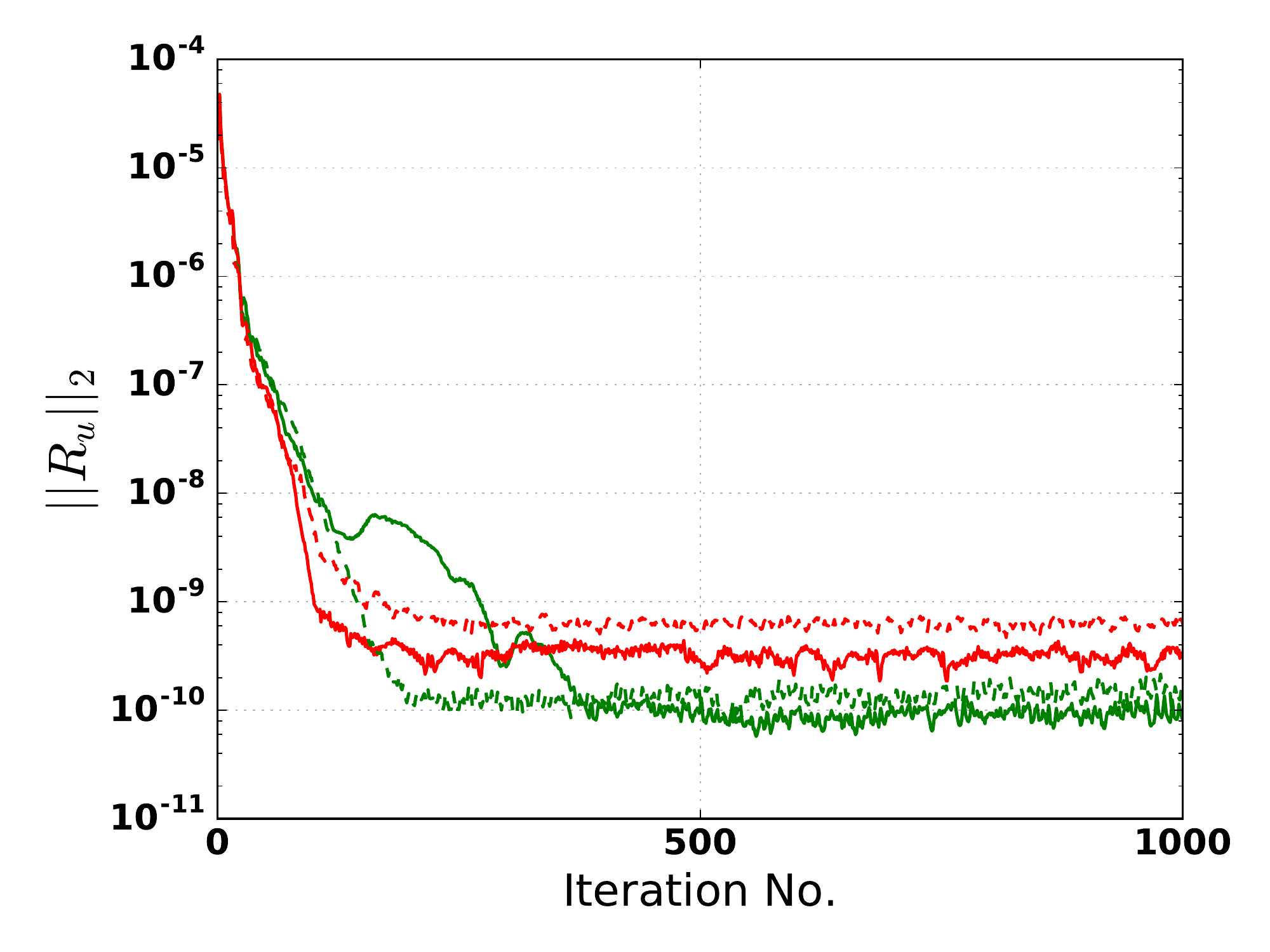}}
\caption{AcuSolve's convergence history for S809 airfoil at $Re = 2 \times 10^6$, $\alpha = 12^{\circ}$ (dashed) and $\alpha = 14^{\circ}$ (solid).}
\label{figures:acusolve:s809_forcehist}
\end{figure*}

Fig. \ref{figures:acusolve:s809_forcehist} shows the rate of convergence for the base SA and the NN-augmented SA for a sample  problem. The initial condition was taken to be uniform free-stream for all the runs.  The NN-augmented model displays comparable convergence characteristics to the baseline model, thus demonstrating the portability of the approach. Additional overhead exists in passing the features ${ \boldsymbol  \eta}$ to the ANN and obtaining $\beta$ at grid locations. This was confirmed to add $<10\%$ of additional compute time compared to the baseline calculation.

%
%
%

\section{Summary and perspectives}
A data-driven framework comprising of full-field inversion and machine learning was used to develop predictive capabilities for the modeling of  turbulent separated flows over airfoils. This framework is embedded in a traditional RANS solver to improve the applicability of the Spalart-Allmaras (SA) turbulence model to strong adverse pressure gradients present in flow past airfoils pre- and post-stall. With a view towards assimilating sparse data from a wide range of flows, the inversion process was formulated as an optimization problem to minimize the difference between the experimentally measured lift coefficient and the model output. In contrast to parametric inversion, the turbulence model discrepancy was inferred as a field (i.e. at every grid point in the solution domain).  The resulting model-correction function was then reconstructed using an artificial neural network (NN) as a function of locally non-dimensional flow quantities such as the ratio of eddy to kinematic viscosity,  vorticity to strain-rate magnitude. 

During the predictive process, the NN is queried at every iteration of the flow solver to obtain model corrections which are embedded into the predictive model. The resulting data-augmented turbulence model was then used for predictive simulations of airfoils and flow conditions that were not part of the neural network training. Extensive tests were made and the following conclusions were observed:

$\bullet$ The data-assisted SA model showed significant improvement over the baseline model in predicting lift and drag coefficients and stall onset angles. 

$\bullet$ The model predictions were confirmed to be significantly improved for airfoil shapes and flow conditions that were not part of the training set.

$\bullet$ No 
deterioration of accuracy was noticed in situations (low angles of attack) in  which the original model was accurate. 

$\bullet$ Though the inference process used only the lift-coefficient data, the NN-augmented model was demonstrated to provide
considerable predictive improvements of surface pressure distributions. This reinforces confidence that the procedure does not overfit the model to the lift data and that predictive improvements are realized for the right reasons.

$\bullet$ An ensemble of predictions based on different training sets was used to assess the sensitivity of the model outputs to the training data. While there was expected variability in the results, the model augmentations brought predictions closer to the experimental results for all training sets.

$\bullet$ Solver convergence was assessed and the cost overhead for the NN augmentations was observed to be minimal.

$\bullet$ Portability of the approach was demonstrated by generating the data-assisted SA model using a structured finite-volume solver and then using it in AcuSolve, a commercial, unstructured finite-element solver. The predictive improvements were confirmed to be preserved across both solvers.

While the present work was focused on demonstrating the {\em potential} of data-driven approaches based on field inference and learning, much work remains to be done in developing turbulence models for application in more general settings.  In such situations, the ensemble of data-sets (as in Fig.~\ref{fig:nnet:samples}) may be utilized as a rule-of-thumb to indicate the variability of the machine learning-augmented model. However, a more formal uncertainty quantification approach that takes into account the uncertainty in the data, variability of the training process and confidence in the baseline model may be desirable. A simplified form of such an approach has been proposed (and demonstrated for much simpler problems) in Ref.~\citenum{paradigm}.

One  immediate avenue to extend this effort is to target specific engineering applications where current CFD predictions are unsatisfactory. In this situation, customized modeling augmentations can be generated to improve predictions. 
Furthermore, industrial applications tend to be focused on a class of problems (for example, wind turbine rotors or turbomachinery blades) in which experimental data may be available in some regimes. These types of problems are most amenable to data-augmented modeling.

The philosophy and formalisms employed in this work are of a general nature and are not restricted to the type of model or the type of model discrepancy that is addressed. The inversion/learning/embedding procedure can be applied, for instance, to address discrepancies in the Reynolds stress anisotropy~\cite{tracey1,xiao2}. In such endeavors, it would be critical to ensure that the model augmentations a) include available experimental data (as one will always be hard-pressed to obtain LES and DNS data in regimes of interest), b) do not influence regions of the flow that are adequately represented by the baseline model (near-wall region in thin boundary layers), and c) do not degrade the convergence properties of the solver. There is much to be gained by carefully exploring a broader set of input features~\cite{ling,xiao2} and alternative machine learning methods~\cite{weatheritt,helen_arxiv}. Finally, respecting realizability limits~\cite{memory1,xiao2} and invariance properties~\cite{ling2} will be necessary to constrain the model, especially when the model is operating in an extrapolatory mode.

%

\begin{center}
\section*{Acknowledgments}
\end{center}
This work was funded by the NASA Aeronautics Research Institute (NARI) under the Leading Edge Aeronautics Research for NASA program (Monitors: Koushik Datta \& Gary Coleman). A portion of this work was performed as part of a summer internship program at Altair Engineering, Inc. ZeJia Zhang and Brendan Tracey assisted with early use of machine learning. The authors acknowledge insightful discussions with Paul Durbin and Juan Alonso.
\vspace{3mm}

\bibliography{main}{}

\begin{thebibliography}{10}
\newcommand{\enquote}[1]{``#1''}

\bibitem{Slotnick}
Slotnick, J., Khodadoust, J., Alonso, J., Darmofal, D., Gropp, W., Lurie, E.,
  and Mavriplis, D., \enquote{{CFD} vision 2030 study: {A} path to
  revolutionary computational aerosciences,} {\em Technical report, NASA
  Langley Research Center, NASA/CR-2014-218178\/}, 2013.

\bibitem{gerolymos}
Gerolymos, G.~A., Sauret, E., and Vallet, I., \enquote{Contribution to
  single-point closure {Reynolds}-stress modelling of inhomogeneous flow,} {\em
  Theoretical and Computational Fluid Dynamics\/}, Vol.~17, No. 5-6, 2004,
  pp.~407--431.

\bibitem{gatski}
Younis, B., Gatski, T., and Speziale, C.~G., \enquote{Towards a rational model
  for the triple velocity correlations of turbulence,} {\em Proceedings of the
  Royal Society of London A: Mathematical, Physical and Engineering
  Sciences\/}, Vol.~456, No. 1996, 2000, pp.~909--920.

\bibitem{poroseva2014velocity}
Poroseva, S. and Murman, S.~M., \enquote{Velocity/Pressure-Gradient
  Correlations in a FORANS Approach to Turbulence Modeling,} {\em 44th AIAA
  Fluid Dynamics Conference, AIAA Aviation, (AIAA 2014-2207)\/}, Jun 2014.

\bibitem{ell1}
Durbin, P.~A., \enquote{Near-wall turbulence closure modeling without
  “damping functions”,} {\em Theoretical and Computational Fluid
  Dynamics\/}, Vol.~3, No.~1, 1991, pp.~1--13.

\bibitem{ell2}
Durbin, P., \enquote{A {Reynolds} stress model for near-wall turbulence,} {\em
  Journal of Fluid Mechanics\/}, Vol.~249, 1993, pp.~465--498.

\bibitem{Milano2002}
Milano, M. and Koumoutsakos, P., \enquote{Neural network modeling for near wall
  turbulent flow,} {\em Journal of Computational Physics\/}, Vol.~182, No.~1,
  2002, pp.~1--26.

\bibitem{Marusic2001}
Marusic, I., Candler, G., Interrante, V., Subbareddy, P., and Moss, A.,
  \enquote{Real time feature extraction for the analysis of turbulent flows,}
  {\em Data Mining for Scientific and Engineering Applications\/}, 2001,
  pp.~223--238.

\bibitem{Yarlanki2012}
Yarlanki, S., Rajendran, B., and Hamann, H., \enquote{Estimation of turbulence
  closure coefficients for data centers using machine learning algorithms,}
  {\em Thermal and Thermomechanical Phenomena in Electronic Systems (ITherm),
  2012 13th IEEE Intersociety Conference on\/}, 2012, pp.~38--42.

\bibitem{japanese}
Kato, H. and Obayashi, S., \enquote{Data Assimilation for Turbulent Flows,}
  {\em 16th AIAA Non-Deterministic Approaches Conference, AIAA SciTech, (AIAA
  2014-1177)\/}, Jan 2014.

\bibitem{bayes1}
Edeling, W., Cinnella, P., Dwight, R.~P., and Bijl, H., \enquote{Bayesian
  estimates of parameter variability in the k--$\varepsilon$ turbulence model,}
  {\em Journal of Computational Physics\/}, Vol.~258, 2014, pp.~73--94.

\bibitem{arunjatesan}
Ray, J., Lefantzi, S., Arunajatesan, S., and DeChant, L.~J., \enquote{Bayesian
  Calibration of a RANS Model with a Complex Response Surface - A Case Study
  with Jet-in-Crossflow Configuration,} {\em 45th AIAA Fluid Dynamics
  Conference Dallas, TX\/}, Jun 2015.

\bibitem{cheung2011}
Cheung, S.~H., Oliver, T.~A., Prudencio, E.~E., Prudhomme, S., and Moser,
  R.~D., \enquote{Bayesian uncertainty analysis with applications to turbulence
  modeling,} {\em Reliability Engineering \& System Safety\/}, Vol.~96, No.~9,
  2011, pp.~1137--1149.

\bibitem{oliver_moser}
Oliver, T.~A. and Moser, R.~D., \enquote{Bayesian uncertainty quantification
  applied to {RANS} turbulence models,} {\em Journal of Physics: Conference
  Series\/}, Vol.~318, No.~4, 2011, pp.~042032.

\bibitem{beck}
Cheung, S.~H. and Beck, J., \enquote{New {Bayesian} updating methodology for
  model validation and robust predictions based on data from hierarchical
  subsystem tests,} {\em Earthquake Engineering Research Laboratory,
  Caltech\/}, , No. CaltechEERL:EERL-2008-04, Jan 2009.

\bibitem{wang}
Dow, E. and Wang, Q., \enquote{Uncertainty Quantification of Structural
  Uncertainties in {RANS} Simulations of Complex Flows,} {\em 20th AIAA
  Computational Fluid Dynamics Conference Honolulu, Hawaii\/}, Jun 2011.

\bibitem{wang2}
Dow, E. and Wang, Q., \enquote{Quantification of Structural Uncertainties in
  the k- $\omega$ Turbulence Model,} {\em 52nd AIAA/ASME/ASCE/AHS/ASC
  Structures, Structural Dynamics and Materials Conference, Denver,
  Colorado\/}, Apr 2011.

\bibitem{memory1}
Emory, M., Pecnik, R., and Iaccarino, G., \enquote{Modeling Structural
  Uncertainties in {Reynolds}-Averaged Computations of Shock/Boundary Layer
  Interactions,} {\em 49th AIAA Aerospace Sciences Meeting including the New
  Horizons Forum and Aerospace Exposition Orlando, Florida\/}, Jan 2011.

\bibitem{gorle1}
Gorle, C., Emory, M., and Iaccarino, G., \enquote{{RANS} modeling of turbulent
  mixing for a jet in supersonic cross flow: model evaluation and uncertainty
  quantification,} {\em Proceedings of the Seventh International Symposium on
  Turbulence, Heat and Mass Transfer (ICHMT Digital Library Online, Palermo,
  Italy, 2012)\/}, 2012.

\bibitem{memory2}
Emory, M., Larsson, J., and Iaccarino, G., \enquote{Modeling of structural
  uncertainties in {Reynolds}-averaged {Navier}-{Stokes} closures,} {\em
  Physics of Fluids\/}, Vol.~25, No.~11, 2013, pp.~110822.

\bibitem{tracey1}
Tracey, B., Duraisamy, K., and Alonso, J., \enquote{Application of Supervised
  Learning to Quantify Uncertainties in Turbulence and Combustion Modeling,}
  {\em 51st AIAA Aerospace Sciences Meeting including the New Horizons Forum
  and Aerospace Exposition Grapevine (Dallas/Ft. Worth Region), Texas\/}, Jan
  2013.

\bibitem{xiao1}
Xiao, H., Wu, J.-L., Wang, J.-X., Sun, R., and Roy, C., \enquote{Quantifying
  and Reducing Model-Form Uncertainties in Reynolds-Averaged Navier-Stokes
  Equations: An Open-Box, Physics-Based, Bayesian Approach,} {\em arXiv
  preprint {\tt arXiv:1508.06315}\/}, 2015.

\bibitem{weatheritt}
Weatheritt, J., {\em The development of data driven approaches to further
  turbulence closures\/}, Ph.D. thesis, University of Southampton, 2015.

\bibitem{ling}
Ling, J. and Templeton, J., \enquote{Evaluation of machine learning algorithms
  for prediction of regions of high Reynolds averaged {Navier} {Stokes}
  uncertainty,} {\em Physics of Fluids\/}, Vol.~27, No.~8, 2015, pp.~085103.

\bibitem{king}
King, R., Hamlington, P.~E., and Dahm, W., \enquote{Autonomic Subgrid-Scale
  Closure for Large Eddy Simulations,} {\em 53rd AIAA Aerospace Sciences
  Meeting, AIAA SciTech, Kissimmee, Florida\/}, Jan 2015.

\bibitem{duraisamy2015new}
Duraisamy, K., Zhang, Z.~J., and Singh, A.~P., \enquote{New Approaches in
  Turbulence and Transition Modeling Using Data-driven Techniques,} {\em 53rd
  AIAA Aerospace Sciences Meeting, AIAA SciTech, Kissimmee, Florida\/}, Jan
  2015.

\bibitem{companion}
Tracey, B.~D., Duraisamy, K., and Alonso, J.~J., \enquote{A Machine Learning
  Strategy to Assist Turbulence Model Development,} {\em 53rd AIAA Aerospace
  Sciences Meeting, AIAA SciTech, Kissimmee, Florida\/}, Jan 2015.

\bibitem{paradigm}
Parish, E.~J. and Duraisamy, K., \enquote{A paradigm for data-driven predictive
  modeling using field inversion and machine learning,} {\em J. Comput.
  Physics\/}, Vol.~305, 2016, pp.~758--774.

\bibitem{PoF_anand}
Singh, A.~P. and Duraisamy, K., \enquote{Using field inversion to quantify
  functional errors in turbulence closures,} {\em Physics of Fluids\/},
  Vol.~28, No.~4, 2016, pp.~045110.

\bibitem{xiao2}
Wang, J.-X., Wu, J.-L., and Xiao, H., \enquote{Physics-Informed Machine
  Learning for Predictive Turbulence Modeling: Using Data to Improve RANS
  Modeled Reynolds Stresses,} {\em arXiv preprint {\tt arXiv:1606.07987}\/},
  2016.

\bibitem{celic2006}
Celic, A. and Hirschel, E.~H., \enquote{Comparison of eddy-viscosity turbulence
  models in flows with adverse pressure gradient,} {\em AIAA Journal\/},
  Vol.~44, No.~10, 2006, pp.~2156--2169.

\bibitem{menter1994}
Menter, F.~R., \enquote{Two-equation eddy-viscosity turbulence models for
  engineering applications,} {\em AIAA Journal\/}, Vol.~32, No.~8, 1994,
  pp.~1598--1605.

\bibitem{wilcox2008}
Wilcox, D.~C., \enquote{Formulation of the k-$\omega$ turbulence model
  revisited,} {\em AIAA Journal\/}, Vol.~46, No.~11, 2008, pp.~2823--2838.

\bibitem{rung2003}
Rung, T., Bunge, U., Schatz, M., and Thiele, F., \enquote{Restatement of the
  Spalart-Allmaras eddy-viscosity model in strain-adaptive formulation,} {\em
  AIAA Journal\/}, Vol.~47, No.~7, 2003, pp.~1396--1399.

\bibitem{spalart}
Spalart, P. and Allmaras, S., \enquote{A one-equation turbulence model for
  aerodynamic flows,} {\em 30th Aerospace Sciences Meeting and Exhibit Reno,
  NV\/}, Jan 1992.

\bibitem{duraisamy_turns1}
Duraisamy, K., McCroskey, W.~J., and Baeder, J.~D., \enquote{Analysis of wind
  tunnel wall interference effects on subsonic unsteady airfoil flows,} {\em
  Journal of {Aircraft}\/}, Vol.~44, No.~5, 2007, pp.~1683--1690.

\bibitem{duraisamy_turns2}
Bremseth, J. and Duraisamy, K., \enquote{Computational analysis of vertical
  axis wind turbine arrays,} {\em Theoretical and Computational Fluid
  Dynamics\/}, Vol.~30, No.~5, 2016, pp.~387--401.

\bibitem{vinod_turns}
Lakshminarayan, V.~K. and Duraisamy, K., \enquote{Adjoint-based estimation and
  control of spatial, temporal and stochastic approximation errors in unsteady
  flow simulations,} {\em Computers \& Fluids\/}, Vol.~121, 2015, pp.~180--191.

\bibitem{muscl}
Van~Leer, B., \enquote{Towards the ultimate conservative difference scheme. V.
  A second-order sequel to Godunov's method,} {\em Journal of computational
  Physics\/}, Vol.~32, No.~1, 1979, pp.~101--136.

\bibitem{roe}
Roe, P.~L., \enquote{Approximate Riemann solvers, parameter vectors, and
  difference schemes,} {\em Journal of computational physics\/}, Vol.~43,
  No.~2, 1981, pp.~357--372.

\bibitem{pulliam}
Pulliam, T.~H. and Chaussee, D., \enquote{A diagonal form of an implicit
  approximate-factorization algorithm,} {\em Journal of Computational
  Physics\/}, Vol.~39, No.~2, 1981, pp.~347--363.

\bibitem{giles}
Giles, M.~B. and Pierce, N.~A., \enquote{An introduction to the adjoint
  approach to design,} {\em Flow, turbulence and combustion\/}, Vol.~65, No.
  3-4, 2000, pp.~393--415.

\bibitem{tikhonov}
Bishop, C.~M., \enquote{Training with noise is equivalent to {Tikhonov}
  regularization,} {\em Neural computation\/}, Vol.~7, No.~1, 1995,
  pp.~108--116.

\bibitem{Note1}
Assuming that the covariance matrices are Gaussian and diagonal.

\bibitem{rumsey2002prediction}
Rumsey, C.~L. and Ying, S.~X., \enquote{Prediction of high lift: review of
  present CFD capability,} {\em Progress in Aerospace Sciences\/}, Vol.~38,
  No.~2, 2002, pp.~145--180.

\bibitem{Note2}
$\delta ^\ast $ is the displacement thickness of the boundary layer and
  $\protect \frac {dP}{ds}$ is the pressure gradient.

\bibitem{mlpaper1}
Zhang, Z.~J. and Duraisamy, K., \enquote{Machine Learning Methods for
  Data-Driven Turbulence Modeling,} {\em 22nd AIAA Computational Fluid Dynamics
  Conference, AIAA Aviation, (AIAA 2015-2460), Dallas, TX\/}, Jun 2015.

\bibitem{helen_arxiv}
Zhang, Z., Duraisamy, K., and Gumerov, N., \enquote{Efficient Multiscale
  Gaussian Process Regression using Hierarchical Clustering,} {\em arXiv
  preprint {\tt arXiv:1511.02258v2}\/}, 2016.

\bibitem{mlbook}
Bishop, C.~M., {\em Pattern recognition and machine learning\/},
  Springer-Verlag New York, 1st ed., 2006.

\bibitem{Note3}
We also appreciate that other techniques such as support vector and polynomial
  regressors can be as scalable as NNs and thus, the choice of NNs is based on
  prior experience rather than on objective considerations.

\bibitem{fann}
Nissen, S., \enquote{Implementation of a fast artificial neural network library
  (fann),} {\em {\tt http://leenissen.dk/fann/}, Department of Computer
  Science, University of Copenhagen\/}, 2003.

\bibitem{somers805}
Somers, D.~M., \enquote{Design and experimental results for the {S805}
  Airfoil,} {\em NREL Report, NREL/SR-440-6917\/}, 1997.

\bibitem{somers809}
Somers, D.~M., \enquote{Design and experimental results for the {S809}
  Airfoil,} {\em NREL Report, NREL/SR-440-6918\/}, 1997.

\bibitem{somers814}
Somers, D.~M., \enquote{Design and experimental results for the {S814}
  Airfoil,} {\em NREL Report, NREL/SR-440-6919\/}, 1997.

\bibitem{Hughes1989}
Hughes, T. J.~R., France, L.~P., and Hulbert, G.~M., \enquote{A new finite
  element formulation for computational fluid dynamics.VIII. The
  Galerkin/least-squares method for advective-diffusive equations,} {\em
  Computer Methods in Applied Mechanics and Engineering\/}, Vol.~73, No.~2,
  1989, pp.~173--189.

\bibitem{Shakib1991}
Shakib, F., Hughes, T. J.~R., and Johan, Z., \enquote{A new finite element
  formulation for computational fluid dynamics. X. The compressible Euler and
  Navier-Stokes equations,} {\em Computer Methods in Applied Mechanics and
  Engineering\/}, Vol.~89, No. 1--3, 1991, pp.~141--219.

\bibitem{Corson2012}
Corson, D.~A., Griffith, D.~T., Ashwill, T., and Shakib, F.,
  \enquote{Investigating aeroelastic performance of multi-megawatt wind turbine
  rotors using CFD,} {\em 53$^{rd}$ AIAA/ASME/ASCE/AHS/ASC Structures,
  Structural Dynamics and Materials Conference, Honolulu, Hawaii\/}, , No.
  2012--1827, 2012.

\bibitem{AcuSolve2016}
Corson, D.~A., Zamora, A., and Medida, S., \enquote{A Comparative Assessment of
  Correlation-based Transition Models for Wind Power Applications,} {\em
  34$^{th}$ AIAA Applied Aerodynamics Conference, Washington. D. C.\/}, , No.
  2016-3129, 2016.

\bibitem{Godo2010}
Godo, M.~N., Corson, D., and Lgensky, S.~M., \enquote{A Comparative Aerodynamic
  Study of Commercial Bicycle Wheels using CFD,} {\em 48$^{th}$ AIAA Aerospace
  Sciences Meeting, FL\/}, , No. 2010-1431, 2010.

\bibitem{Lyons2009}
Lyons, D.~C., Peltier, L.~J., Zajaczkowski, F.~J., and Paterson, E.~G.,
  \enquote{Assessment of DES Models for Separated Flow From a Hump in a
  Turbulent Boundary Layer,} {\em Journal of Fluids Engineering\/}, Vol.~131,
  No.~11, 2009, pp.~111203--1--111203--9.

\bibitem{Bagwell2006}
Bagwell, T.~G., \enquote{CFD Simulation of Flow Tones From Grazing Flow Past a
  Deep Cavity,} {\em Proceedings of 2006 ASME International Mechanical
  Engineering Congress and Exposition, Chicago, IL\/}, , No. 2006-15633, 2006,
  pp.~105--114.

\bibitem{Johnson2002}
Johnson, K. and Bittorf, K., \enquote{Validating the Galerkin Least Square
  Finite Element Methods in Predicting Mixing Flows in Stirred Tank Reactors,}
  {\em Proceedings of CFD 2002, The 10th Annual Conference of the CFD Society
  of Canada\/}, , No. 40048419, 2002, pp.~490--496.

\bibitem{Jansen2000}
Jansen, K.~E., Whiting, C.~H., and Hulbert, G.~M., \enquote{A generalized-alpha
  method for integrating the filtered Navier-Stokes equations with a stabilized
  finite element method,} {\em Computer Methods in Applied Mechanics and
  Engineering\/}, Vol.~190, No. 3--4, 2000, pp.~305--319.

\bibitem{ling2}
Ling, J., Jones, R., and Templeton, J., \enquote{Machine learning strategies
  for systems with invariance properties,} {\em Journal of Computational
  Physics\/}, Vol.~318, 2016, pp.~22--35.

\bibitem{BFGS}
Dennis, Jr, J.~E. and Mor{\'e}, J.~J., \enquote{Quasi-Newton methods,
  motivation and theory,} {\em SIAM review\/}, Vol.~19, No.~1, 1977,
  pp.~46--89.

\bibitem{Note4}
${\protect \ensuremath {\protect \bf { R}}}$ and ${\protect \ensuremath
  {\protect \bf { \psi }}}$ are of dimension $5 N_m$.

\bibitem{tapenade}
Hascoet, L. and Pascual, V., \enquote{The Tapenade Automatic Differentiation
  tool: principles, model, and specification,} {\em ACM Transactions on
  Mathematical Software (TOMS)\/}, Vol.~39, No.~3, 2013, pp.~20.

\bibitem{langleybump}
Naughton, J.~W., Viken, S.~A., and Greenblatt, D., \enquote{Skin-Friction
  Measurements on the {NASA} Hump Model,} {\em AIAA Journal\/}, Vol.~44, No.~6,
  2016, pp.~1255--1265.

\end{thebibliography}
\bibliographystyle{aiaa}

\section*{Appendix A : Discrete adjoint method for field inversion}
\label{appendix:a}
The formulation and application of field inversion techniques to turbulence modeling problems is provided in Refs.~\citenum{paradigm,PoF_anand}. A brief description is presented herein for completeness. The optimization procedure to minimize Eq.~\ref{eq:obj_cp} or Eq.~\ref{eq:obj_cl} uses a gradient-based Quasi-Newton method employing the limited memory Broyden-
Fletcher-Goldfarb-Shanno  algorithm~\cite{BFGS}. Since the optimization problem is extremely high dimensional (as the number of parameters equals the number of control volumes $N_m$), an adjoint approach is required to efficiently compute
gradients. In the adjoint technique, given an objective function $\mathcal{J}$ that we wish to minimize, the total derivative with respect to the parameter vector $\be = \{\be(x_1), \be(x_2),...,\be(x_{N_m})\}^T$  is given by
\begin{eqnarray}
\frac{d \mathcal{J}}{d\be} =  \frac{\partial\mathcal{J}}{\partial \be} + \mat{\psi}^T\frac{\partial \mat{R}}{ \partial \be}.
\label{eq:appendix:1}
\end{eqnarray}
In the above equation~\footnote{ $\mat{R}$ and $\mat{\psi}$ are of dimension $5  N_m$}, $\mat{R}$ represents the governing equations and $\mat{\psi}$ is the vector of adjoint variables, which is determined by
\begin{eqnarray}
\left[\frac{\partial \mat{R}}{\partial \mat{{U}}}\right]^T \mat{\psi} = -\left[\frac{\partial \mathcal{J}}{\partial \mat{{U}}}\right]^T.
\label{eq:appendix:2}
\end{eqnarray}
The partial derivatives in Eq. \ref{eq:appendix:2} are calculated using the tapenade~\cite{tapenade} tool for automatic differentiation and the system is solved using pseudo time stepping. Since \be is explicitly present only as a multiplier to the production term $\mat{P}$,
the expression for the gradient is given by
\begin{eqnarray}
        \frac{d \mathcal{J}}{d\be} =  \frac{\partial\mathcal{J}}{\partial \be} - \mat{\psi}_{\tilde{\nu}}^T \mat{P},
\end{eqnarray}
where $\mat{\psi}_{\tilde{\nu}}$ represents the adjoint variable corresponding to the working variable of the SA model.

\section*{Appendix B: Inverse modeling for separated flows}
\label{appendix:b}
\begin{figure}
\centering
\subfigure[Wall pressure coefficient]{\includegraphics[width=0.4\textwidth,trim={.25cm .25cm .25cm .25cm},clip]{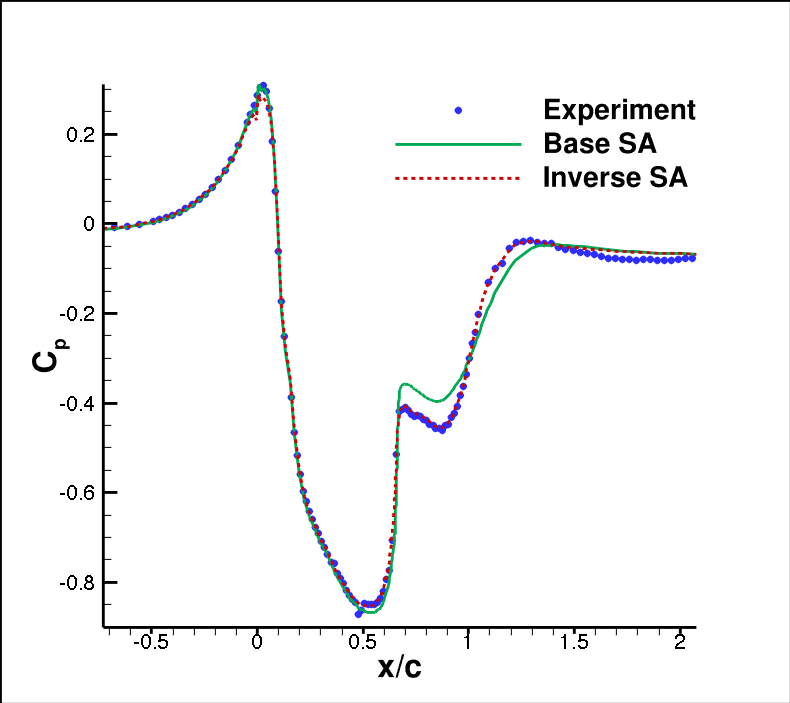}}
\subfigure[Reynolds shear stress contours $\overline{u_1^\prime u_2^\prime}/U_\infty^2$]{\includegraphics[width=0.58\textwidth,trim={.5cm .5cm .5cm .5cm},clip]{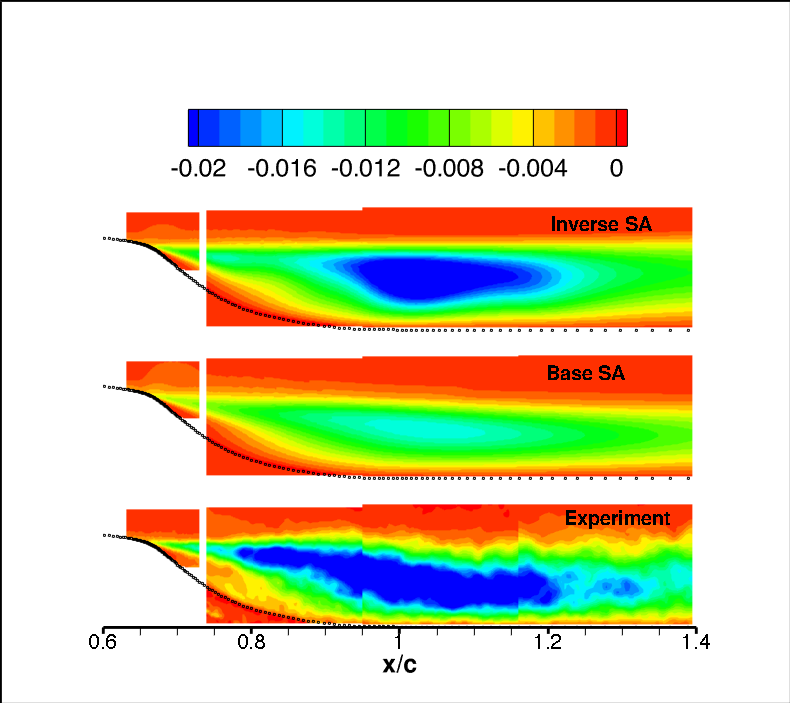}}
\vspace{-4mm}
\caption{ Application of inverse modeling to separated flow over a smooth surface.}
\label{langleybump}
\end{figure}

As mentioned in section III, inverse modeling is used to provide quantitative
information about the model discrepancy. Machine learning-based augmentations were based on inverse fields which were generated using lift data. 
This was confirmed to also result in good surface pressure predictions.  
To assess the generality of the procedure and its impact on other field quantities, the inverse problem was applied 
to a NASA benchmark test~\cite{langleybump}  (Fig.~\ref{langleybump}), which involves separated flow over a smooth hump.
In this problem, $\be(x,y)$ was inferred with the objective of matching the wall pressure distribution in the region $0.5 \leq x/c \leq 1.5$. 
Even though the objective function was only taken to be the surface pressure,  improvement is seen in the Reynolds stress predictions
(Fig.~\ref{langleybump}c). 
As a consequence of the overall improvement in the field solution, the  the predicted length of the separation bubble was found be 15\% more accurate compared to the baseline solution.
Results can be improved by considering more information from the experiment~\cite{PoF_anand}, 
but this exercise offers further evidence that for separated flows, pressure data is valuable in inferring model discrepancy.

\section*{Appendix C: Spalart--Allmaras Model}
The one-equation Spalart--Allmaras (S--A) turbulence model \cite{spalart} is used for the work presented in this paper. The S-A model solves for the modified Eddy viscosity, $\tilde{\nu}$, which relates to the kinematic Eddy viscosity $\nu_t$ as follows:

\begin{equation}\nu_t = \tilde{\nu}f_{v1}; \quad f_{v1} = \dfrac{\chi^3}{\chi^3 + c_{v1}^3}; \quad \chi= \dfrac{\tilde{\nu}}{\nu}\end{equation}

\noindent The governing equation of the S-A model without the trip terms is given by:
\begin{equation} \label{samodel} \dfrac{D\tilde{\nu}}{Dt} = P - D + \dfrac{1}{\sigma}\left[\nabla . ((\nu + \tilde{\nu})\nabla \tilde{\nu}) + c_{b2}(\nabla \tilde{\nu})^2\right]  \end{equation}
where, $P$ and $D$ are the production and destruction terms of $\tilde{\nu}$, given by:
\begin{equation} P = c_{b1}\tilde{\Omega}\tilde{\nu} \quad \mathrm{and} \quad  D = c_{w1}f_w [\dfrac{\tilde{\nu}}{d}]^2 \end{equation}
\(\tilde{\Omega}\) is a function of the vorticity magnitude, $\Omega$, and is defined as:
\begin{equation} \label{omega_tilde} \tilde{\Omega} = \Omega + \dfrac{\tilde{\nu}}{\kappa^2 d^2}f_{v2}, \quad f_{v2} = 1 - \dfrac{\chi}{1+\chi f_{v1}} \end{equation}
The function \(f_w\) is defined as:
\begin{equation} \label{f_w} f_w = g\left[\dfrac{1+c_{w3}^6}{g^6 + c_{w3}^6}\right]^{\frac{1}{6}}, \quad g = r + c_{w2}(r^6-r), \quad r = \dfrac{\tilde{\nu}}{\tilde{\Omega}\kappa^2 d^2} \end{equation}

\noindent The model constants are: \(c_{b1} = 0.1355, \sigma = 2/3, c_{b2} = 0.622, \kappa = 0.41, c_{w1} = c_{b1}/\kappa^2 + (1+c_{b2})/\sigma, c_{w2} = 0.622, c_{w3} = 2.0, c_{v1} = 7.1.\)

\end{document}